%% file: ms.tex
\PassOptionsToPackage{usenames,dvipsnames}{xcolor}

\documentclass[sigconf,authorversion]{acmart}
\AtBeginDocument{%
  \providecommand\BibTeX{{%
    \normalfont B\kern-0.5em{\scshape i\kern-0.25em b}\kern-0.8em\TeX}}}

\acmDOI{}
\acmConference[I3D'22]{SIGGRAPH I3D}{May 2022}{} 
\acmYear{2022}
\copyrightyear{2022}

\usepackage{multicol}
\usepackage{tikz}
\usepackage[export]{adjustbox}
\usetikzlibrary{positioning}
\usepackage{comment}
\usepackage{mathtools}

\usepackage[linesnumbered,ruled,vlined]{algorithm2e} % For algorithms

\SetAlFnt{\small}
\SetAlCapFnt{\small}
\SetAlCapNameFnt{\small}
\SetAlCapHSkip{0pt}

\definecolor{myblue}{rgb}{0.0,0.0,0.75}
\definecolor{myred}{rgb}{0.9,0,0}
\definecolor{mygreen}{rgb}{0,0.6,0}

\SetCommentSty{mycommfont}

\SetKwInput{KwInput}{Input}                % Set the Input
\SetKwInput{KwOutput}{Output}              % set the Output

\begin{document}

\title{Bringing Linearly Transformed Cosines to Anisotropic GGX}

\settopmatter{authorsperrow=4}
\author{Aakash KT}
\affiliation{CVIT, KCIS, IIIT-Hyderabad \country{}}

\author{Eric Heitz}
\affiliation{Unity Technologies \country{}}

\author{Jonathan Dupuy}
\affiliation{Unity Technologies \country{}}

\author{P. J. Narayanan}
\affiliation{CVIT, KCIS, IIIT-Hyderabad \country{}}

\renewcommand{\shortauthors}{KT et al.}

\begin{abstract}
Linearly Transformed Cosines (LTCs) are a family of distributions that are used for real-time area-light shading thanks to their analytic integration properties. 
Modern game engines use an LTC approximation of the ubiquitous GGX model, but currently this approximation only exists for isotropic GGX and thus anisotropic GGX is not supported. 
While the higher dimensionality presents a challenge in itself, we show that several additional problems arise when fitting, post-processing, storing, and interpolating LTCs in the anisotropic case.
Each of these operations must be done carefully to avoid rendering artifacts.
We find robust solutions for each operation by introducing and exploiting invariance properties of LTCs. 
As a result, we obtain a small $8^4$ look-up table that provides a plausible and artifact-free LTC approximation to anisotropic GGX and brings it to real-time area-light shading.
\end{abstract}

%%\begin{CCSXML}
%%<ccs2012>
%%<concept>
%%<concept_id>10010147.10010371.10010372.10010376</concept_id>
%%<concept_desc>Computing methodologies~Reflectance modeling</concept_desc>
%%<concept_significance>500</concept_significance>
%%</concept>
%%</ccs2012>
%%\end{CCSXML}

%%\ccsdesc[500]{Computing methodologies~Reflectance modeling}

%%\keywords{real-time rendering, area lighting, BRDF}

\begin{teaserfigure}
\begin{center}
\begin{tabular}{@{\hspace{-0.5mm}} c @{\hspace{1mm}} c @{}}
Area lighting with isotropic materials~\cite{heitz2016} & Area lighting with anisotropic materials (ours) \\
\includegraphics[width=0.5\linewidth,frame]{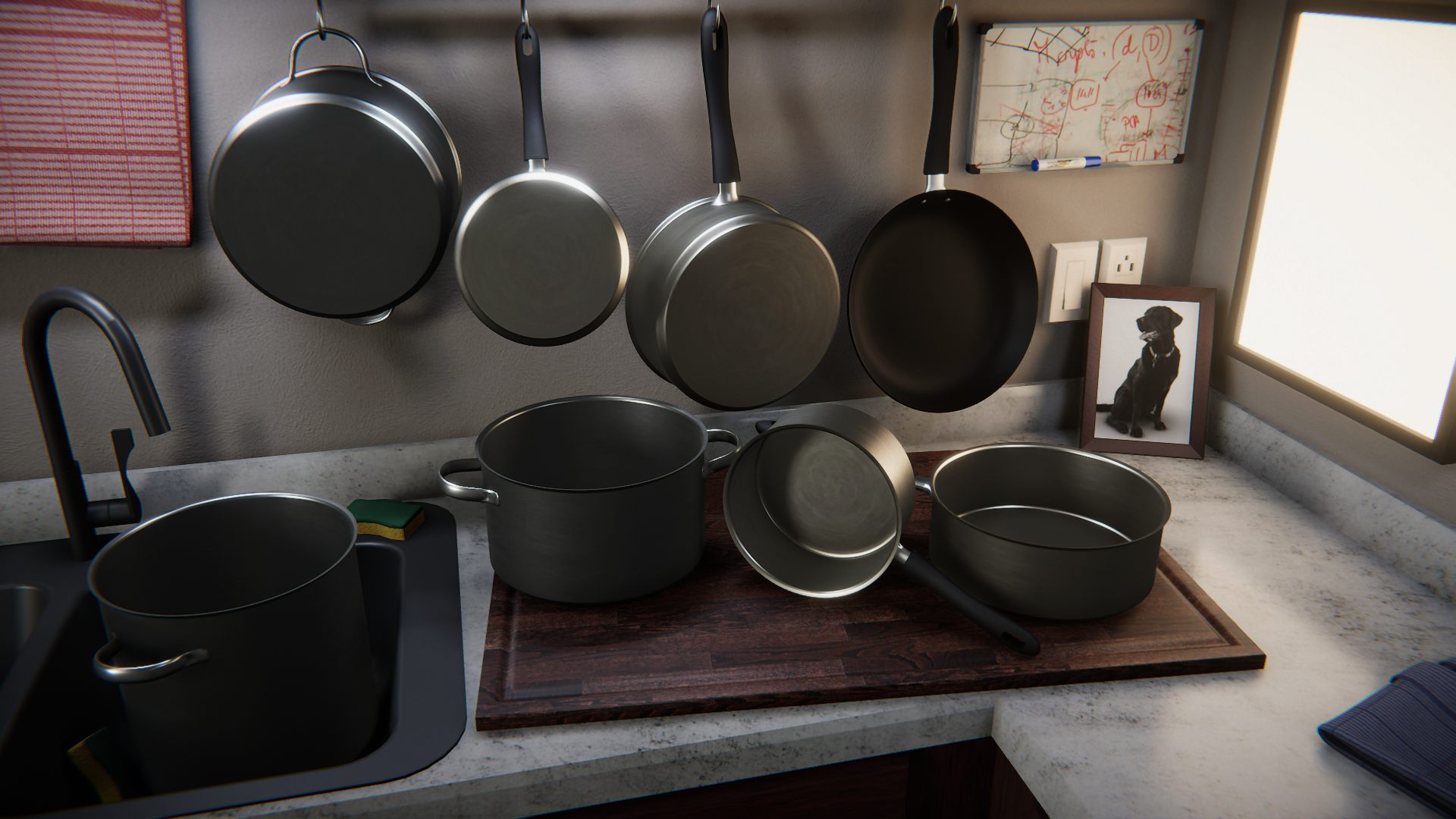} &
\includegraphics[width=0.5\linewidth,frame]{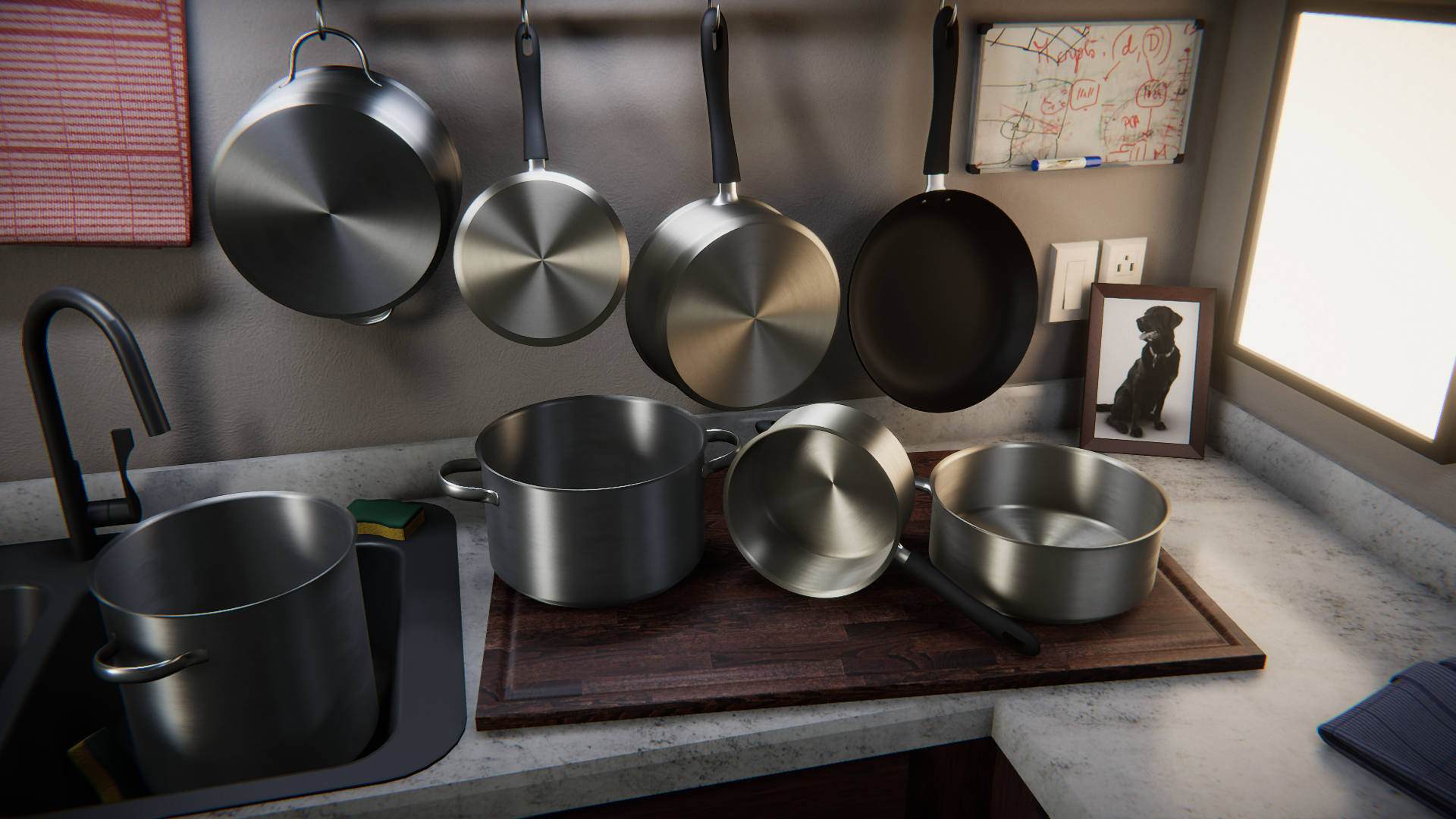}
\end{tabular}
\end{center}
\vspace{-5mm}
\caption{\label{fig:teaser} Real-time area lighting with Linearly Transformed Cosines (LTCs) in a commercial game engine. 
\emph{Our LTC approximation for anisotropic materials takes 0.7 ms at 1080p resolution on an NVIDIA GeForce RTX 2080 GPU.
}
}
\vspace{3mm}
\end{teaserfigure}

\maketitle

\textbf{Acknowledgements.} Thanks to Thomas Deliot for making a prototype in the Unity engine and to Stephen Hill for his feedback. Aakash KT was funded by the "Kohli Fellowhip" of KCIS.

\input{section_introduction.tex}

\input{section_related_work.tex}

\input{section_background.tex}

\input{section_fitting.tex}

\clearpage

\input{section_interpolation.tex}

\input{section_symmetry.tex}

\input{section_inversion.tex}

\input{section_discretization.tex}

\input{section_implementation.tex}

\input{section_results.tex}

\input{section_conclusion.tex}

\clearpage

% \begin{acks}
% \end{acks}

\bibliographystyle{ACM-Reference-Format}
\bibliography{ms.bbl}

% \appendix

\end{document}

%% file: section_introduction.tex
\section{Introduction}
\label{sec:introduction}

Today's physically based shading models are largely based on the GGX \textit{Bidirectional Reflectance Distribution Function} (BRDF)~\cite{walter2007,hill2020}.
In real-time engines, computing direct illumination requires integrating the BRDF-light product. 
Dedicated techniques have been developed to integrate the GGX BRDF against different kinds of lights (probes, area lights, volumes, etc.).
In this paper, we focus on \textit{Linearly Transformed Cosines} (LTCs), which have been widely adopted as a means to integrate the GGX BRDF against area lights of various shapes~\cite{heitz2016,heitz2017a,heitz2017b}.
For instance, LTCs are used in the Unity and Unreal engines for this purpose~\cite{benyoub2019,wassmer2018}.
However, their support is currently limited to the isotropic GGX BRDF, and thus anisotropic materials such as brushed metals cannot be shaded under area lighting with LTCs (see Figure~\ref{fig:teaser}).
The objective of this work is to alleviate this limitation and bring real-time area lighting to the anisotropic GGX BRDF via LTCs. 

\begin{figure}[!h]
    \centering
    \begin{tikzpicture}
        \draw (0.0, 0) node[anchor=west] {\includegraphics[height=0.2\linewidth]{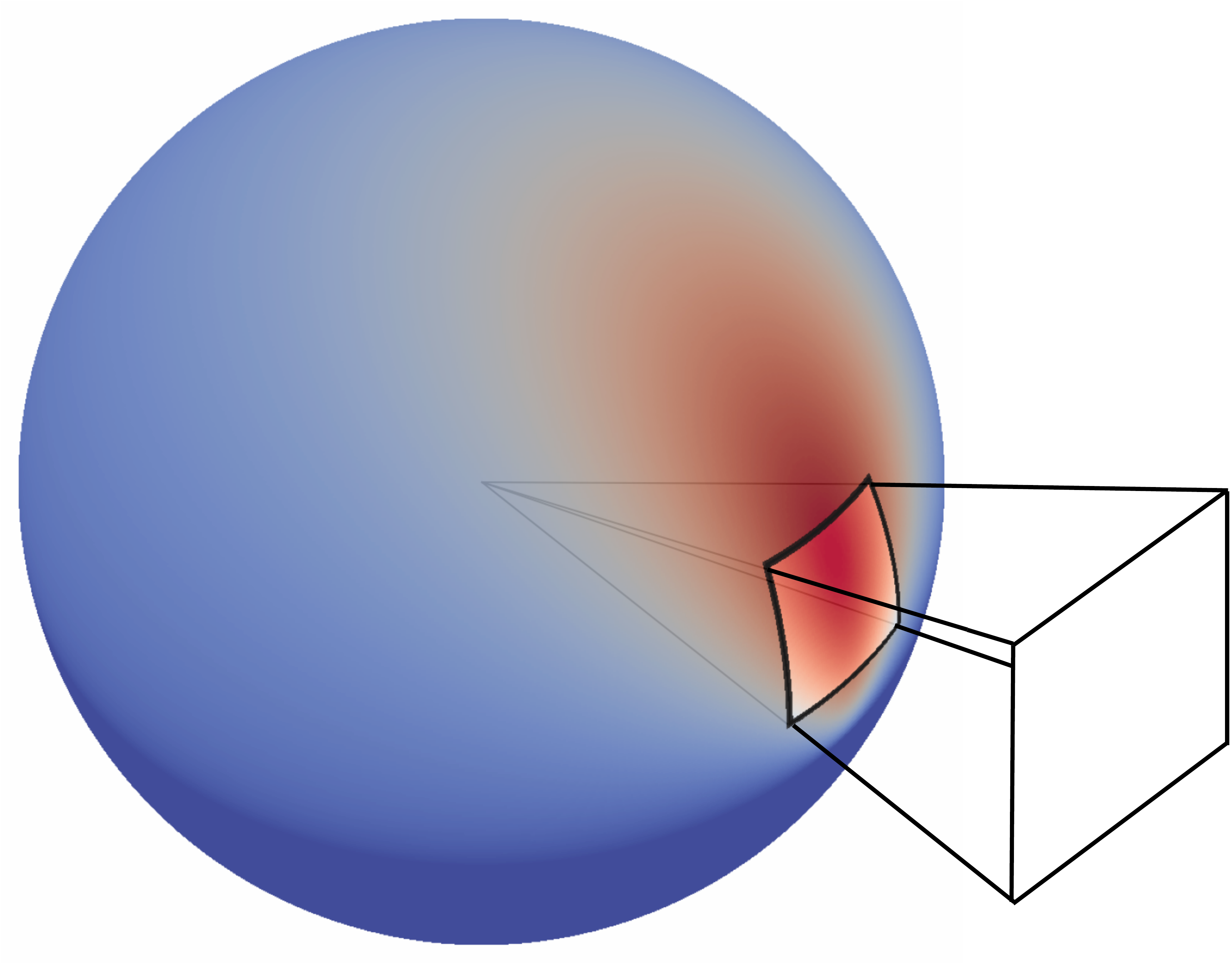}};
        \draw (3.0, 0) node[anchor=west] {\includegraphics[height=0.2\linewidth]{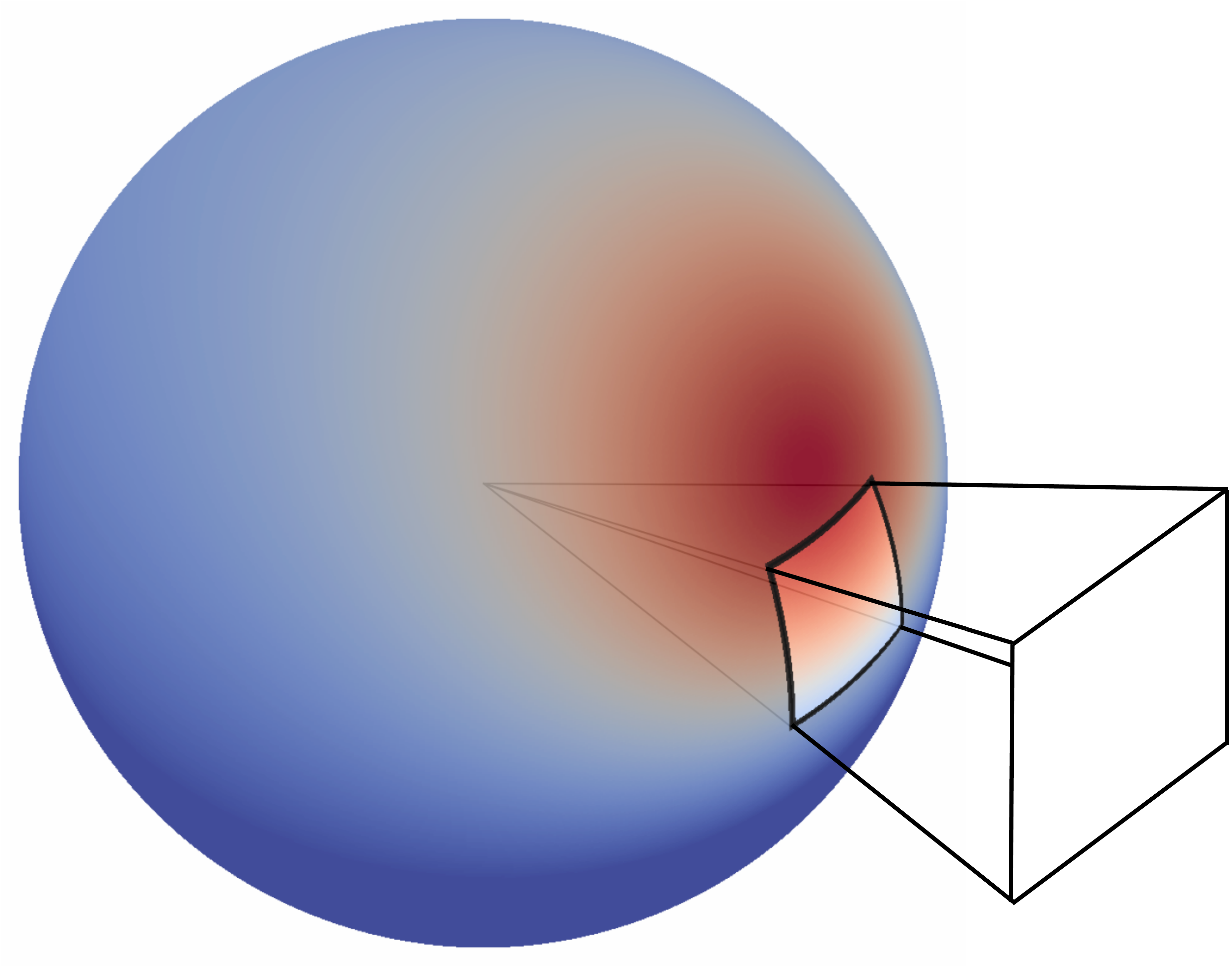}};
        \draw (6.0, 0) node[anchor=west] {\includegraphics[height=0.2\linewidth]{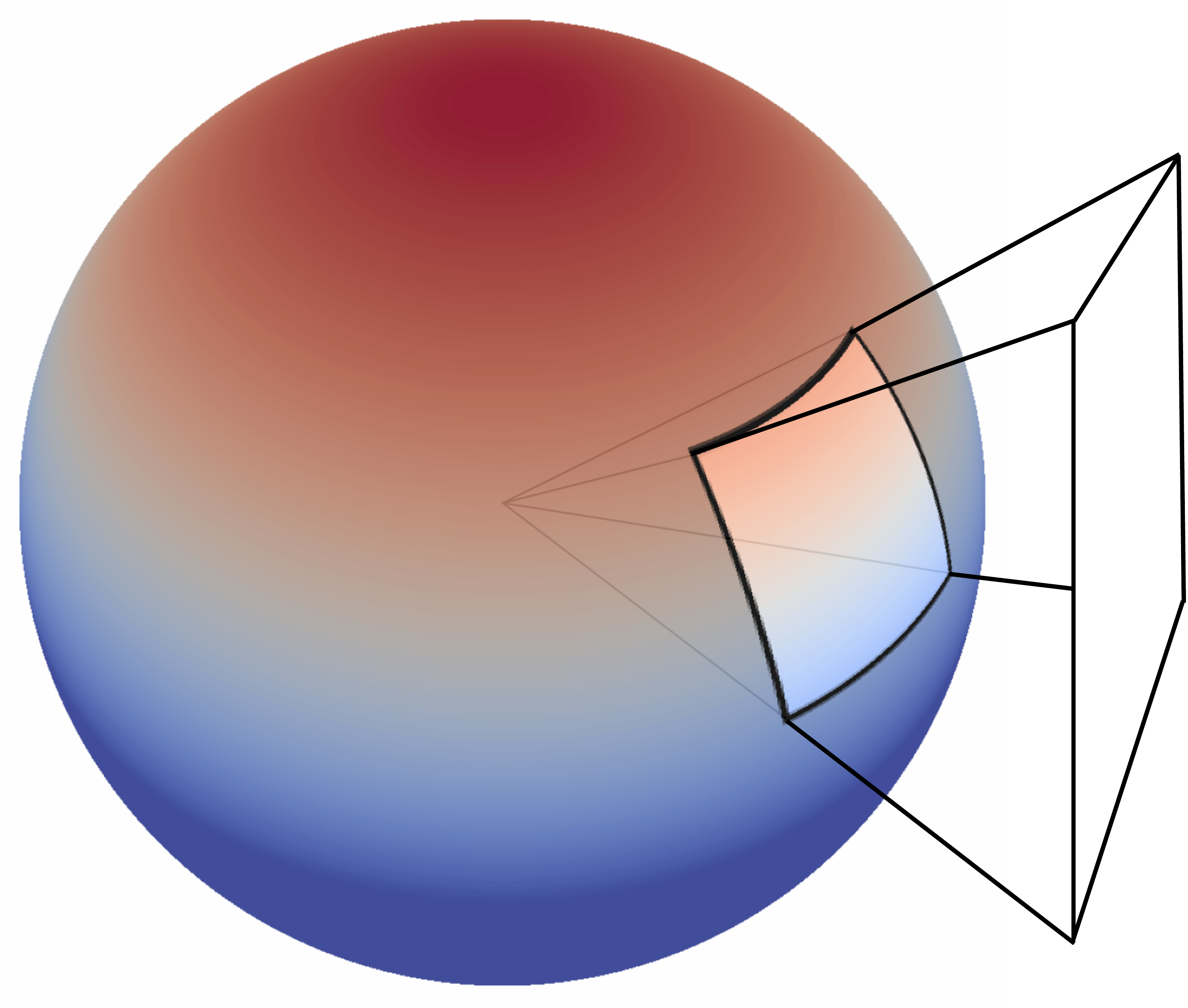}};
        \draw (2.7, 0.0) node {$\approx$};
        \draw (5.7, 0.5) node {\footnotesize $M$};
        \draw (5.7, 0.2) node {$\longleftarrow$};
        \draw (5.7, -0.4) node {$\longrightarrow$};
        \draw (5.7, -0.7) node {\footnotesize $M^{-1}$};
        \draw (0.3, 1.1) node[anchor=west] {\small (a) GGX};
        \draw (3.3, 1.1) node[anchor=west] {\small (b) LTC};
        \draw (6.3, 1.1) node[anchor=west] {\small (c) Cosine};
        \draw (0.4, -1.1) node[anchor=west] {\tiny \color{myred}not analytic};
        \draw (3.6, -1.1) node[anchor=west] {\tiny \color{mygreen}analytic};
        \draw (6.6, -1.1) node[anchor=west] {\tiny \color{mygreen}analytic};
    \end{tikzpicture}
    \vspace{-4mm}
    \caption{\label{fig:ltc_intro}
    \emph{(a) A GGX lobe cannot be analytically integrated over the spherical domain covered by the area light.
    (b) An LTC represented by a matrix $M$ provides a good approximation to the GGX lobe, and the integral equals (c) the analytic integral of a cosine lobe over the light transformed by $M^{-1}$.
    }
    }   
     \vspace{-20mm}
\end{figure}
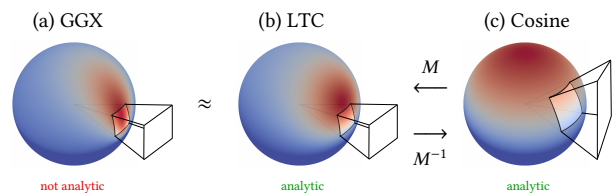

More specifically, LTCs are spherical distributions with analytic integration properties over specific spherical domains. 
Thanks to the LTC approximation of GGX, the integral of the BRDF over the spherical domain covered by an area light can be computed analytically in real time (Fig.~\ref{fig:ltc_intro}).
LTCs are represented by $3\times3$ matrices $M$ fitted to isotropic GGX lobes and stored in a small 2D look-up table~\cite{heitz2016}.
Computing a similar look-up table for anisotropic GGX raises new challenges, which is the focus of this work.

\clearpage
\paragraph{Objective.}

The crux of the problem is to obtain the $3\times3$ matrix $M$ of the LTC that best approximates a given GGX lobe.
Previously, Heitz et al.~\cite{heitz2016} proposed a fitting approach to compute a 2D look-up table 
\begin{align}
\label{eq:lut_iso}
M = T_{\text{isoGGX}}(\theta, \alpha)
\end{align}
that approximates GGX lobes defined by the incidence angle $\theta$ and a roughness coefficient $\alpha$.
This is sufficient to cover the full isotropic GGX BRDF.
Our objective is to compute a similar 4D look-up table 
\begin{align}
\label{eq:lut_aniso}
M = T_{\text{anisoGGX}}(\theta, \phi, \alpha_x, \alpha_y)
\end{align}
that takes an additional azimuthal angle $\phi$ and anisotropic roughness coefficients $(\alpha_x, \alpha_y)$.
Note that once the matrix $M$ is obtained, all of the existing applications of LTCs (area-light integration with various shapes, importance sampling, etc.) can be used without further modification.
Thus, the only problem to solve is the precomputation of the 4D look-up table $T_{\text{anisoGGX}}$.

\paragraph{Contributions.}

\begin{figure}[!h]
\centering
\begin{tabular}{@{} c @{\hspace{5mm}} c @{}}
(a) {\color{myred} broken fitted entry} (Sec.~\ref{sec:fitting}) &
(b) {\color{myred} ill-defined interpolation} (Sec.~\ref{sec:interpolation}) \\
\includegraphics[height=0.4\linewidth]{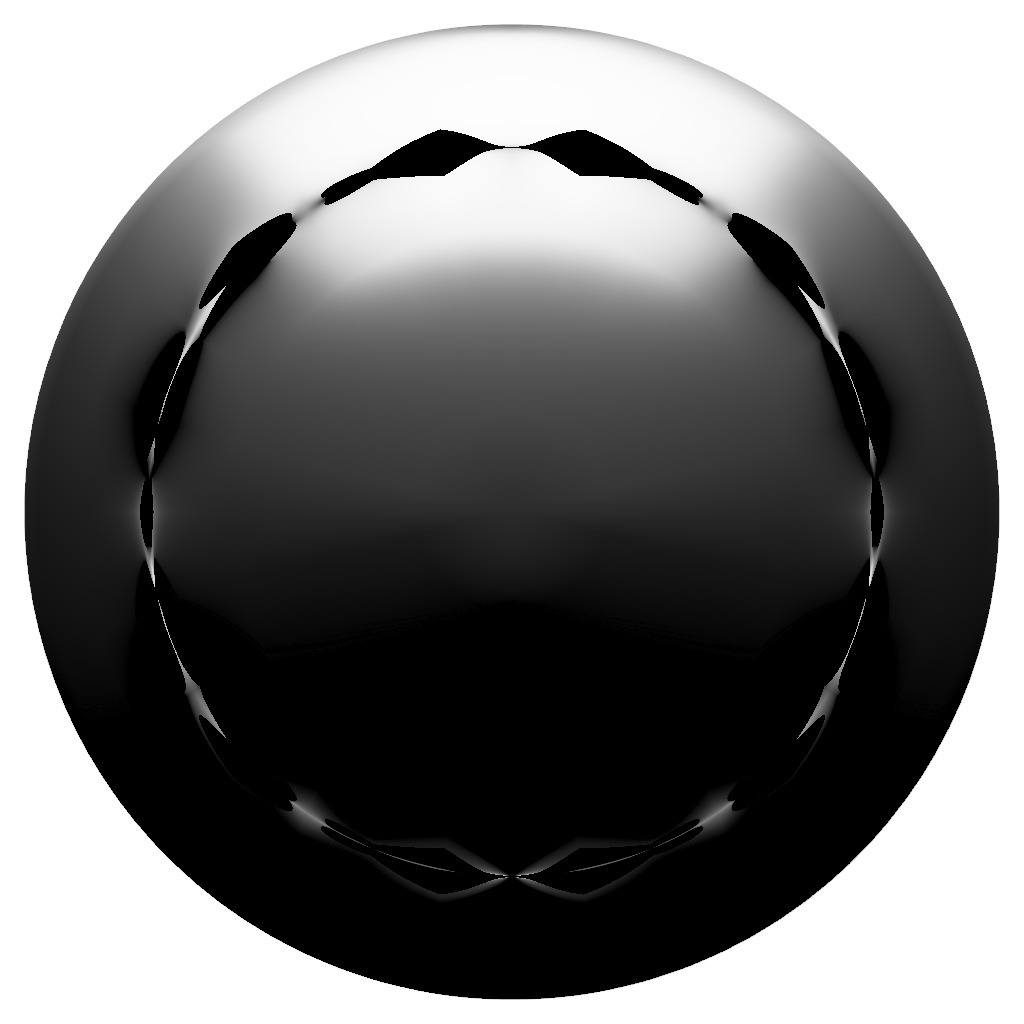} &
\includegraphics[height=0.4\linewidth]{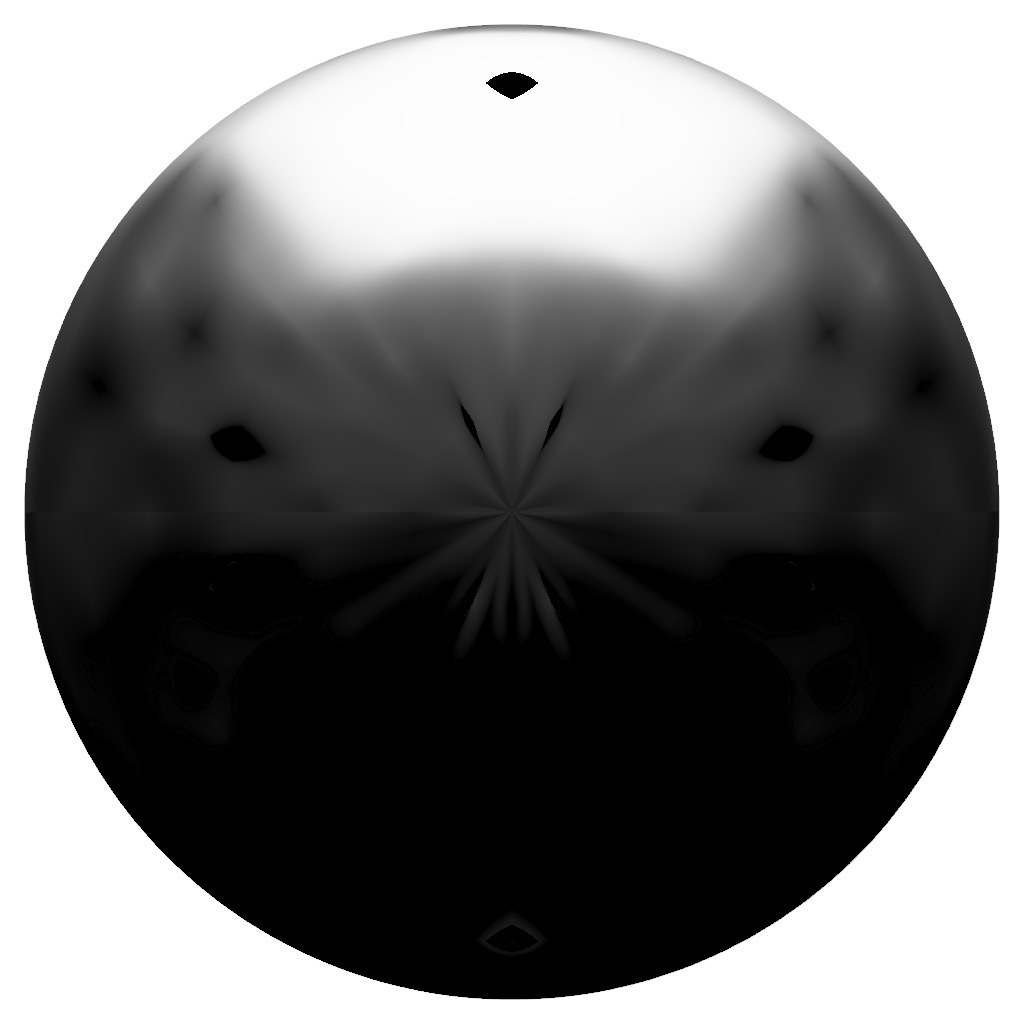} \\
(c) {\color{myred} broken symmetry} (Sec.~\ref{sec:symmetries}) &
(d) {\color{myred} early inversion} (Sec.~\ref{sec:inversion}) \\
\includegraphics[height=0.4\linewidth]{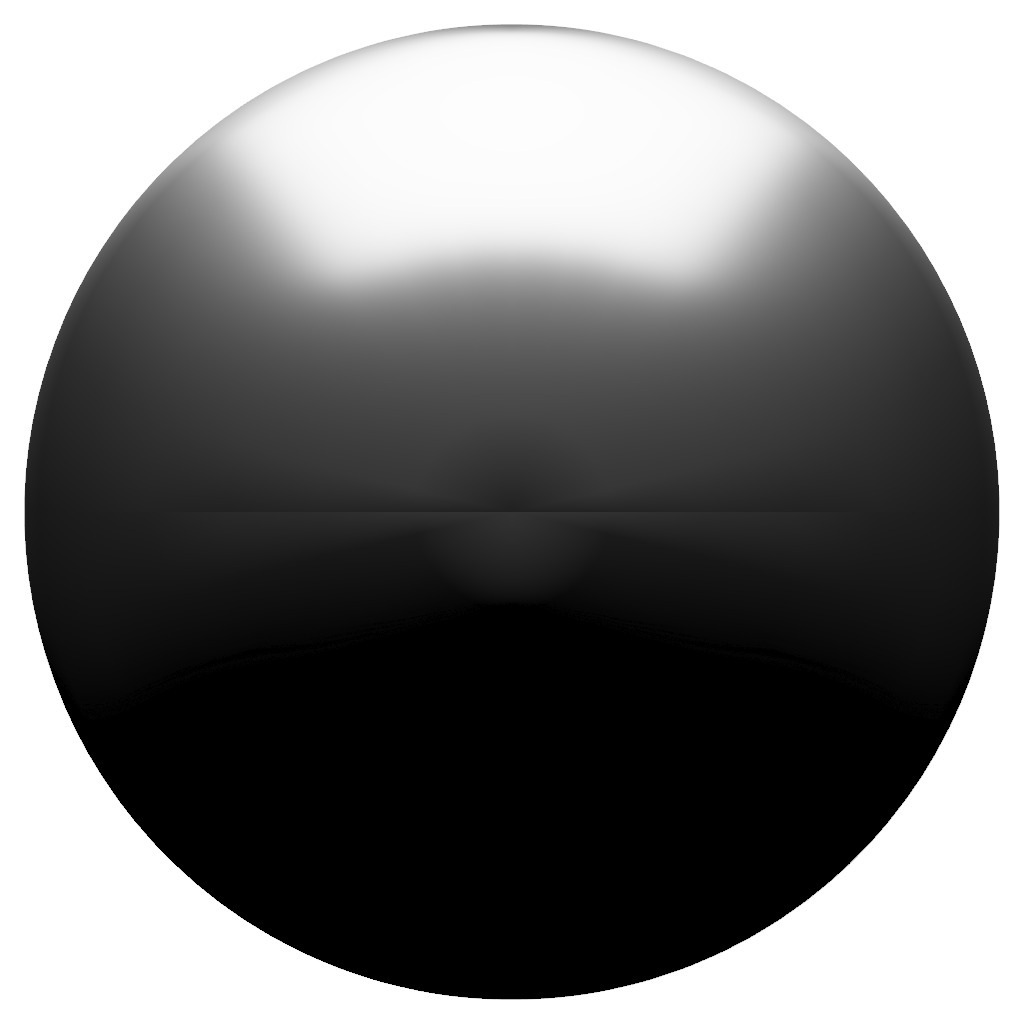} &
\includegraphics[height=0.4\linewidth]{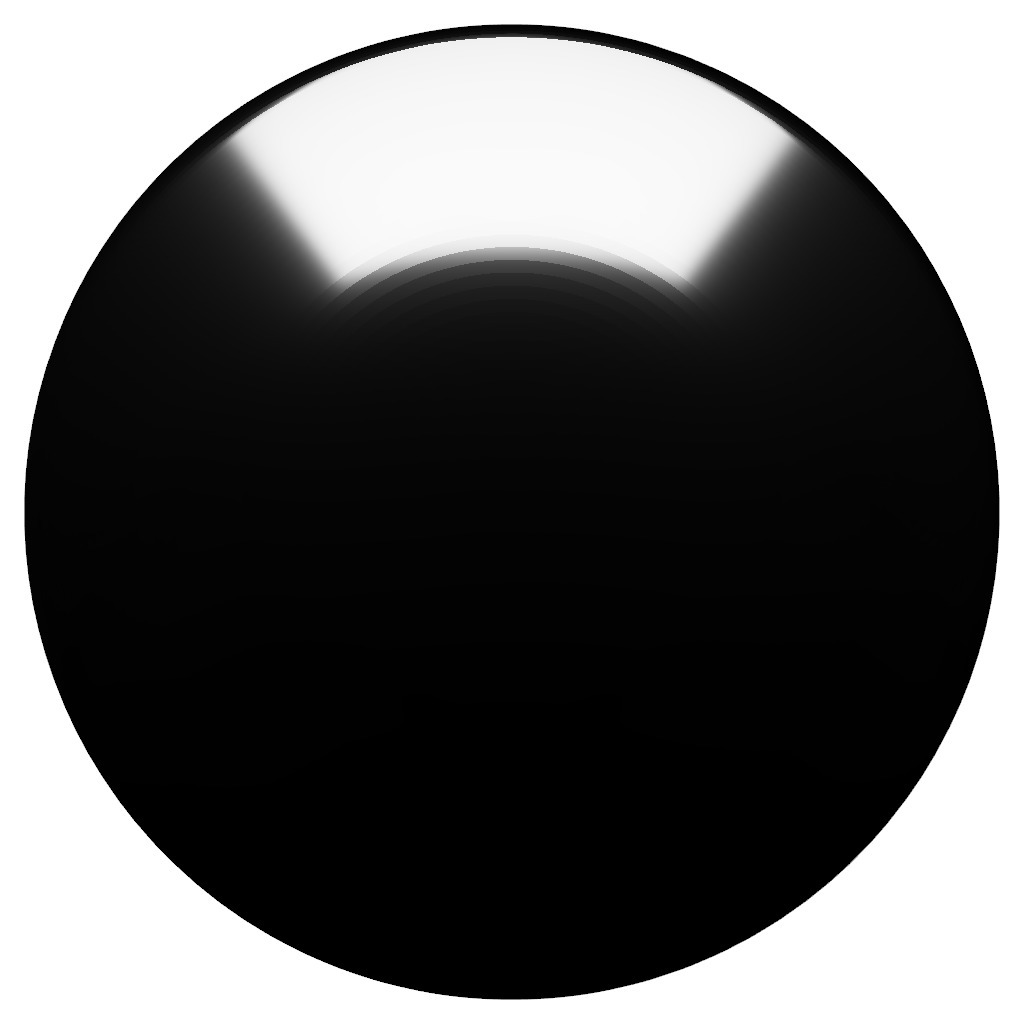}
\end{tabular}
\vspace{-4mm}
\caption{\label{fig:problems_intro} Illustration of the problems to overcome.}
\end{figure}

The difficulty is that the fitting approach employed by Heitz et al. to compute $T_{\text{isoGGX}}$ cannot simply be extended to compute $T_{\text{anisoGGX}}$. 
Indeed, in the isotropic case, the full dimensionality of LTCs is not used and this avoids several problems that arise in the anisotropic case. 
The artifacts highlighted in Figure~\ref{fig:problems_intro} show that successfully bringing LTCs to anisotropic GGX requires \textit{robust fitting} (Sec.~\ref{sec:fitting}), \textit{well-defined interpolation} (Sec.~\ref{sec:interpolation}), \textit{valid symmetries} (Sec.~\ref{sec:symmetries}) and \textit{accurate storage} (Sec.~\ref{sec:inversion} and \ref{sec:discretization}).
We introduce new mathematical properties of LTCs, such as non-uniqueness and axial symmetries, that are required to understand and overcome these failure cases.
The final outcome of our method is a 4D look-up table that yields a plausible and artifact-free LTC approximation to anisotropic GGX and is small enough to be used in real time. 
We validate this table in the context of area-light shading with anisotropic GGX materials.

%% file: section_related_work.tex
\section{Related Work}
\label{sec:related_work}

\paragraph{Real-time stochastic techniques.}

Recent graphics hardware makes it possible to use purely stochastic techniques such as reservoir sampling~\cite{bitterli2020}.
With stochastic approaches, integrating area lighting with arbitrary materials is simple but leads to noisy results.
In the context of this paper, we instead aim to provide an \emph{analytic shading} technique that produces a clean (noise-free) image.

\paragraph{Real-time analytic shading techniques.}

Even though stochastic techniques are appealing for the future, analytic methods remain important for today's real-time graphics.
The first analytic solution to area lighting dates back to Lambert, who derived the irradiance from a polygonal light~\cite{lambert1760}.
This early formula was brought to graphics for radiosity by Baum et al.~\cite{baum1989} and was later extended by Arvo~\shortcite{arvo1995}, who derived the integral over spherical polygons of cosines of arbitrary integer exponents (i.e., \textit{Phong distributions}).
Despite the detailed implementation of this technique provided by Snyder~\cite{snyder1996}, it had limited practical impact due to the algorithmic complexity of the integration.
In practice, real-time methods involved cheap approximations, mainly based on punctual evaluations~\cite{wang2008,drobot2014,lagarde2014}.
As GPUs became more powerful, more accurate techniques arose, such as the approach of Lecocq et al., which was the first method to provide an accurate approximation for physically based materials while still being fast enough for real-time rendering~\cite{lecocq2015,lecocq2016}.
This method was later outperformed by \emph{Linearly Transformed Cosines} (LTCs)~\cite{heitz2016}, which remain today's leading approach for real-time area-light shading.
We build on the state-of-the-art LTC method by adding support of anisotropic materials.

\paragraph{Applications of Linearly Transformed Cosines (LTCs).}

While LTCs were initially proposed for real-time polygonal-light shading, they have since found uses in many applications that will benefit from our contribution. 
The LTC analytic integration has been extended to other types of light, such as line lights~\cite{heitz2017a} and sphere/disk lights~\cite{heitz2017b}.
Another important addition to LTC integration is shadowing. 
Though LTCs do not include visibility in the analytic shading integral, a low-variance ratio estimator has been proposed to incorporate shadows on top of the analytic shading integral~\cite{heitz2018}.
An alternative approach consists of incorporating visibility by removing the edges of occluders in the LTC integral~\cite{zhou2021,kt2021}. 
The analytic integration property of LTCs has also proven useful in offline rendering, in the context of path guiding~\cite{Diolatzis2020}. 
Besides integration, LTCs can also be importance sampled to provide noise-free ray tracing with very low samples per pixel~\cite{Peters2021}. 
Furthermore, LTCs have also found uses in differentiable rendering. 
Specifically, they have been used to select points on edges for efficient differentiable rendering~\cite{li2018} and to analytically compute gradients of the rendering equation~\cite{zhou2021}. 
Note that all of the aforementioned applications leverage properties of LTC distributions and work independently of how these distributions were fitted to a given material, and most of them use the isotropic GGX look-up table originally provided by Heitz et al.~\cite{heitz2016}.
We provide a look-up table for anisotropic GGX.

\paragraph{BRDF fitting.}

There is a significant amount of work on the problem of fitting parametric models to BRDFs~\cite{ngan2005,dupuy2018} but the problem we address is different.
BRDF fitting means fitting a 4D function with a simpler one.
In our case, we fit the 2D outgoing-radiance lobe of the BRDF in each view-roughness configuration separately.

%% file: section_background.tex
\section{Background}
\label{sec:background}

Here we review the mathematical background related to GGX and LTCs used in the subsequent sections.
Note that the implementation of our method only requires Algorithms~\ref{alg:ggx_sampling} and \ref{alg:ltc_sampling}, with the rest used for plots, reference comparisons and technical proofs.

\subsection{Background on the GGX BRDF}
\label{sec:background_ggx}

The GGX (``\textit{Ground Glass Unknown}'') microfacet BRDF was introduced by Walter et al.~\cite{walter2007} and its anisotropic extension by Heitz~\cite{heitz2014}.
The equations of this model are as follows:

\paragraph{Normalized directions.}

$\omega_v$ denotes the view direction and $\omega_l$ the light direction, with the following parameterizations:
\begin{align}
\omega = (x,y,z) = (\sin\theta\cos\phi, \sin\theta\sin\phi, \cos\theta).
\end{align}

\paragraph{Normal Distribution Function (NDF)}

The NDF represents the statistical distribution of specular microfacets that reflect the incident light.
It is parameterized by two roughness parameters $(\alpha_x, \alpha_y)$:
\begin{align}
\label{eq:ggx_ndf}
D_{\text{ggx}}(\omega) = \frac{1}{\pi\,\alpha_x\,\alpha_y\left(\frac{x^2}{\alpha_x^2}+\frac{y^2}{\alpha_y^2}+z^2\right)^2}.
\end{align}
%
%This term is used in Equation~(\ref{eq:ggx_brdf}).

\paragraph{Masking-shadowing.}

The masking-shadowing function $G_2$ computes the attenuation due to the microsurface's self-shadowing:
\begin{align}
\label{eq:ggx_g2}
G_2(\omega_v, \omega_l) = \frac{1}{1 + \Lambda(\omega_v) + \Lambda(\omega_l)} \text{ with } \Lambda(\omega) = \frac{-1+\sqrt{1 + \frac{\alpha_x^2\,x^2+\alpha_y^2\,y^2}{z^2}}}{2}.
\end{align}
%
%This term is used in Equation~(\ref{eq:ggx_brdf}) and Algorithm~\ref{alg:ggx_sampling}.

\paragraph{Bidirectional Reflectance Distribution Function (BRDF)}

%The direct illumination is integral of the cosine-weighted BRDF lobe over the area light. 
The cosine-weighted GGX BRDF is defined as
\begin{align}
\label{eq:ggx_brdf}
\rho(\omega_v, \omega_l)\, \cos\theta_l = \frac{F(\omega_v, \omega_h) \, D_{\text{ggx}}(\omega_h) \, G_2(\omega_v, \omega_l)}{4 \, \cos\theta_v},
\end{align}
where $\omega_h = \frac{\omega_v+\omega_l}{\|\omega_v+\omega_l\|}$ is the half vector and $F$ is a Fresnel term.
In the following, we do as Hill et al.~\cite{hill2016} and always consider $F=1$, since it can be reintroduced after the LTC approximation via a separate table.
Equation~(\ref{eq:ggx_brdf}) is used in the fitting approach of Heitz et al.~\cite{heitz2016}.
We use this formula to make reference comparisons, but not in the implementation of our method.

\paragraph{Sampling.}

Sampling the Visible Normals Distribution Function (VNDF)~\cite{heitz2018b} produces an approximate sampling of the cosine-weighted BRDF.
The remaining weight of the samples is $\frac{1 + \Lambda(\omega_l)}{1 + \Lambda(\omega_v)+ \Lambda(\omega_l)}$.
In Section~\ref{sec:fitting}, we use the rejection-sampling Algorithm~\ref{alg:ggx_sampling} to produce samples from the density that are perfectly proportional to the GGX cosine-weighted BRDF.
\begin{algorithm}[!h]
\DontPrintSemicolon
\While{true}
{
    sample $\omega_h$ from the GGX VNDF \tcc{Heitz's procedure~\cite{heitz2018b}}
    $\omega_l = \text{reflect}\left(\omega_v, \omega_h\right)$ \\
    $U \leftarrow \text{rand}()$ \tcc{Uniform random number in [0, 1)}
    \If{$U$ < $\frac{1 + \Lambda(\omega_l)}{1 + \Lambda(\omega_v)+ \Lambda(\omega_l)}$}
    {
    \Return $\omega_l$ \\
    }
}
\caption{\label{alg:ggx_sampling} Sampling the cosine-weighted GGX BRDF.}
\end{algorithm}
%

%%%%%%%%%%%%%%%%%%%%%
%%%%%%%%%%%%%%%%%%%%%
%%%%%%%%%%%%%%%%%%%%%
%%%%%%%%%%%%%%%%%%%%%
\subsection{Background on LTCs}
\label{sec:background_ltc}

We now review properties of Linearly Transformed Cosines introduced by Heitz et al.~\cite{heitz2016} and illustrated in Figure~\ref{fig:ltc_intro}-(b, c):

\paragraph{Definition.}

An LTC is defined as a matrix $M$ that maps a clamped cosine distribution $D_o$ to a spherical distribution defined as
\begin{align}
    \label{eq:ltc_definition}
    D(\omega) = D_o(\omega_o) \frac{\partial \omega_o}{\partial \omega} = D_o\left (\frac{M^{-1}\omega}{|| M^{-1} \omega||} \right) \frac{|M^{-1}|}{||M^{-1}\omega||^3}.
\end{align}
This equation is used by Heitz et al.~\cite{heitz2016} in their fitting procedure to precompute the look-up table of Equation~(\ref{eq:lut_iso}).
We use this formula to make a proof in Section~\ref{sec:interpolation}, but not in the implementation of our method. 

\paragraph{Area-light integration.}

The integral of an LTC $D$ over the spherical domain $\mathcal{A}$ covered by an area light is the integral of the clamped cosine distribution $D_o$ over the spherical domain $\mathcal{A}_o$ covered by the area light linearly transformed by $M^{-1}$:
\begin{align}
    \label{eq:ltc_integration}
    \int_{\mathcal{A}} D(\omega)\,\mathrm{d}\omega = \int_{\mathcal{A}_o} D_o(\omega_o)\,\mathrm{d}\omega_o.
\end{align}
Like Heitz et al., we use this property at run time in the fragment shader to evaluate the integral of an LTC over an area light.  
The integration procedure depends on the shape of the light~\cite{heitz2016,heitz2017a,heitz2017b}.
Note that our contribution relates to how the matrix $M$ is obtained,  which is independent of how the integration is computed.

\paragraph{Sampling.}

An LTC can be sampled by generating samples $\omega_o$ from the clamped cosine distribution and transforming them with the LTC matrix $M$, as shown in Algorithm~\ref{alg:ltc_sampling}.
We use this algorithm in our fitting procedure, which is introduced in Section~\ref{sec:fitting}.
~\vspace{-2mm}
\begin{algorithm}[!h]
\DontPrintSemicolon
sample $\omega_o$ from a clamped cosine \\
$\omega = \frac{M \, \omega_o}{\|M \, \omega_o\|}$ \\
\Return $\omega$ 
\caption{\label{alg:ltc_sampling} Sampling an LTC.}
\end{algorithm}
~\vspace{-6mm}

%% file: section_fitting.tex
\section{Fitting}
\label{sec:fitting}

\begin{figure}[h!]
\vspace{-2mm}
    \centering
    \begin{tabular}{c @{\hspace{2mm}} c @{\hspace{2mm}} c @{\hspace{2mm}} c @{\hspace{2mm}} c @{}}
    & initialization & $L^3$ fit & $L_{\textit{SW}}$ fit & target \vspace{-1mm}\\
    & {\tiny LTC} & {\tiny LTC} & {\tiny LTC } & {\tiny GGX} \\
    \raisebox{7mm}{\small (a)}&
    \includegraphics[width=0.19\linewidth]{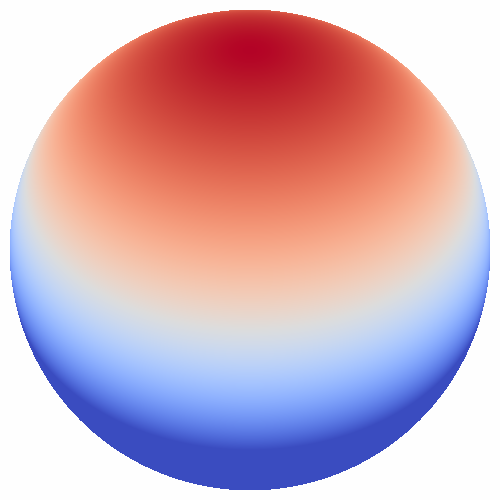}&
    \includegraphics[width=0.19\linewidth]{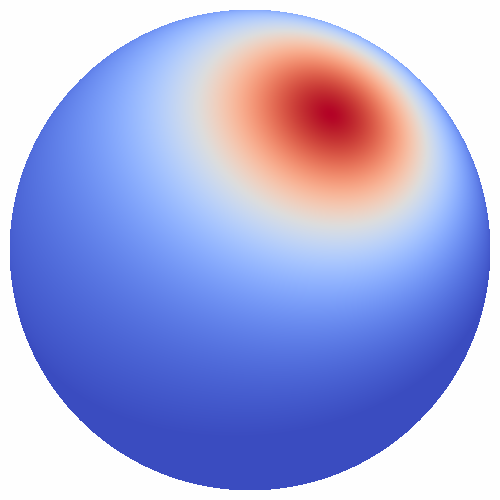}&
    \includegraphics[width=0.19\linewidth]{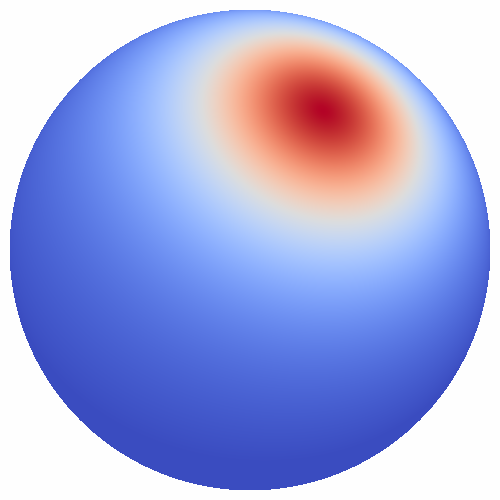}&
    \includegraphics[width=0.19\linewidth]{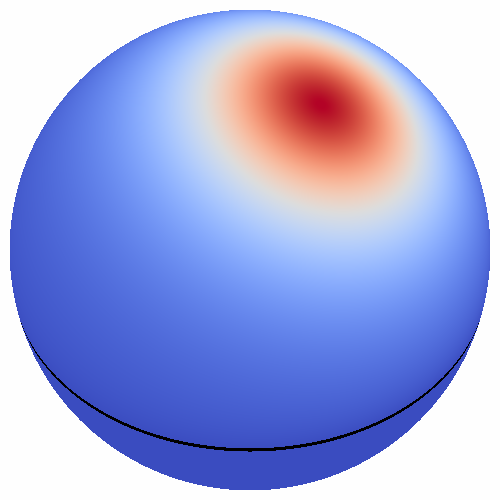} \\
    \raisebox{7mm}{\small (b)}&
    \includegraphics[width=0.19\linewidth]{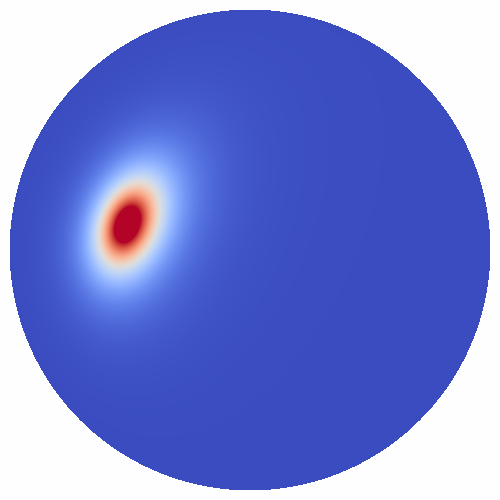}&
    \includegraphics[width=0.19\linewidth]{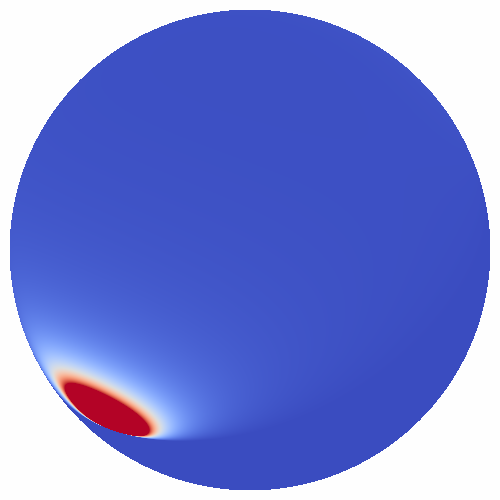}&
    \includegraphics[width=0.19\linewidth]{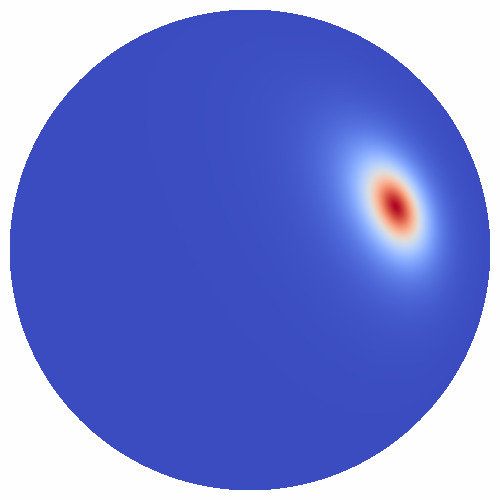}&
    \includegraphics[width=0.19\linewidth]{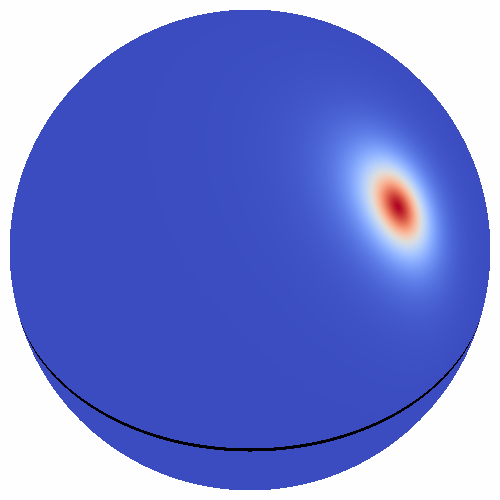} \\
    \raisebox{7mm}{\small (c)}&
    \includegraphics[width=0.19\linewidth]{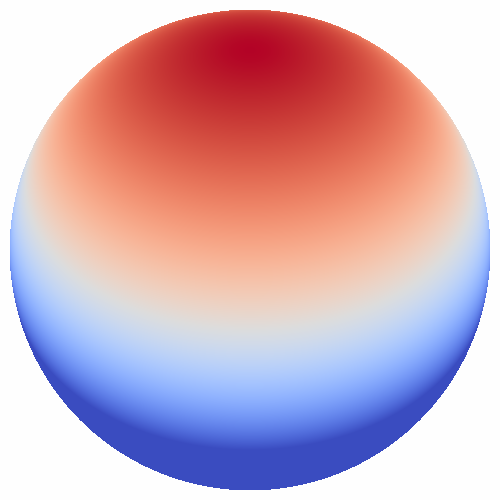}&
    \includegraphics[width=0.19\linewidth]{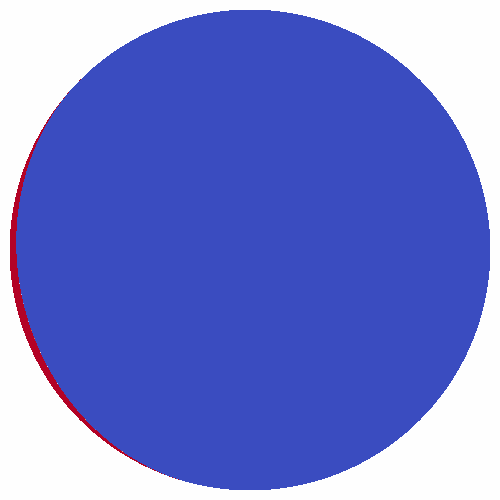}&
    \includegraphics[width=0.19\linewidth]{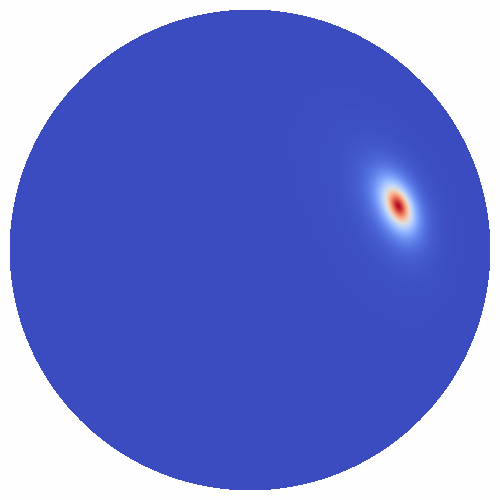}&
    \includegraphics[width=0.19\linewidth]{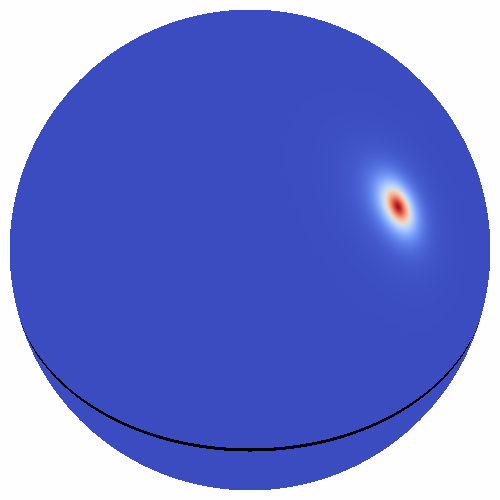} 
    \end{tabular}
    \vspace{-4mm}
    \caption{\label{fig:sw_and_l3_loss} Fitting an LTC to a GGX lobe.
    \textit{We compare the $L^3$ fit of Heitz et al. to the $L_{\textit{SW}}$ fit we propose.}
    }
    \vspace{-10mm}
\end{figure}

In this section, we address the problem of fitting an LTC represented by a matrix $M$ to a GGX lobe, as shown in Figure~\ref{fig:sw_and_l3_loss}.

\subsection{Experimenting with the Previous Approach}
\label{sec:fitting_previous}

Heitz et al. minimize the point-wise $L_3$ error between the cosine-weighted GGX BRDF (Eq.~(\ref{eq:ggx_brdf})) and the LTC (Eq.~(\ref{eq:ltc_definition})).
Their approach works in simple cases such as in Figure~\ref{fig:sw_and_l3_loss}-(a), but we observed two main issues that make it unstable in more challenging configurations, leading to broken fits in our look-up table (Fig.~\ref{fig:problems_intro}-(a)).

\paragraph{Problem 1: null gradients.}

In Figure~\ref{fig:sw_and_l3_loss}-(b), the LTC provided as a starting point does not overlap with the target distribution. 
The gradients of the $L^3$ error metric are thus close to 0 and the optimizer diverges.

\paragraph{Problem 2: high values.}

To avoid problem 1, in Figure~\ref{fig:sw_and_l3_loss}-(c) we use a diffuse LTC (represented by an identity matrix $M$) as the starting point, such that there is significant overlap with the target GGX distribution.
However, the target is sharp and evaluates to high values at the center of its lobe.
These high values produce extremely high $L^3$ error gradients, which cause the optimizer to overshoot and stay trapped in a divergent configuration with null gradients (back to problem 1).

\paragraph{Discussion.}

Despite these issues, the $L_3$ optimization of Heitz et al. is successful because of the \textit{accuracy of their starting points}. 
They use a diffuse LTC (an identity matrix $M$) for high roughnesses and initialize the matrix parameters with the already-optimized neighboring entries of the look-up table as the roughness decreases. 
The resolution of their table ($64\times64$) ensures neighboring entries are close enough to provide accurate starting points.
However, we need to aggressively reduce the resolution to store a 4D table (Sec.~\ref{sec:discretization}), so neighboring entries do not always overlap, especially with sharp distributions (low roughness). 
This is why we need an optimization process that is robust even with poor initialization.

\subsection{Our Approach}
\label{sec:fitting_ours}

Our objective is to find an optimization metric that is not subject to vanishing gradients or numerical instabilities with sharp distributions and works regardless of the accuracy of the initialization.

\paragraph{The Sliced Wasserstein loss.}

We use the \textit{Sliced Wasserstein (SW)} loss~\cite{rabin2012,bonneel2015} between the direction samples of the target GGX lobe and the samples of the LTC distribution. 
This sample-wise loss approximates the optimal transport between two distributions and has shown several benefits in the machine learning community. 
The advantage over point-wise losses such as $L_3$ used by Heitz et al. is that it always provides smooth and stable gradients~\cite{kolouri18}.

\paragraph{Definition.}

Consider two Probability Density Functions (PDFs) $f$ and $g$ and their respective marginals $f_\omega$ and $g_\omega$ over a random direction $\omega \in \Omega$.
The SW distance between $f$ and $g$ is the expected difference between their respective Inverse Cumulative Distribution Functions (iCDF) $F_\omega^{-1}$ and $G_\omega^{-1}$ over all random directions: 
\begin{equation}
    \label{eq:sw_loss}
    L_{\textit{SW}}(f, g) = \mathop{{}\mathbb{E}}_{\omega \in \Omega}\left[ \int_0^1 \left|F^{-1}_\omega(u) - G^{-1}_\omega(u)\right| \, \mathrm{d}u \right].
\end{equation}
Despite this formulation appearing complicated at first sight, its implementation is extremely simple, as we shall see in Algorithm~\ref{alg:fitting}.

\paragraph{Discretization.}

An inverse CDF can be approximated by a list of sorted samples from the corresponding density. 
Hence, if we consider two sets of random samples $(f_1, ..., f_n)$ and $(g_1, ..., g_n)$ from respectively $f$ and $g$ and their \textit{sorted} projections $(f_{1, \omega}, ..., f_{n, \omega})$ and $(g_{1, \omega}, ..., g_{n, \omega})$ onto direction $\omega$, Equation~(\ref{eq:sw_loss}) can be written as
\begin{equation}
    \label{eq:sw_loss_stochastic}
    L_{\textit{SW}}(f, g) = \lim_{n\rightarrow\infty} \mathbb{E}_{\omega \in \Omega}\left[ \frac{1}{n}\sum_{i=1}^n \left|f_{i, \omega} - g_{i, \omega}\right| \right],
\end{equation}
i.e., the average of the differences between the sorted projections over the set of projection directions.

\paragraph{Optimization.}

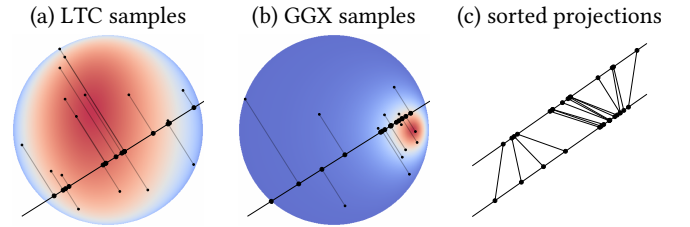
\begin{figure}[!h]
    \centering
    \begin{tikzpicture}
        \draw (0.0, 0) node {\includegraphics[width=0.3\linewidth]{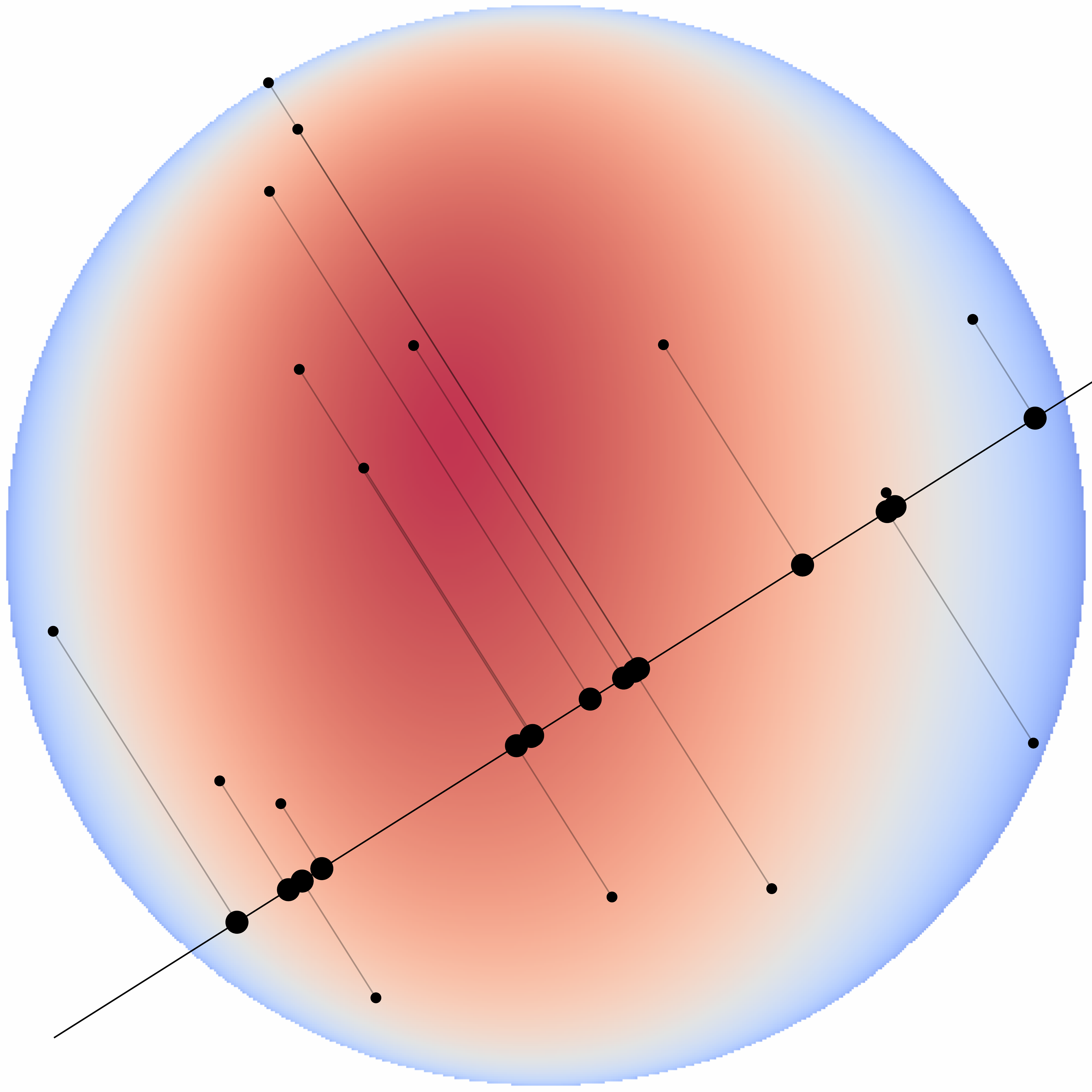}};
        \draw (3.0, 0) node {\includegraphics[width=0.3\linewidth]{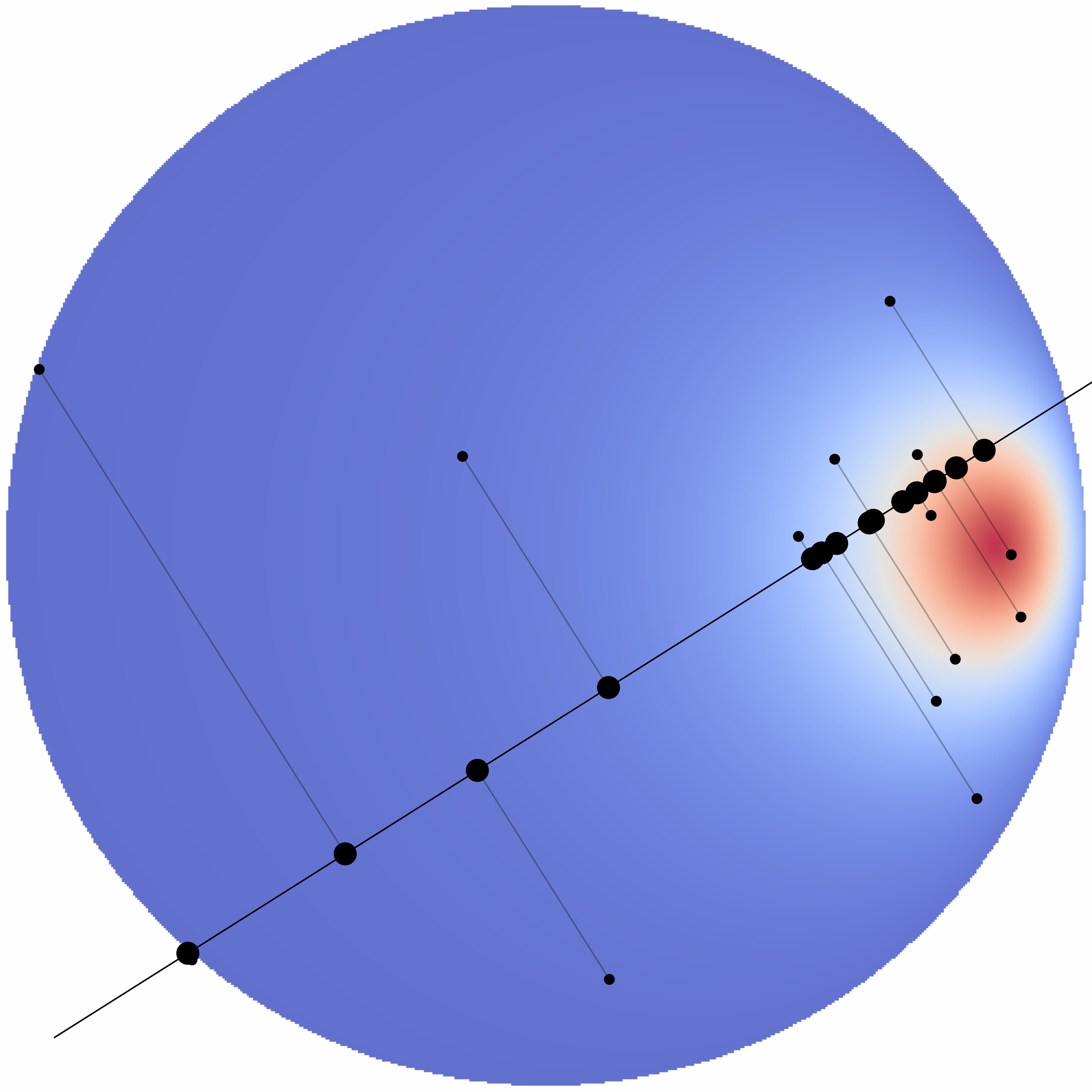}};
        \draw (6.0, 0) node {\includegraphics[width=0.3\linewidth]{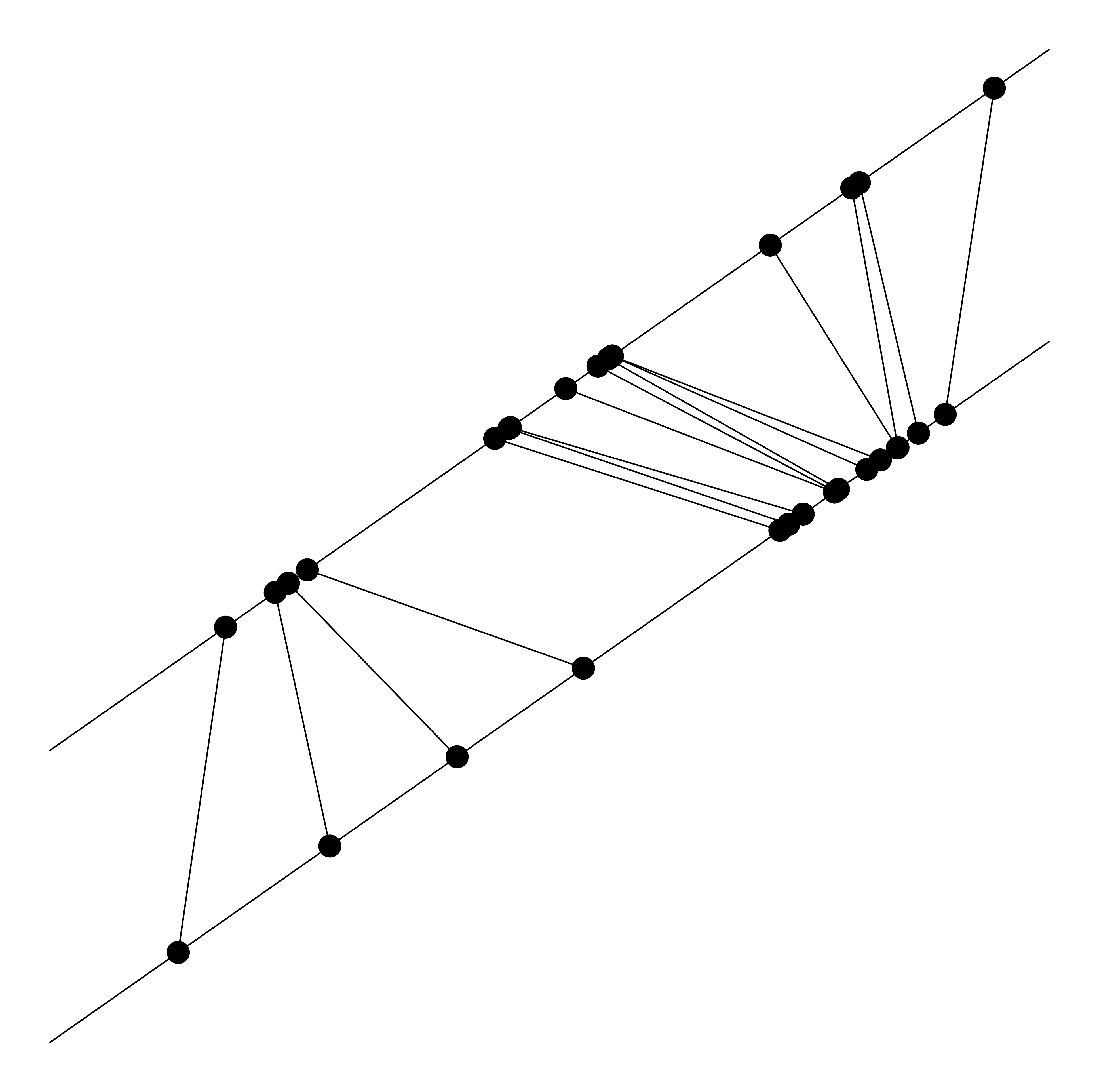}};
        \draw (0.0, 1.5) node {(a) LTC samples};
        \draw (3.0, 1.5) node {(b) GGX samples};
        \draw (6.0, 1.5) node {(c) sorted projections};
    \end{tikzpicture}
    \vspace{-6mm}
    \caption{\label{fig:fitting_sw} Illustration of Algorithm~\ref{alg:fitting}. \textit{We project random samples from the LTC (a) and the GGX lobe (b) onto a random direction and average the absolute differences between the sorted projections (c).}}   
     \vspace{-3mm}
\end{figure}

In Algorithm~\ref{alg:fitting}, we compute a stochastic estimator of Equation~(\ref{eq:sw_loss_stochastic}) by sampling $n$ random samples from the densities, projecting the samples onto a random direction $\omega$, sorting the projections and averaging the absolute differences.
Finally, we propagate the gradient of the loss back to the matrix $M$ and do a gradient descent step.
The variables highlighted in blue are the ones through which the gradients are propagated from $L_{\textit{SW}}$ back to $M$.
Figure~\ref{fig:fitting_sw} illustrates the calculations. 
It is because this algorithm computes a mapping between samples rather than a difference between the densities that it is numerically stable even with sharp or non-overlapping densities, as shown in Figure~\ref{fig:sw_and_l3_loss}.
With this algorithm, we successfully fit all the entries of our look-up table $T_{\text{anisoGGX}}$.

\begin{algorithm}
\DontPrintSemicolon
\KwInput{\color{myblue}LTC matrix $M$}
\KwInput{GGX lobe view direction $\omega_v$ and roughnesses $(\alpha_x, \alpha_y)$}
generate $n$ random samples ($g_1, ..., g_n$) from the GGX lobe \tcc{Algorithm~\ref{alg:ggx_sampling}} 
generate $n$ random samples {\color{myblue}($f_1, ..., f_n$)} from the LTC distribution \tcc{Algorithm~\ref{alg:ltc_sampling}} 
generate a random direction $\omega$ \\
{\color{myblue}$(f_{1, \omega}, ..., f_{n, \omega}) = \text{sort}(f_1\cdot\omega, ..., f_n\cdot\omega)$} \\
$(g_{1, \omega}, ..., g_{n, \omega}) = \text{sort}(g_1\cdot\omega, ..., g_n\cdot\omega)$ \\
{\color{myblue}$L_{\textit{SW}} = \frac{1}{n} \, \sum_{i=1}^n \left|f_{i, \omega} - g_{i, \omega}\right|$} \\
backpropagate gradient from {\color{myblue}$L_{\textit{SW}}$} to {\color{myblue}$M$} \\
update $M = M - \epsilon \, \frac{\nabla L_{\textit{SW}}}{\nabla M}$ \\
\caption{\label{alg:fitting} Optimization step over the LTC matrix $M$ with the Sliced-Wasserstein distance (blue variables depend on $M$).}
\end{algorithm}

\paragraph{Implementation.}

We implement this stochastic estimator in a differentiable calculus library, PyTorch~\cite{pytorch}, which provides automatic gradient backpropagation, and use the \textit{Stochastic Gradient Descent (SGD)}~\cite{sgd} optimizer for $M$.
We compute 10000 gradient descent steps and for each step we use $n=2048$ random samples and average the estimator over 64 random directions.

%% file: section_interpolation.tex
\section{Interpolation}
\label{sec:interpolation}

Even with the robust fitting approach described in the previous section, we noticed that our rendered results still suffered from jiggling artifacts shown in Figure~\ref{fig:problems_intro}-(b) when interpolating the entries of our fitted look-up table $T_{\text{anisoGGX}}$. 
In this section, we explain that the \emph{non-uniqueness} of LTCs~(\ref{sec:interpolation_nonuniqueness}) makes interpolation ill-defined, and we propose a solution to this problem (\ref{sec:interpolation_solution}).

\subsection{Non-Uniqueness of LTCs}
\label{sec:interpolation_nonuniqueness}

The essential point of this section is that different matrices $M$ can produce the same LTC distribution.

\paragraph{Intuitive explanation.}

\begin{figure}[!h]
    \centering
    \begin{tikzpicture}
        \node at (-4, 0) (cosine){\includegraphics[height=0.35\linewidth]{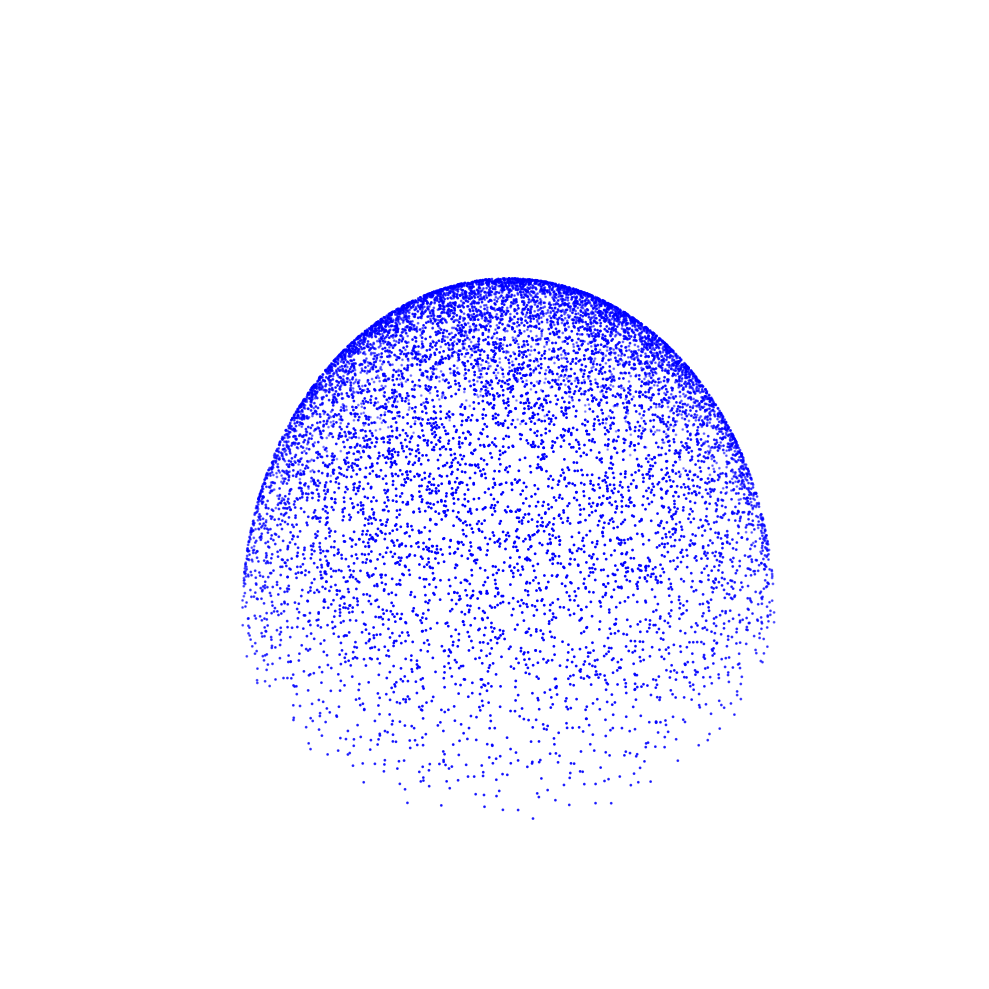}};
        \node at (-4, -3+0.5) (ltc){\includegraphics[height=0.35\linewidth]{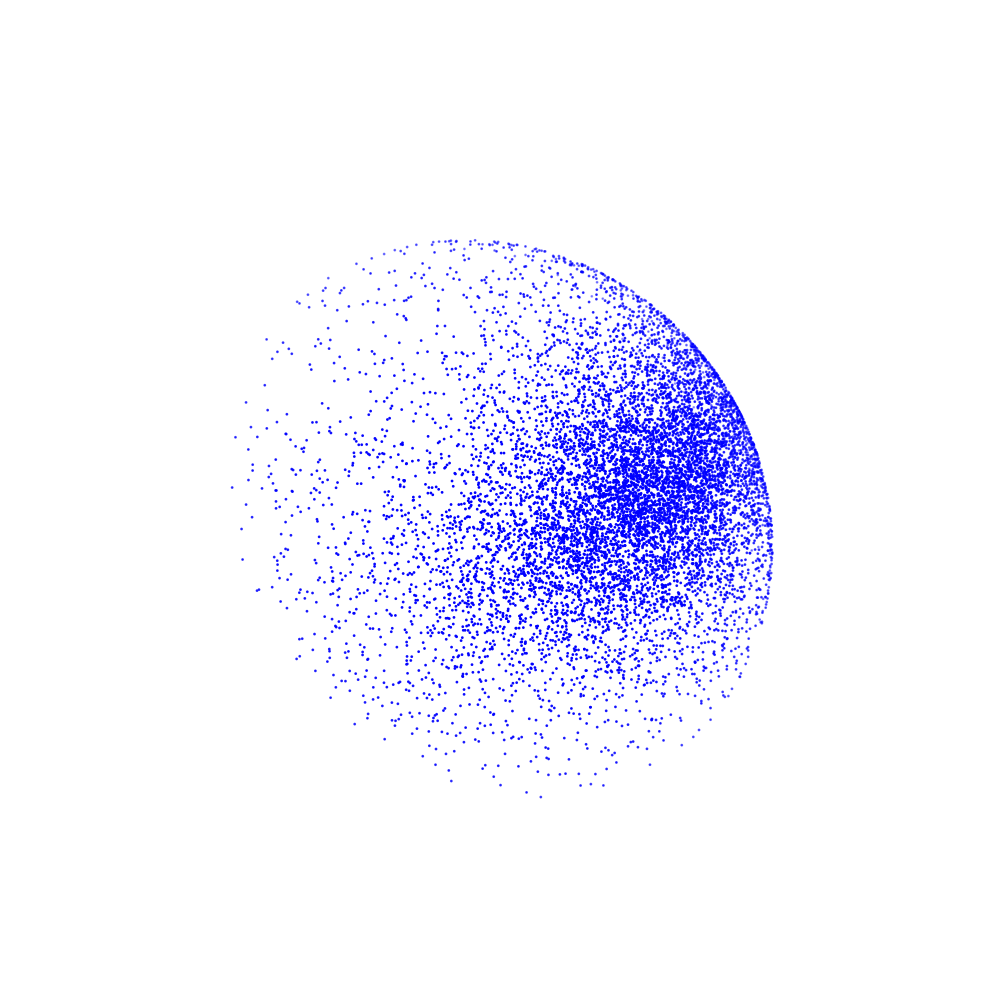}};
        \node at (-1, 0) (cosinerot){\includegraphics[height=0.35\linewidth]{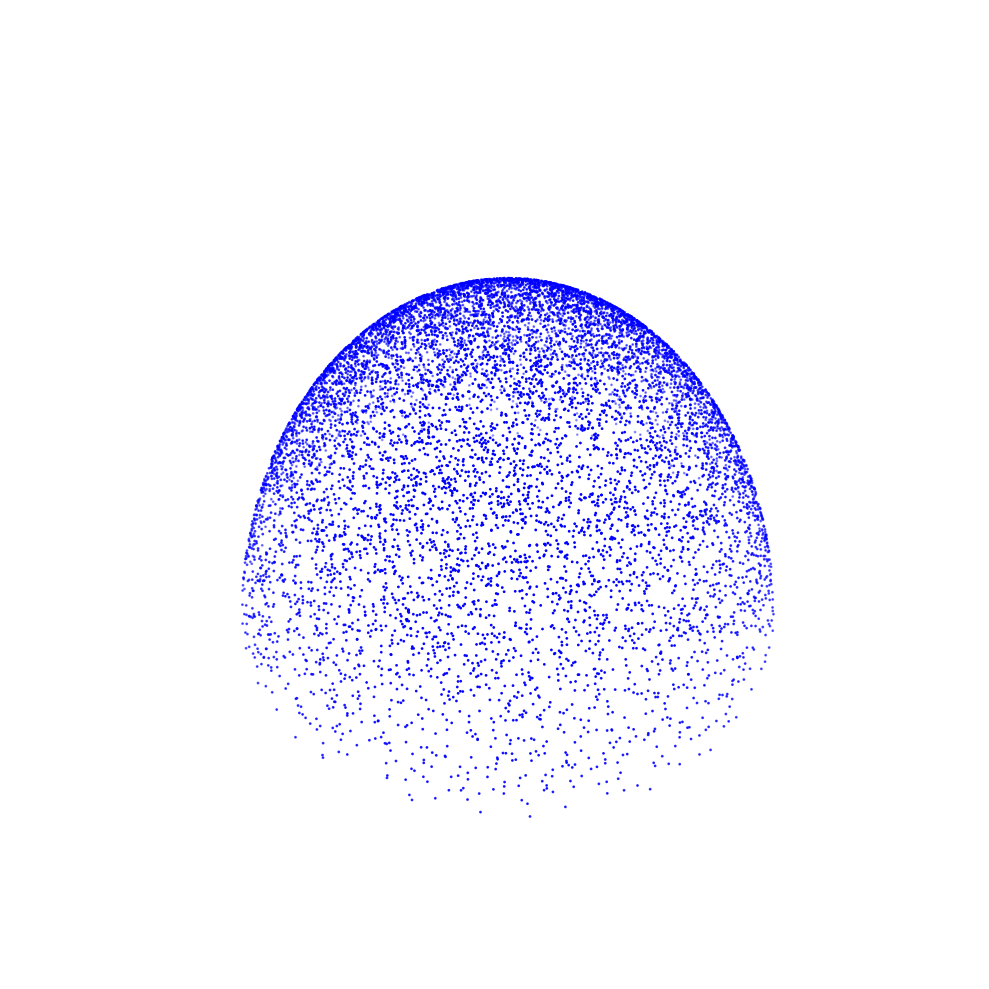}};
        \node at (2, 0) (cosinerotflip){\includegraphics[height=0.35\linewidth]{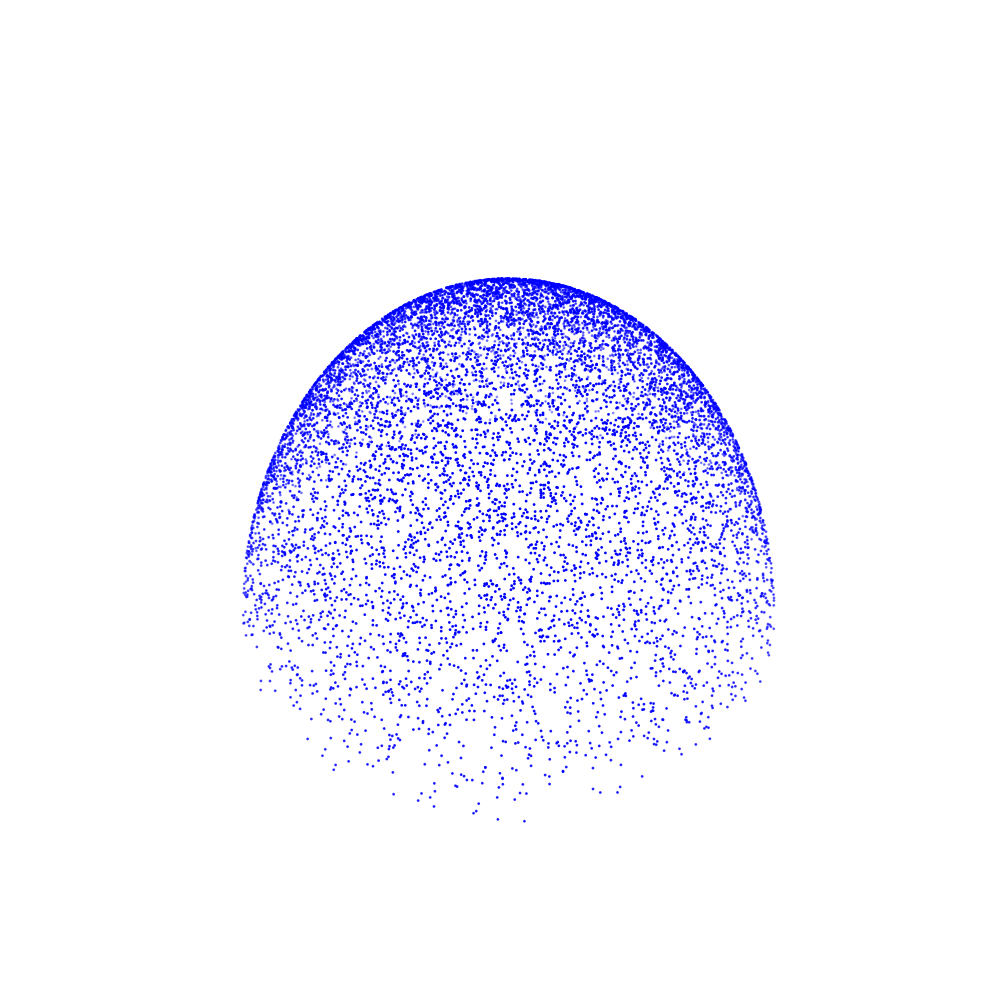}};
        \node at (2, -3+0.5) (ltcrotflip){\includegraphics[height=0.35\linewidth]{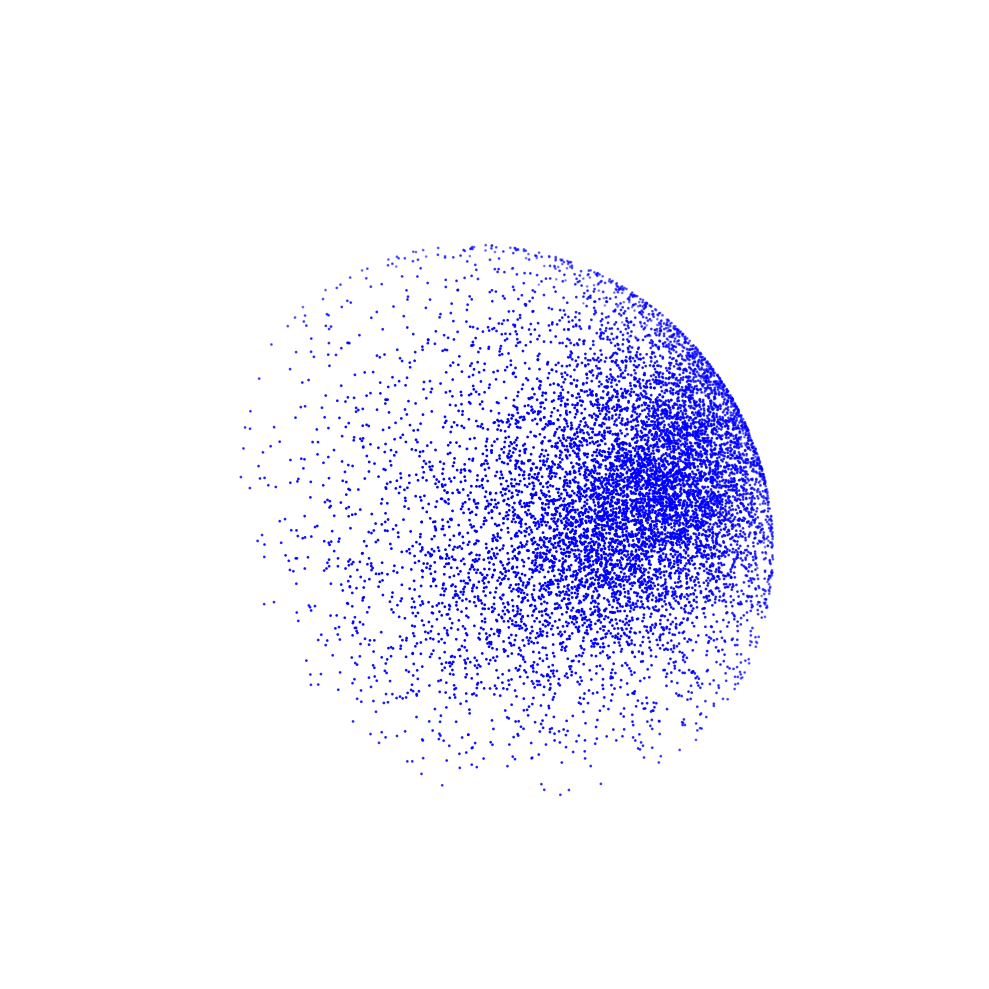}};
        
        \node at (-5, -1) (cosineplot){\includegraphics[height=0.1\linewidth]{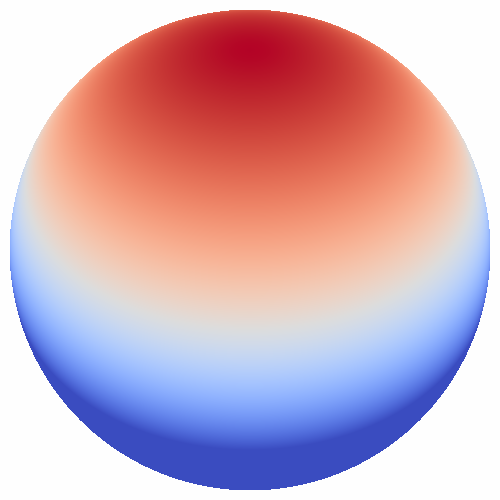}};
        \node at (-5, -4+0.5) (ltcplot){\includegraphics[height=0.1\linewidth]{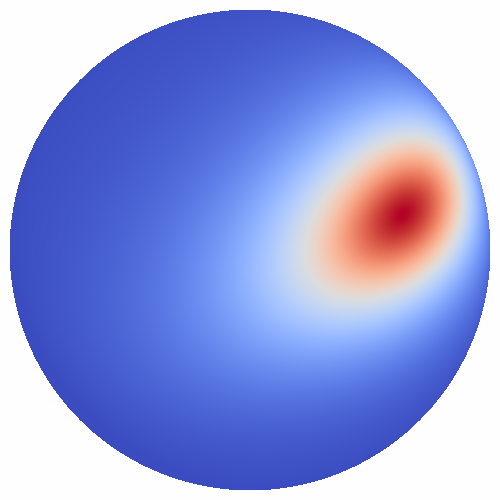}};
        \node at (-2, -1) (cosinerotplot){\includegraphics[height=0.1\linewidth]{figures/rotation_flip_example/cosine_plot.png}};
        \node at (1, -1) (cosinerotflipplot){\includegraphics[height=0.1\linewidth]{figures/rotation_flip_example/cosine_plot.png}};
        \node at (1, -4+0.5) (ltcrotflipplot){\includegraphics[height=0.1\linewidth]{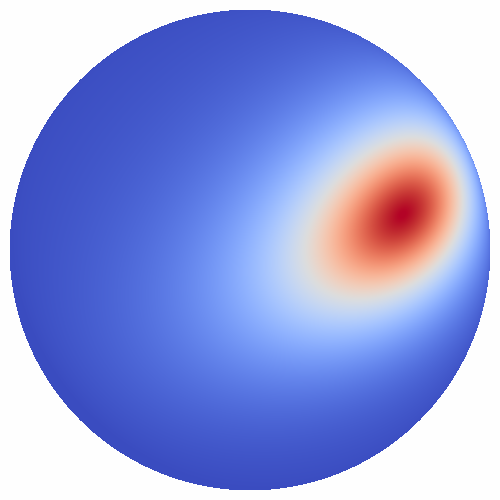}};
        
        \draw[<-](cosine.south) -- (ltc.north);
        \draw[<-](cosine.east) -- (cosinerot.west);
        \draw[<-](cosinerot.east) -- (cosinerotflip.west);
        \draw[<-](cosinerotflip.south) -- (ltcrotflip.north);
        
        \draw (-3.7, -1.5) node {$M$};
        \draw (+2.3, -1.5) node {$M$};
        \draw (-2.5, -2.25+2.5) node {$R_z$};
        \draw (+0.5, -2.25+2.5) node {$F_{xy}$};

    \end{tikzpicture}
    \vspace{-8mm}
    \caption{ \label{fig:rotation_flip_example} Non-uniqueness of LTCs.
    \textit{Cosine distributions remain unchanged under z-axis rotation $R_z$ and $xy$ flipping $F_{xy}$. 
    Linearly transformed cosines inherit this invariance.}}
    \vspace{4mm}
\end{figure}

An LTC is a cosine distribution that has undergone a linear transformation $M$ (and a normalization). 
But the cosine distribution can also be \emph{invariant} under some linear transformations: rotations $R_z$ aligned with the $z$-axis and flipping matrices $F_{xy}$ that flip the $x$- and/or $y$-axis.
Hence, these linear operations can be appended to the matrix $M$ without changing the resulting LTC distribution, as shown in Figure~\ref{fig:rotation_flip_example}.

\paragraph{Property.}

For any rotation and flipping matrices
\begin{align}
R_z = \begin{bmatrix*}[r]
\cos\alpha & -\sin\alpha & 0\\ 
\sin\alpha & \cos\alpha & 0\\ 
0 & 0 & 1
\end{bmatrix*}, \hspace{5mm}
F_{xy} = \begin{bmatrix*}[r]
\pm 1 & 0 & 0\\ 
0 & \pm 1 & 0\\ 
0 & 0 & 1
\end{bmatrix*},
\end{align}
the LTC distributions associated with matrices $M$ and $M \, R_z \, F_{xy}$ are the same.

\paragraph{Proof.}

We show that the evaluation of Equation~(\ref{eq:ltc_definition}) remains the same if we replace $M$ by $M \, R_z \, F_{xy}$.
First, the value of the Jacobian $\frac{\partial \omega_o}{\partial \omega}$ is unchanged because rotations or flips are area-preserving transformations (i.e., their Jacobians are 1).
Second, the variable $\omega_o$ is mapped to another location where $D_o$ (the cosine distribution) evaluates to the same value, since rotations around $z$, or $x,y$-axis flipping do not change the value of the cosine distribution.

\subsection{Well-Defined Interpolation with Alignment}
\label{sec:interpolation_solution}

The obvious way to interpolate LTCs consists of interpolating their matrices $M$.
The non-uniqueness of the matrix $M$ of a given LTC therefore has a direct impact on the interpolation behavior.

\begin{figure}[!h]
    \vspace{-4mm}
    \centering
    \begin{tabular}{c @{\hspace{0.9mm}} c @{\hspace{0.9mm}} c @{\hspace{0.9mm}} c @{\hspace{0.9mm}} c @{\hspace{0.9mm}} c}
    ~ & \multicolumn{5}{c}{(a) $M_1$ and $M_2$ represent the same LTC distribution}
     \\
    \raisebox{3mm}{\rotatebox{90}{\tiny (1) unaligned}} &
    \includegraphics[width=0.18\linewidth]{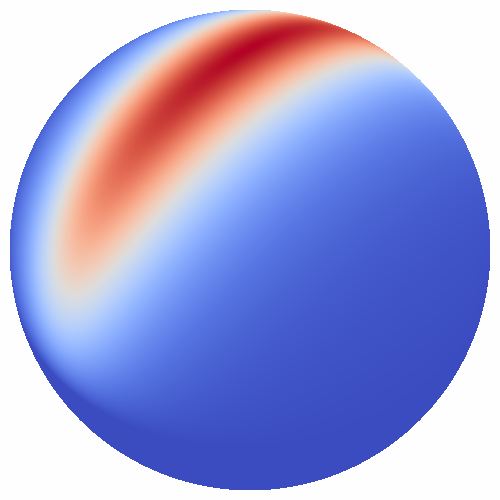}&
    \includegraphics[width=0.18\linewidth]{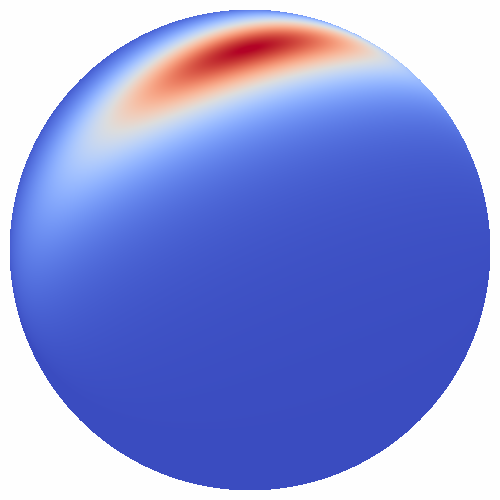}&
    \includegraphics[width=0.18\linewidth]{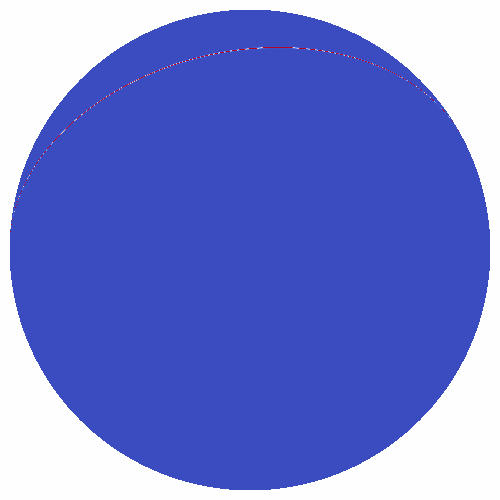}&
    \includegraphics[width=0.18\linewidth]{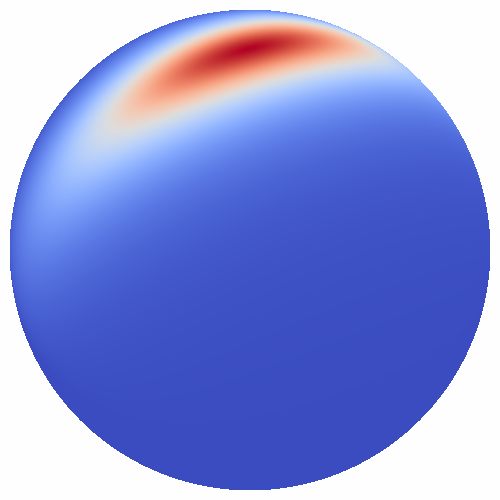}&
    \includegraphics[width=0.18\linewidth]{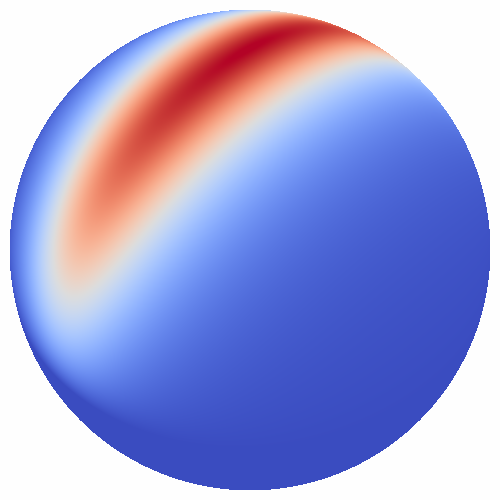}
    \\
    \raisebox{4mm}{\rotatebox{90}{\tiny (2) aligned}} &
    \includegraphics[width=0.18\linewidth]{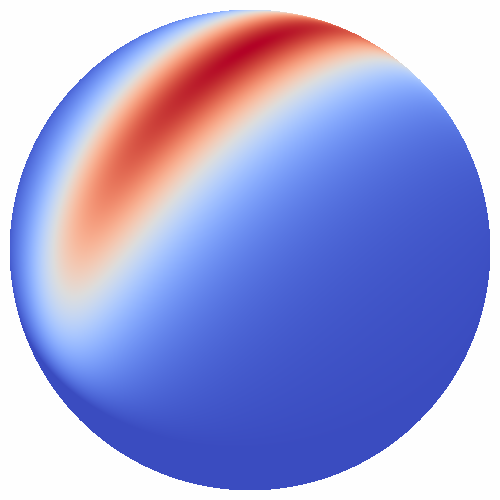}&
    \includegraphics[width=0.18\linewidth]{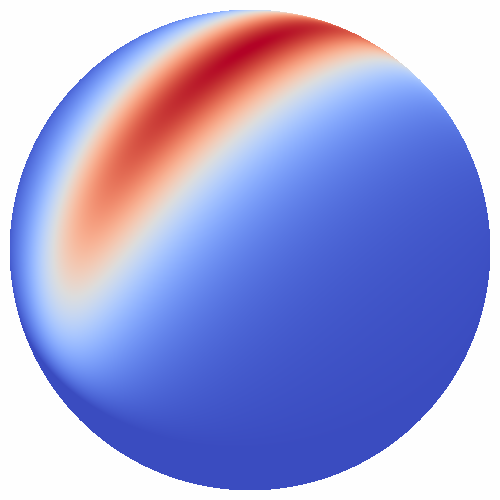}&
    \includegraphics[width=0.18\linewidth]{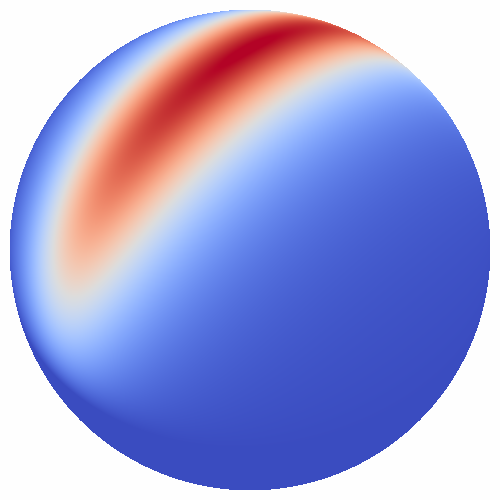}&
    \includegraphics[width=0.18\linewidth]{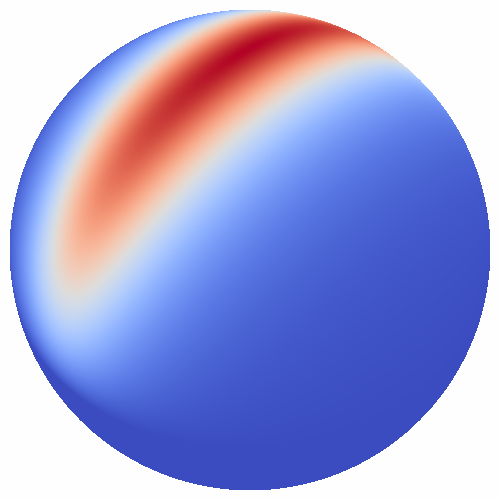}&
    \includegraphics[width=0.18\linewidth]{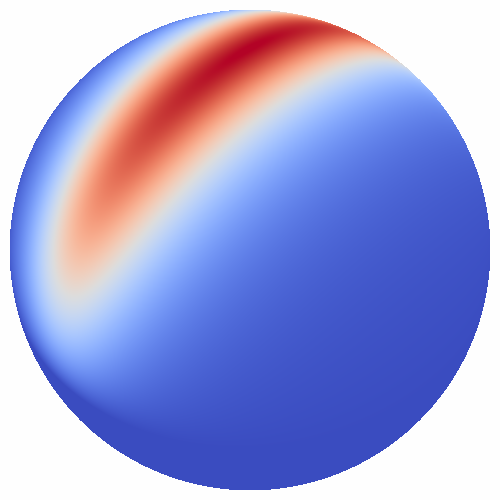} 
    \\
    ~ & \multicolumn{5}{c}{(b) $M_1$ and $M_2$ represent different LTC distributions}
    \\
    \raisebox{3mm}{\rotatebox{90}{\tiny (1) unaligned}} &
    \includegraphics[width=0.18\linewidth]{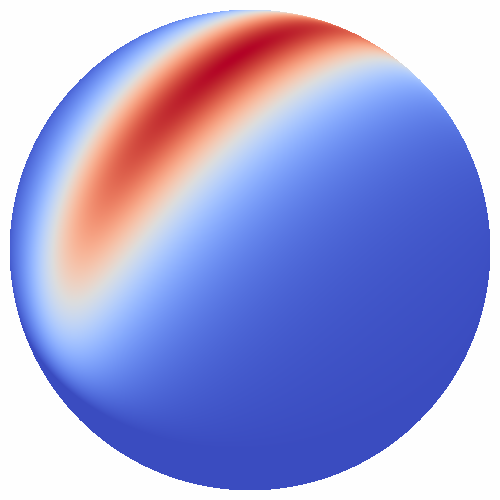}&
    \includegraphics[width=0.18\linewidth]{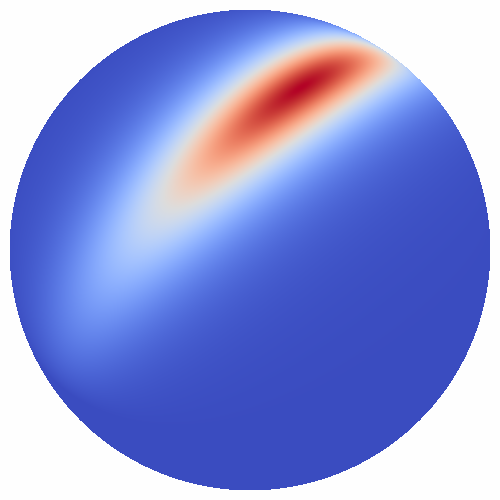}&
    \includegraphics[width=0.18\linewidth]{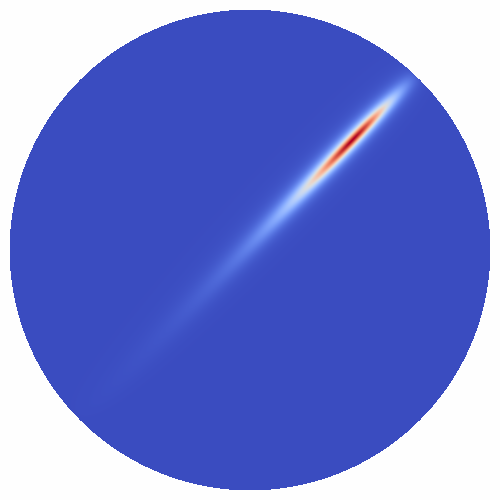}&
    \includegraphics[width=0.18\linewidth]{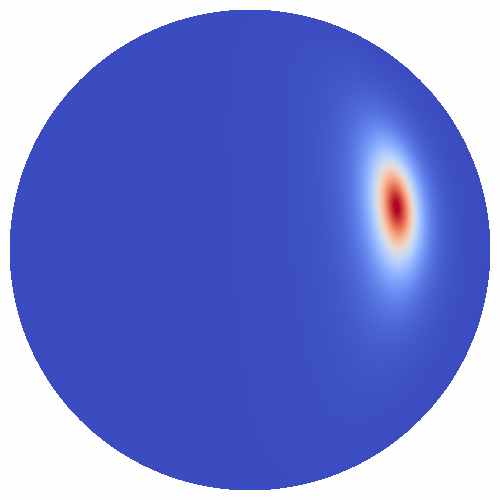}&
    \includegraphics[width=0.18\linewidth]{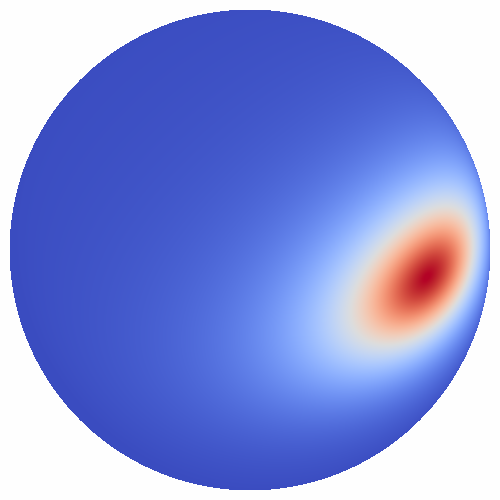}
    \\
    \raisebox{4mm}{\rotatebox{90}{\tiny (2) aligned}} &
    \includegraphics[width=0.18\linewidth]{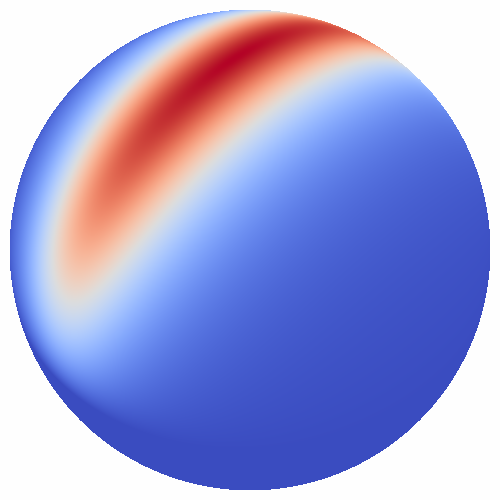}&
    \includegraphics[width=0.18\linewidth]{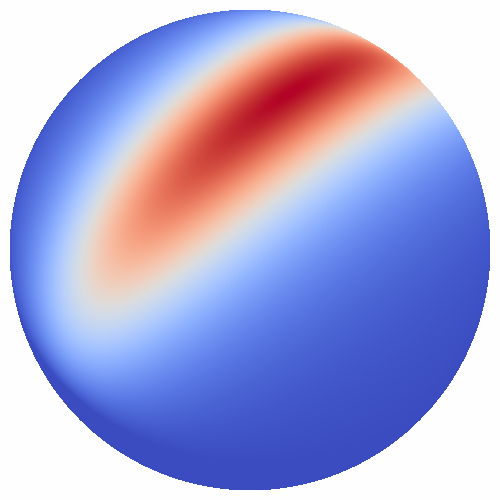}&
    \includegraphics[width=0.18\linewidth]{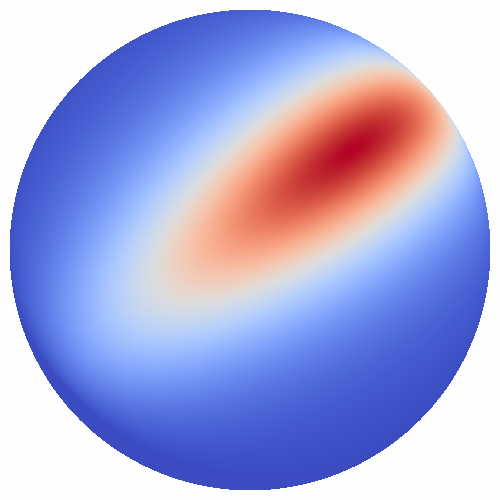}&
    \includegraphics[width=0.18\linewidth]{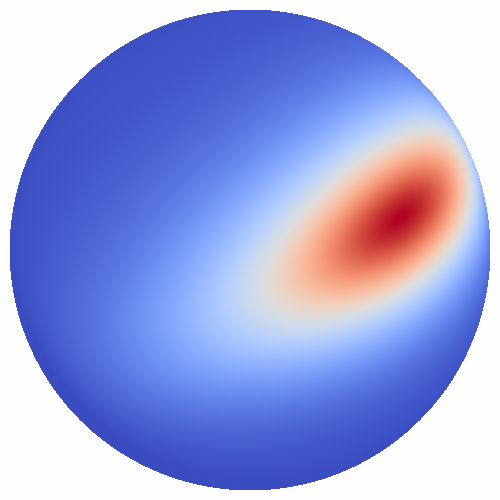}&
    \includegraphics[width=0.18\linewidth]{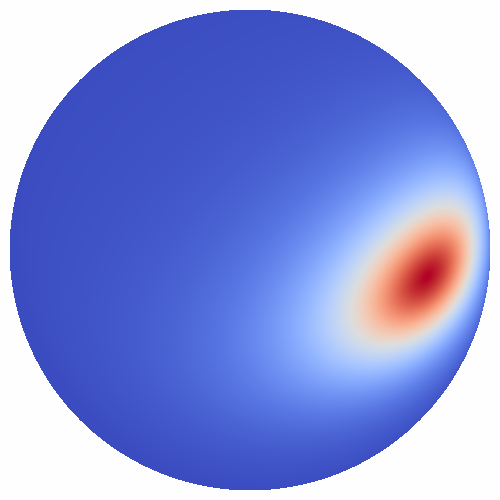} 
    \\
    & $M_1$ & $\frac{3\,M_1 + M_2}{4}$ & $\frac{M_1 + M_2}{2}$ & $\frac{M_1 + 3\,M_2}{4}$ & $M_2$ 
    \end{tabular}
    \vspace{-4mm}
    \caption{\label{fig:interpolation} Interpolating LTC matrices with(out) alignment.} 
    \vspace{-4mm}
\end{figure}

\paragraph{Naive interpolation.}

In Figure~\ref{fig:interpolation}-(a1), we show that two different matrices that represent the same LTC distribution yield another unexpected distribution under interpolation.
In the general case of Figure~\ref{fig:interpolation}-(b1), where $M_1$ and $M_2$ represent different LTC distributions, interpolating the matrices does not produce a smooth transition between their respective distributions. 
Intuitively, this is because the matrices are \textit{misaligned} due to the degrees of freedom introduced by their arbitrary rotation or flipping. This is what causes the interpolation artifacts shown in Figure~\ref{fig:problems_intro}-(b).

\paragraph{Aligned LTCs.}

\begin{figure}[!h]
\vspace{-2mm}
\centering
\begin{tabular}{@{} c c @{\hspace{10mm}} c c @{}}
\raisebox{10mm}{$M$} 
&  
\includegraphics[width=0.25\linewidth]{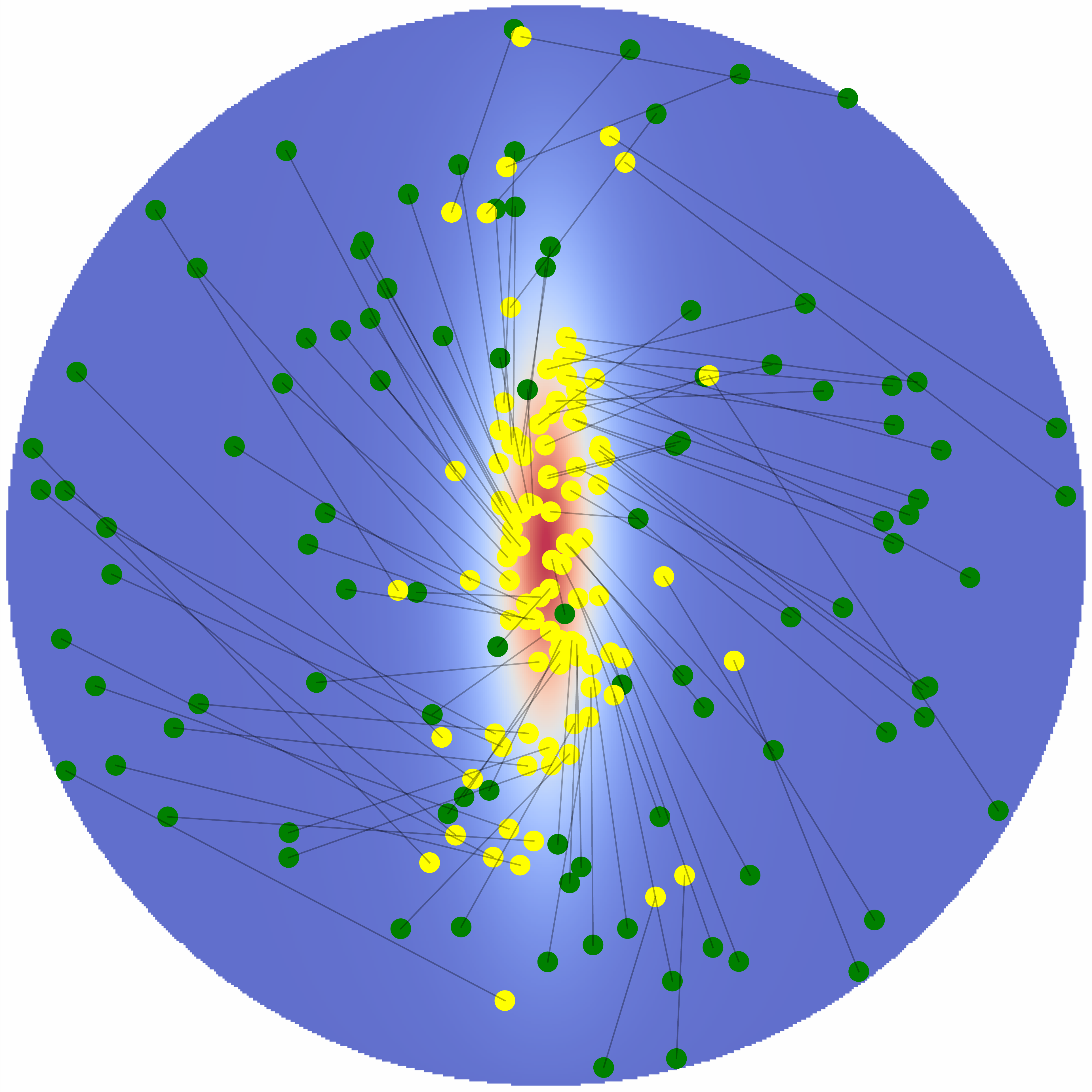} 
&
\raisebox{10mm}{$M_{\text{aligned}}$}
& 
\includegraphics[width=0.25\linewidth]{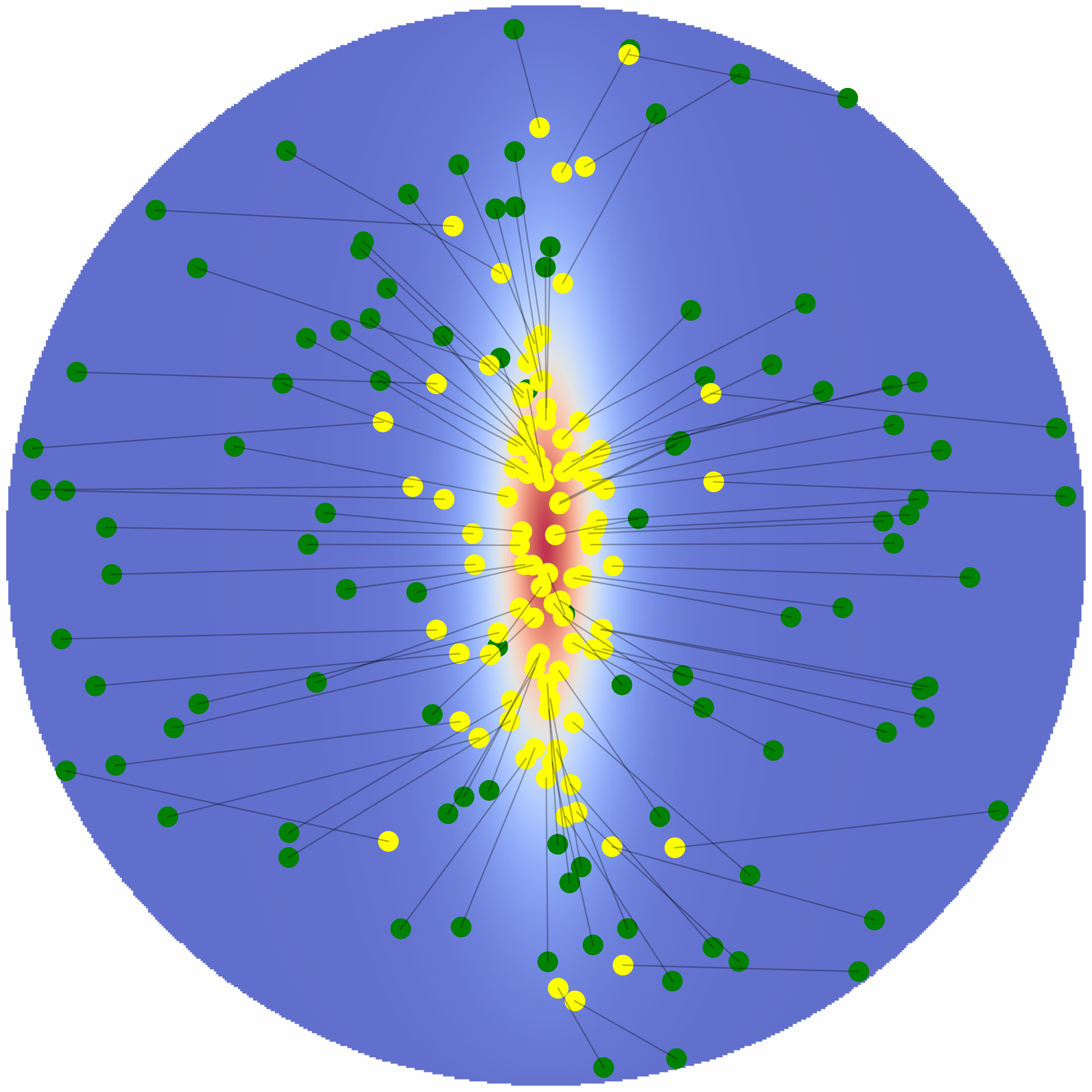}
\end{tabular}
\vspace{-4mm}
\caption{\label{fig:align} LTC alignment.
\textit{We minimize the average squared distance between the cosine (yellow) and LTC (green) samples.}
}
\vspace{-10mm}
\end{figure}

Our idea is to \textit{align} the LTC matrices $M$ to cancel out rotation and flipping of the transformed cosine samples by aligning them with the original cosine samples.
To do that, for a given $M$, we find the LTC matrix $M_{\text{aligned}}=M \, R_z \, F_{xy}$ that minimizes the average squared distance between the original cosine samples and their transformed counterparts, i.e. we compute
\begin{align}
\label{eq:optim_align}
   \displaystyle\min_{F_{x y}} \min_{R_z}~~\mathop{{}\mathbb{E}}_{\omega_o \sim D_o} \left[ \left\| \frac{M \, R_z \, F_{x y} \cdot \omega_o}{\|M \, R_z \, F_{x y} \cdot \omega_o \|} - \omega_o \right\|^2 \right].
\end{align}
Intuitively, this optimization minimizes the average squared lengths of the lines in Figure~\ref{fig:align}.
We implement it as a linear search over the rotation angle $\alpha$ and the four flipping cases. 

\paragraph{Robust interpolation.}

By aligning the LTC matrices before interpolating them, we obtain robust interpolation behavior. 
As expected, interpolating between the same distribution leaves it unchanged (Fig.~\ref{fig:interpolation}-(a2)) and interpolating between different distributions produces smooth transitions (Fig.~\ref{fig:interpolation}-(b2)). 
Aligning the entries of our fitted look-up table $T_{\text{anisoGGX}}$ removes the interpolation artifacts of Figure~\ref{fig:problems_intro}-(b).

\paragraph{Discussion.}

Heitz et al.~\cite{heitz2016} do not report interpolation problems despite using a naive interpolation without alignment. 
This is because they only need four parameters of the LTC matrix to fit isotropic GGX:
$M = {\tiny\begin{bmatrix}
a & 0 & b\\ 
0 & c & 0\\ 
d & 0 & 1
\end{bmatrix}}$.
A side effect of this special case is that the null entries of the matrix force its alignment and thus make the interpolation well-defined.
The interpolation problem that we solve arises in the general case where all nine parameters of the LTC matrix are used, which is necessary for fitting anisotropic GGX.
This is why we are, to our knowledge, the first to face the problem of  LTC interpolation misbehaving, and to investigate the non-uniqueness property of LTCs.

%% file: section_symmetry.tex
\section{Symmetries}
\label{sec:symmetries}

In this section, we leverage symmetries of the anisotropic GGX BRDF to remove numerical errors and reduce the storage induced by the dimensionality of our 4D fitted look-up table $T_{\text{anisoGGX}}(\theta, \phi, \alpha_x, \alpha_y)$.

\subsection{Parameterization of the Look-Up Table}

\paragraph{Azimuthal symmetry.} 

\begin{figure}[!h]
\centering
\begin{tabular}{@{} c @{\hspace{0.5mm}} c @{\hspace{0.5mm}} c @{\hspace{0.5mm}} c @{}}
$0 \le \phi < \frac{\pi}{2}$ &
$\frac{\pi}{2} \le \phi < \pi$ &
$\pi \le \phi < \frac{3\pi}{2}$ &
$\frac{3\pi}{2} \le \phi < 2\pi$ 
\\
\includegraphics[width=0.24\linewidth]{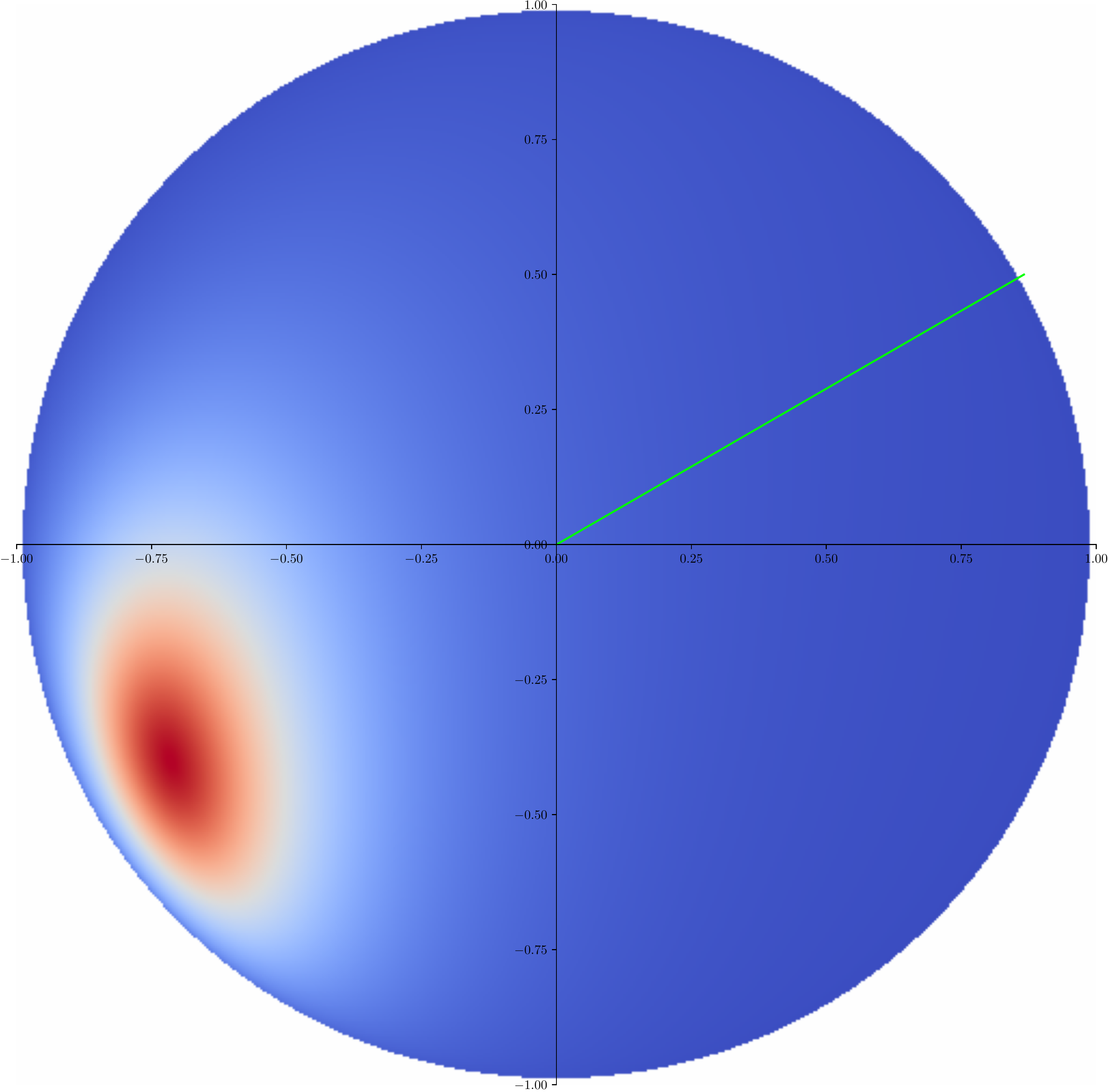}&
\includegraphics[width=0.24\linewidth]{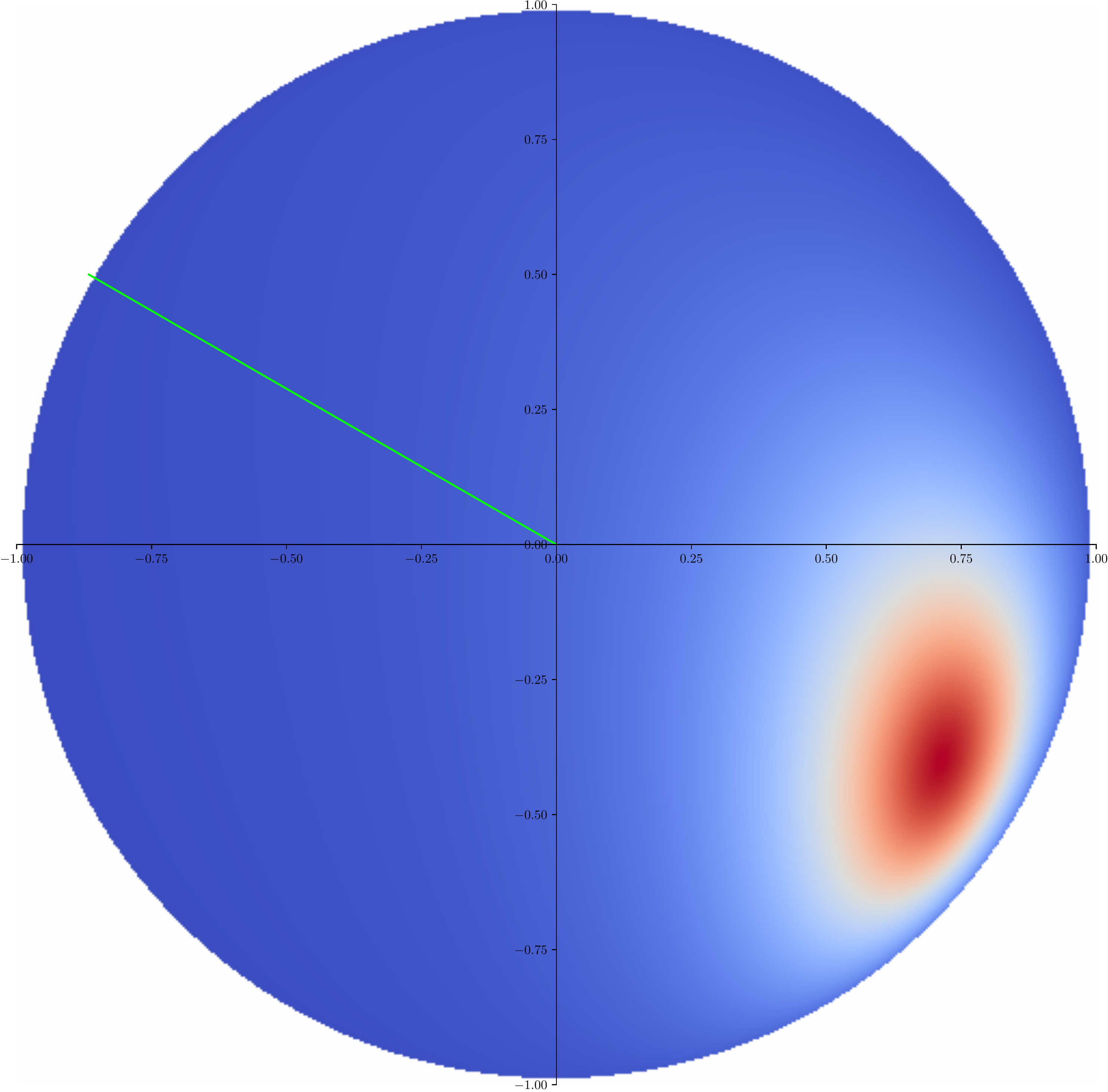}&
\includegraphics[width=0.24\linewidth]{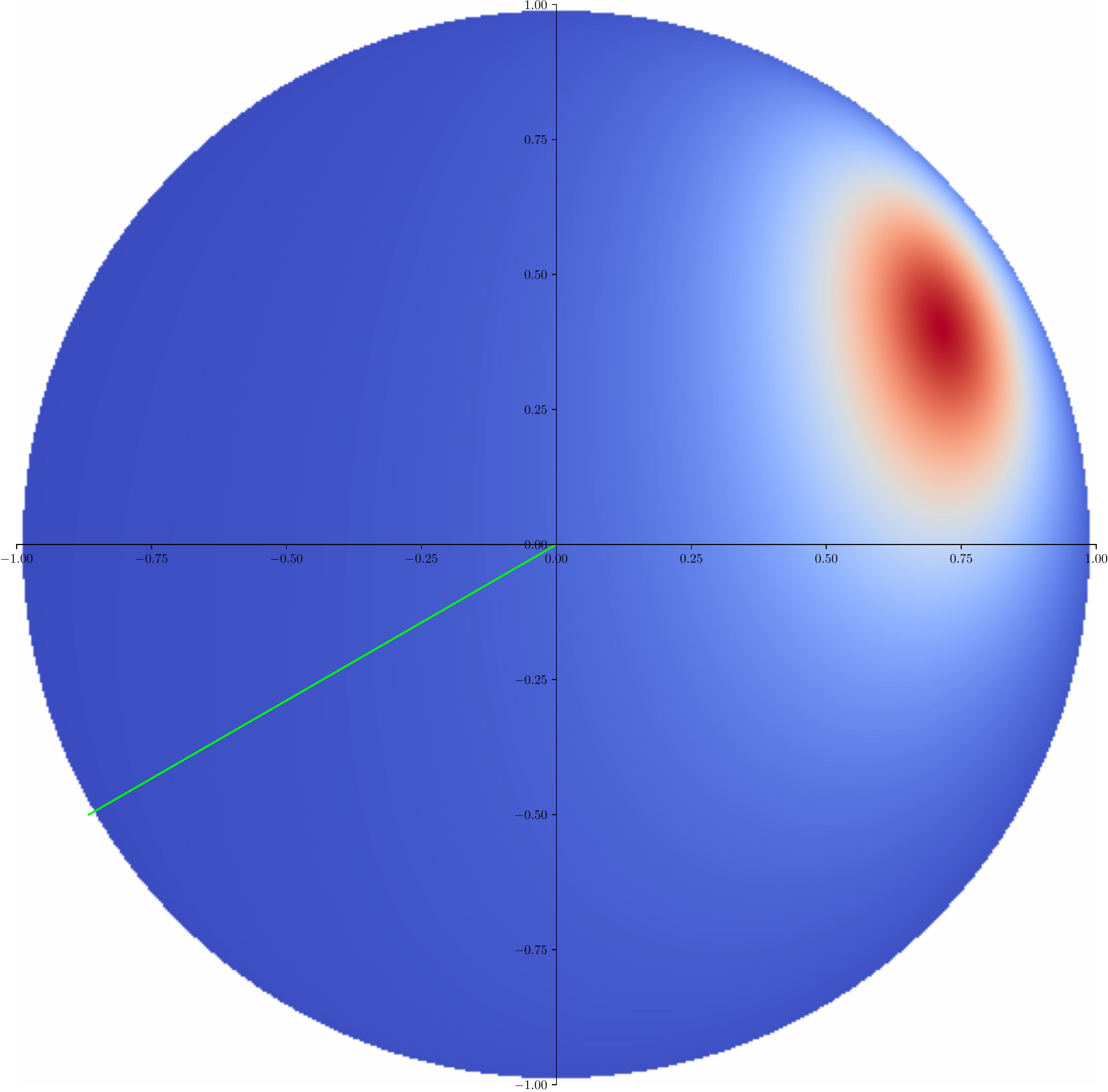}&
\includegraphics[width=0.24\linewidth]{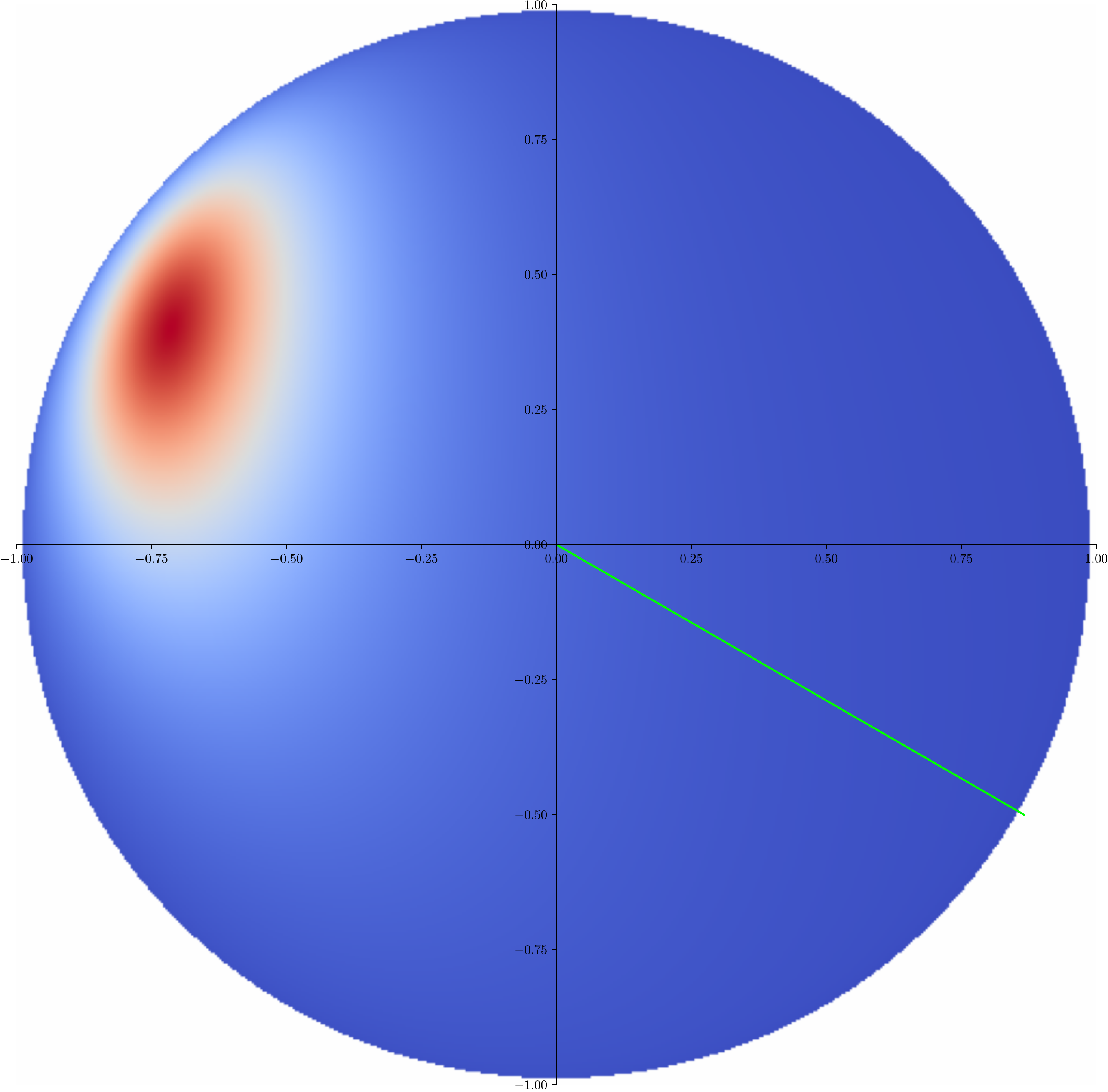}
\end{tabular}
\vspace{-4mm}
\caption{\label{fig:symmetry_azimuthal} Azimuthal symmetries of the GGX BRDF. 
\textit{When the view vector (green) is symmetrized over the $x$ and $y$ axes, the GGX lobe undergoes the same symmetry.}
}
\vspace{-4mm}
\end{figure}

The GGX BRDF has axial symmetries over the $x$ and $y$ axes with respect to the view vector, as shown in Figure~\ref{fig:symmetry_azimuthal}.
We leverage this property to fit our look-up table over $\phi \in [0, \frac{\pi}{2}]$ rather than $\phi \in [0, 2\pi]$.
We recover the full range $[0, 2\pi]$ in the following manner:
\begin{align}
\label{eq:symmetry_azimuthal}
M = 
\begin{cases}
{\tiny\begin{bmatrix*}[r]
+1 & \phantom{-}0 & 0\\ 
\phantom{-}0 & +1 & 0\\ 
0 & 0 & 1
\end{bmatrix*}}
\cdot T_{\text{anisoGGX}}(\phi) & \text{if } 0 \le \phi < \frac{\pi}{2}, \vspace{1mm}\\ 
{\tiny\begin{bmatrix*}[r]
-1 & \phantom{-}0 & 0\\ 
0 & +1 & 0\\ 
0 & 0 & 1
\end{bmatrix*}}
\cdot T_{\text{anisoGGX}}(\pi - \phi) & \text{if } \frac{\pi}{2} \le \phi < \pi, \vspace{1mm}\\
{\tiny\begin{bmatrix*}[r]
-1 & 0 & 0\\ 
0 & -1 & 0\\ 
0 & 0 & 1
\end{bmatrix*}}    
\cdot T_{\text{anisoGGX}}(\phi - \pi) & \text{if } \pi \le \phi < \frac{3\pi}{2}, \vspace{1mm}\\ 
{\tiny\begin{bmatrix*}[r]
+1 & 0 & 0\\ 
\phantom{-}0 & -1 & 0\\ 
0 & 0 & 1
\end{bmatrix*}} 
\cdot T_{\text{anisoGGX}}(2\pi - \phi) & \text{if } \frac{3\pi}{2} \le \phi < 2\pi.
\end{cases} 
\end{align}
This reduces the size of the look-up table by a factor of four for the same angular resolution.

\paragraph{Roughness symmetry.} 

\begin{figure}[!h]
\centering
\begin{tabular}{@{} c @{\hspace{5mm}} c @{}}
$\phi=15°, \alpha_x=0.20, \alpha_y=0.10$ &
$\phi=75°, \alpha_x=0.10, \alpha_y=0.20$
\\
\includegraphics[width=0.24\linewidth]{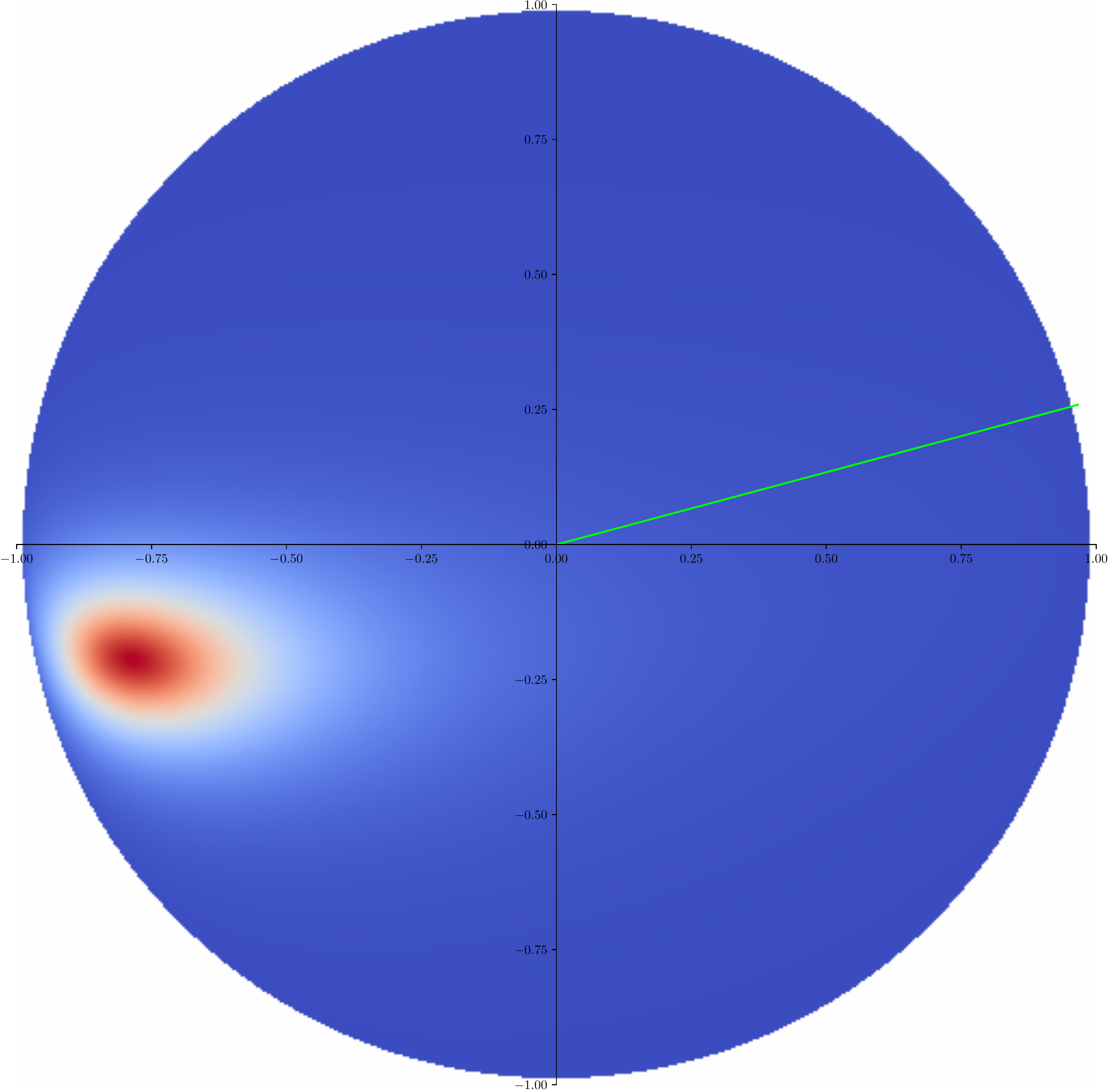}&
\includegraphics[width=0.24\linewidth]{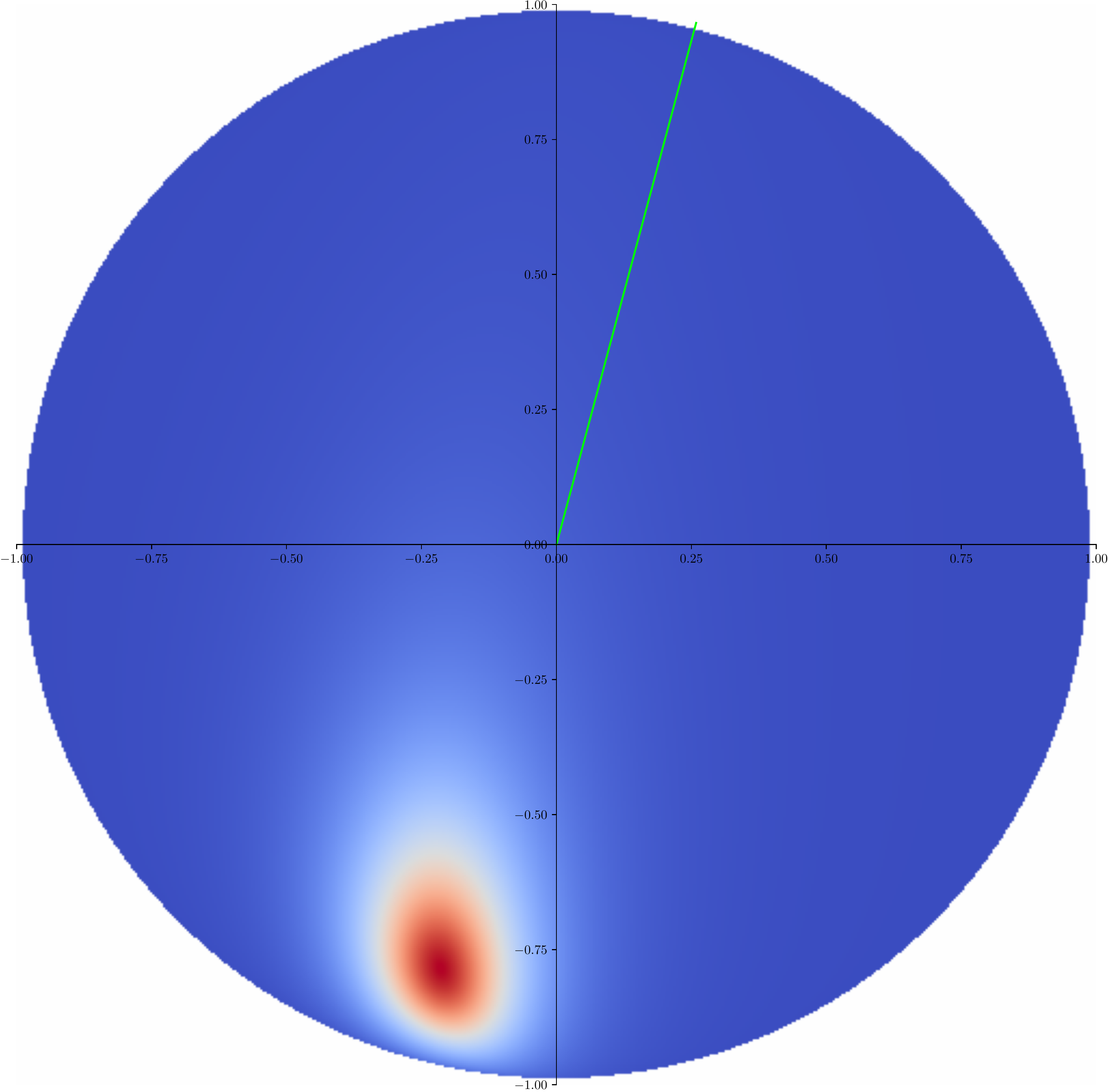}
\end{tabular}
\vspace{-4mm}
\caption{\label{fig:symmetry_roughness} Roughness symmetries of the GGX BRDF. 
\textit{Permuting the $x$ and $y$ coordinates of the view vector and the roughnesses $\alpha_x$ and $\alpha_y$ produces the same permutation in the lobe's shape.}
}
\vspace{-4mm}
\end{figure}

The GGX BRDF has roughness symmetries shown in Figure~\ref{fig:symmetry_roughness} that can be written in the following manner: 
\begin{align}
T_{\text{anisoGGX}}(\theta, \phi, \alpha_x, \alpha_y) = 
{\tiny\begin{bmatrix}
0 & 1 & 0\\ 
1 & 0 & 0\\ 
0 & 0 & 1
\end{bmatrix}}
\cdot T_{\text{anisoGGX}}(\theta, \frac{\pi}{2}-\phi, \alpha_y, \alpha_x).
\end{align}
Storing the data directly with the $(\alpha_x, \alpha_y)$ parameterization virtually duplicates the data, so instead we use an alternative parameterization $T_{\text{anisoGGX}}(\theta,\phi,\alpha, \lambda)$, where $\alpha \in [0,1]$ is the largest roughness and $\lambda \in [0,1]$ is the roughness ratio:
\begin{align}
\label{eq:symmetry_roughness}
M = 
\begin{cases}
{\tiny\begin{bmatrix}
1 & 0 & 0\\ 
0 & 1 & 0\\ 
0 & 0 & 1
\end{bmatrix}}
\cdot T_{\text{anisoGGX}}(\theta, \phi, \alpha_x, \frac{\alpha_y}{\alpha_x}) & \text{if } \alpha_x \geq \alpha_y, \vspace{1mm}\\ 
{\tiny\begin{bmatrix}
0 & 1 & 0\\ 
1 & 0 & 0\\ 
0 & 0 & 1
\end{bmatrix}}
\cdot T_{\text{anisoGGX}}(\theta, \frac{\pi}{2}-\phi, \alpha_y, \frac{\alpha_x}{\alpha_y}) & \text{otherwise}.
\end{cases} 
\end{align}
This increases the roughness resolution by a factor of two for the same storage.

\subsection{Fixing Residual Errors in the Look-Up Table}

Other symmetries of the GGX BRDF imply that certain entries of our fitted look-up table should be null.
This would be the case if the fitting and alignment optimizations were perfect, but even small residual errors are sufficient to break these symmetries and produce the discontinuity and singularity artifacts present in Figure~\ref{fig:problems_intro}-(c).
We can eliminate these artifacts by enforcing the expected symmetries in the entries of the look-up table:

\paragraph{Axial symmetries.}

\begin{figure}[!h]
\centering
\begin{tabular}{@{} c @{\hspace{0.5mm}} c @{\hspace{0.5mm}} c @{\hspace{0.5mm}} c @{}}
(a) $\phi=0$ &
(b) $\phi=\frac{\pi}{2}$ &
(c) $\theta=0, \phi=0$ &
(d) $\theta=0, \phi=\frac{\pi}{2}$ 
\\
\includegraphics[width=0.24\linewidth]{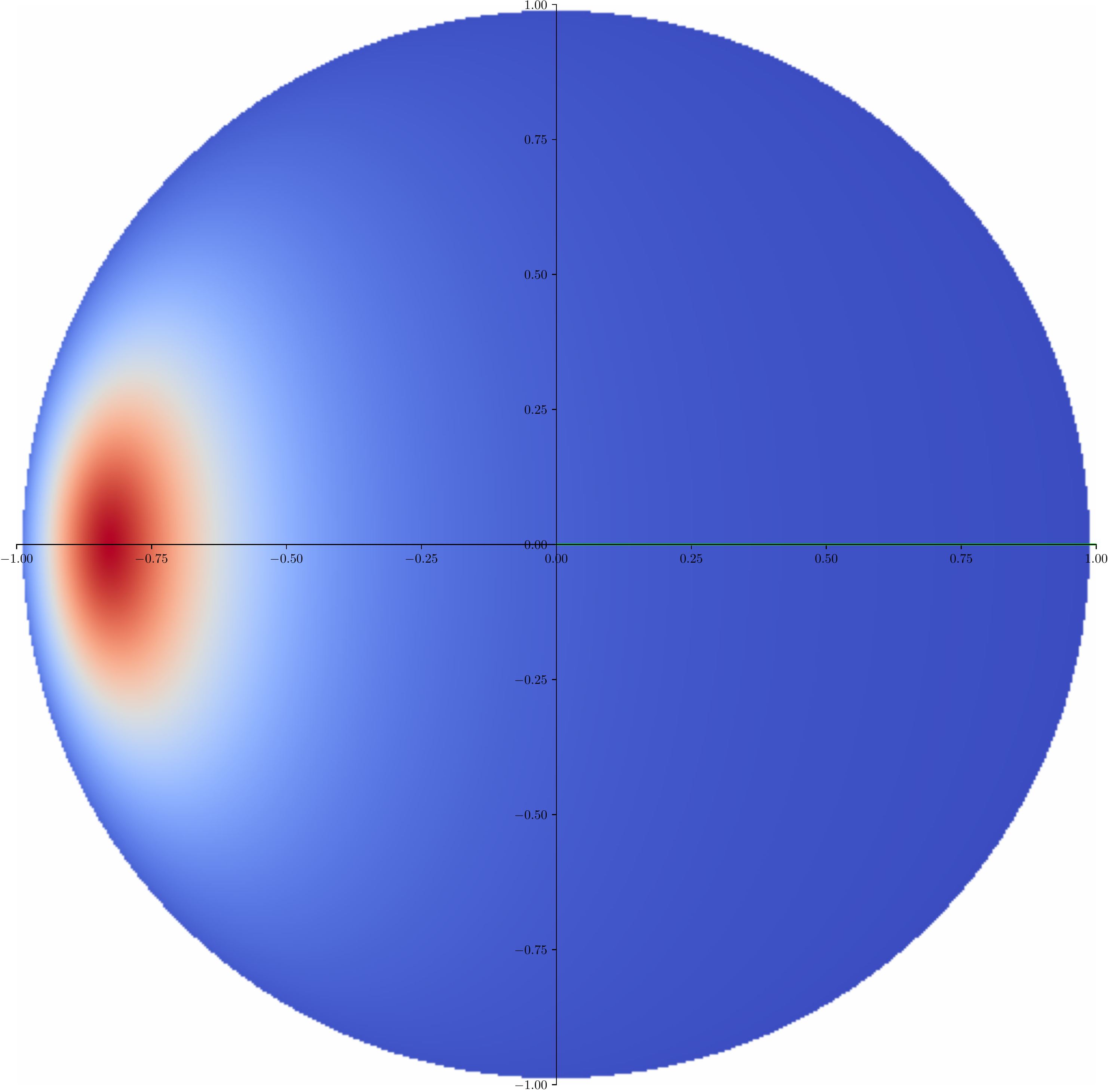}&
\includegraphics[width=0.24\linewidth]{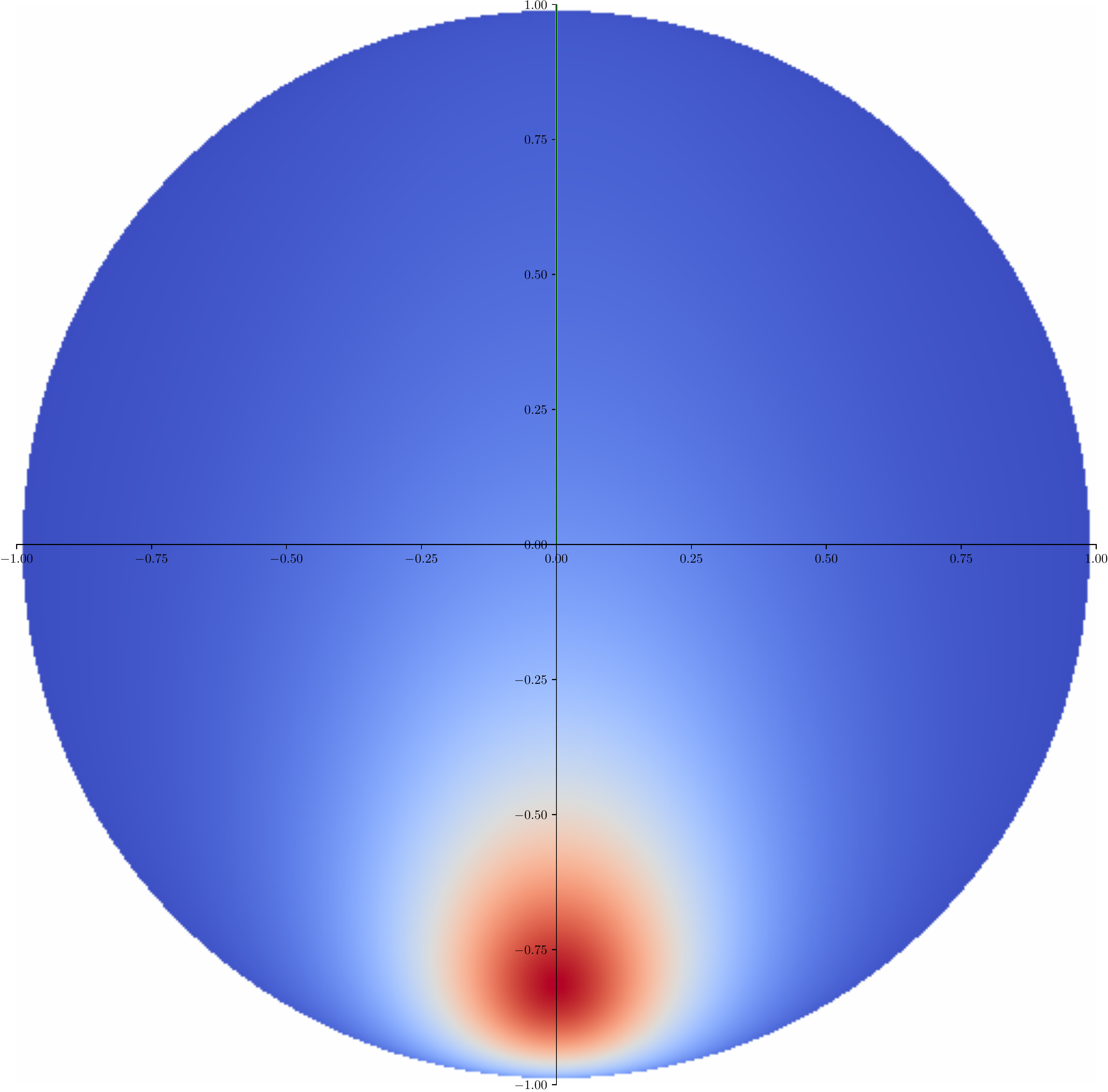}&
\includegraphics[width=0.24\linewidth]{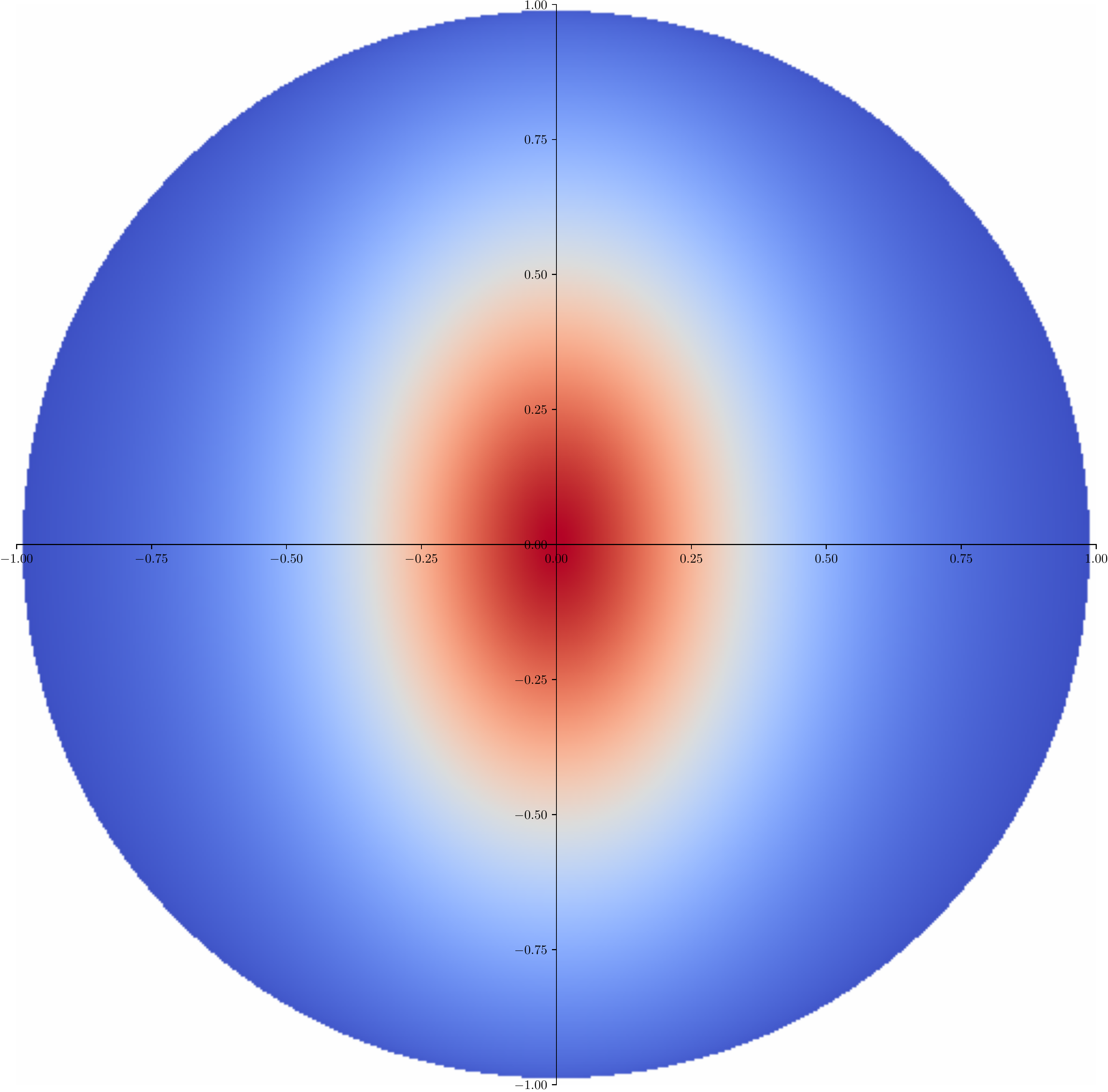}&
\includegraphics[width=0.24\linewidth]{figures/symmetry/quadrant_theta=0.pdf}
\end{tabular}
\vspace{-4mm}
\caption{\label{fig:symmetry2} Symmetries of the GGX lobes with axis-aligned view directions.
}
\vspace{-4mm}
\end{figure}

The GGX lobe has axial symmetry over the $x$ and $y$ axes when, respectively, $\phi=0$ and $\phi=\frac{\pi}{2}$ (Fig.~\ref{fig:symmetry2}-(a, b)).

\paragraph{Rotational symmetry.}

In upward views, where $\theta=0$, the azimuthal angle $\phi$ does not contribute to the shape of the GGX lobe. 
Furthermore, the GGX lobe is centered on $0$ and symmetric over the axes $x$ and $y$ (Fig.~\ref{fig:symmetry2}-(c, d)).

\paragraph{Look-up table post-processing.}

We post-process our look-up table to set the following entries (blue) to zero, and ensure that when $\theta=0$ all of the entries match for the different values of $\phi$:
\begin{align}
\label{eq:postprocessing}
T_{\text{anisoGGX}}(\theta, \phi) = 
\begin{cases}
{\tiny\begin{bmatrix*}[r]
m_{00} & {\color{blue}0} & m_{02}\\ 
{\color{blue}0} & m_{11} & {\color{blue}0}\\ 
m_{20} & {\color{blue}0} & m_{22}
\end{bmatrix*}} & \text{for } \phi=0, \vspace{1mm}\\ 
{\tiny\begin{bmatrix*}[r]
m_{00} & {\color{blue}0} & {\color{blue}0}\\ 
{\color{blue}0} & m_{11} & m_{12} \\ 
{\color{blue}0} & m_{21} & m_{22}
\end{bmatrix*}} & \text{for } \phi=\frac{\pi}{2}, \vspace{1mm}\\ 
{\tiny\begin{bmatrix*}[r]
m_{00} & {\color{blue}0} & {\color{blue}0}\\ 
{\color{blue}0} & m_{11} & {\color{blue}0}\\ 
{\color{blue}0} & {\color{blue}0} & m_{22}
\end{bmatrix*}} & \text{for } \theta=0 \text{ and } \phi=0, \vspace{1mm}\\ 
T_{\text{anisoGGX}}(\theta, 0), & \text{for } \theta=0. 
\end{cases}
\end{align}
\newpage % Sorry :(
\noindent This post-processing step fixes the artifacts present in Figure~\ref{fig:problems_intro}-(c).

%% file: section_inversion.tex
\section{Matrix Inversion}
\label{sec:inversion}

The LTC integration shown in Figure~\ref{fig:ltc_intro} actually uses the inverse matrix $M^{-1}$. 
Heitz et al. spare the inversion by storing and interpolating the inverse in their look-up table.
However, interpolating the inverse creates severe distortions with our coarse resolution. 
This is because the diagonal coefficients of $M$ that control the aperture of the LTC behave linearly with respect to $\alpha_x$ and $\alpha_y$ at low roughness while the diagonal coefficients of $M^{-1}$ correlate with $\frac{1}{\alpha_x}$ and $\frac{1}{\alpha_y}$.
We can therefore use a more aggressive discretization by storing and interpolating $M$ instead of $M^{-1}$, and performing the inversion at run time in the fragment shader. 
To bring all of the entries of the look-up table into the same precision range, we divide $M$ by the length of its third column, i.e., we constrain $M \cdot [0,0,1]$ to be a unit-length vector.

%% file: section_discretization.tex
\section{Discretization}
\label{sec:discretization}

\begin{figure}[!h]
    \centering
    
    \begin{tikzpicture}
    
        \node at (-4, 0.9) {\footnotesize $\alpha=0.06$};
        \node at (-2.2, 0.9) {\footnotesize $\alpha=0.31$};
        \node at (-0.4, 0.9) {\footnotesize $\alpha=0.56$};
        \node at (1.4, 0.9) {\footnotesize $\alpha=0.81$};

        \node at (-5, -0.2) {\rotatebox{90}{\footnotesize $8^4$ LUT}};
        
        \node at (-4, -0.2) {\includegraphics[width=0.17\linewidth]{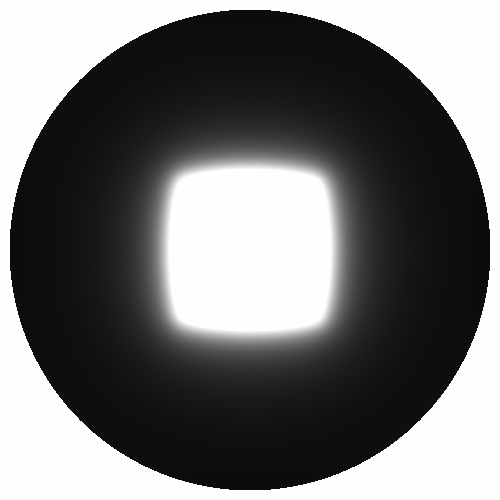}};
        \node at (-3.4, -0.7) {\includegraphics[width=0.07\linewidth]{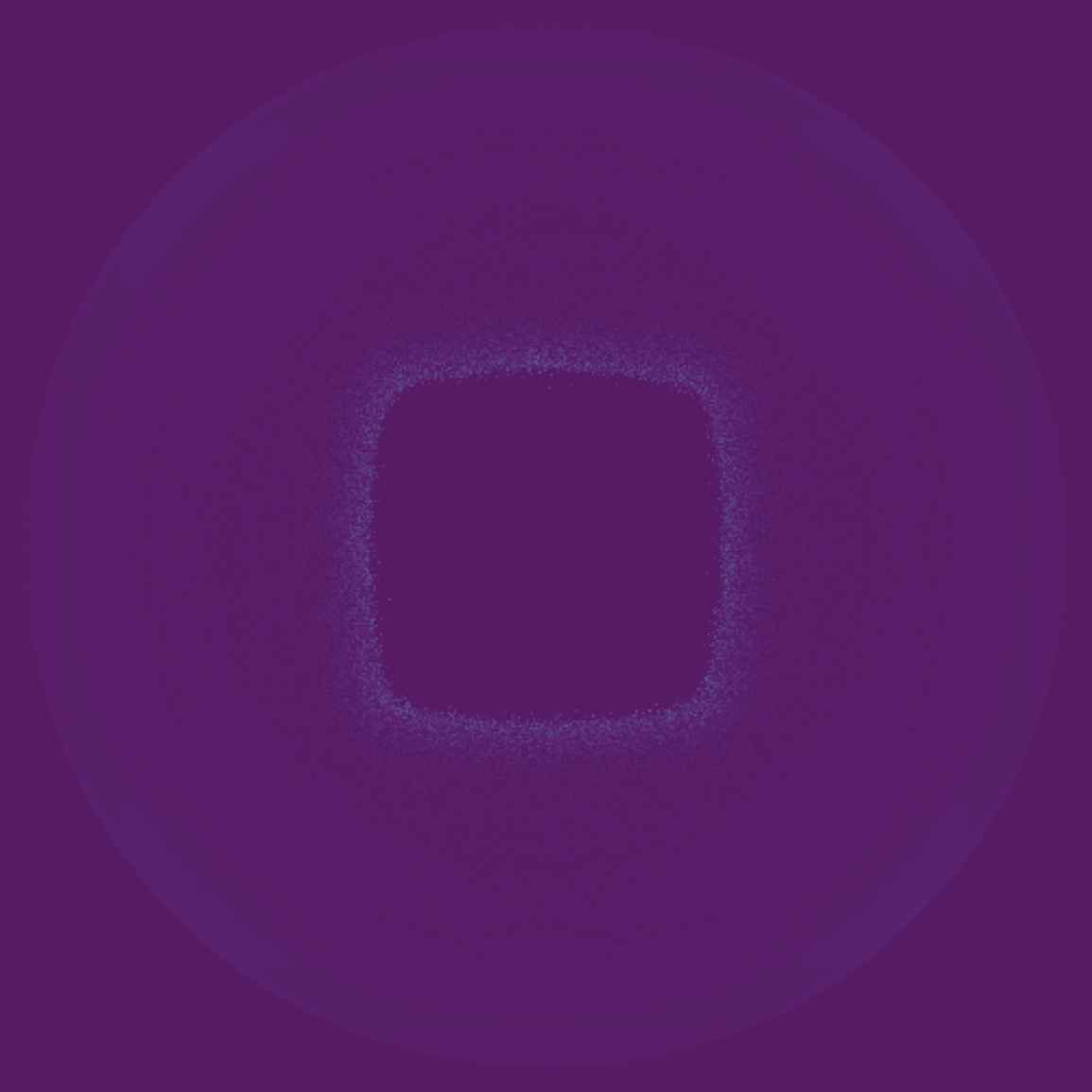}};
        
        \node at (-2.2, -0.2) {\includegraphics[width=0.17\linewidth]{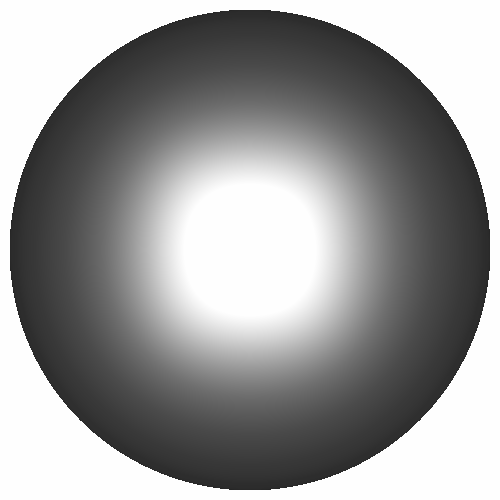}};
        \node at (-1.6, -0.7) {\includegraphics[width=0.07\linewidth]{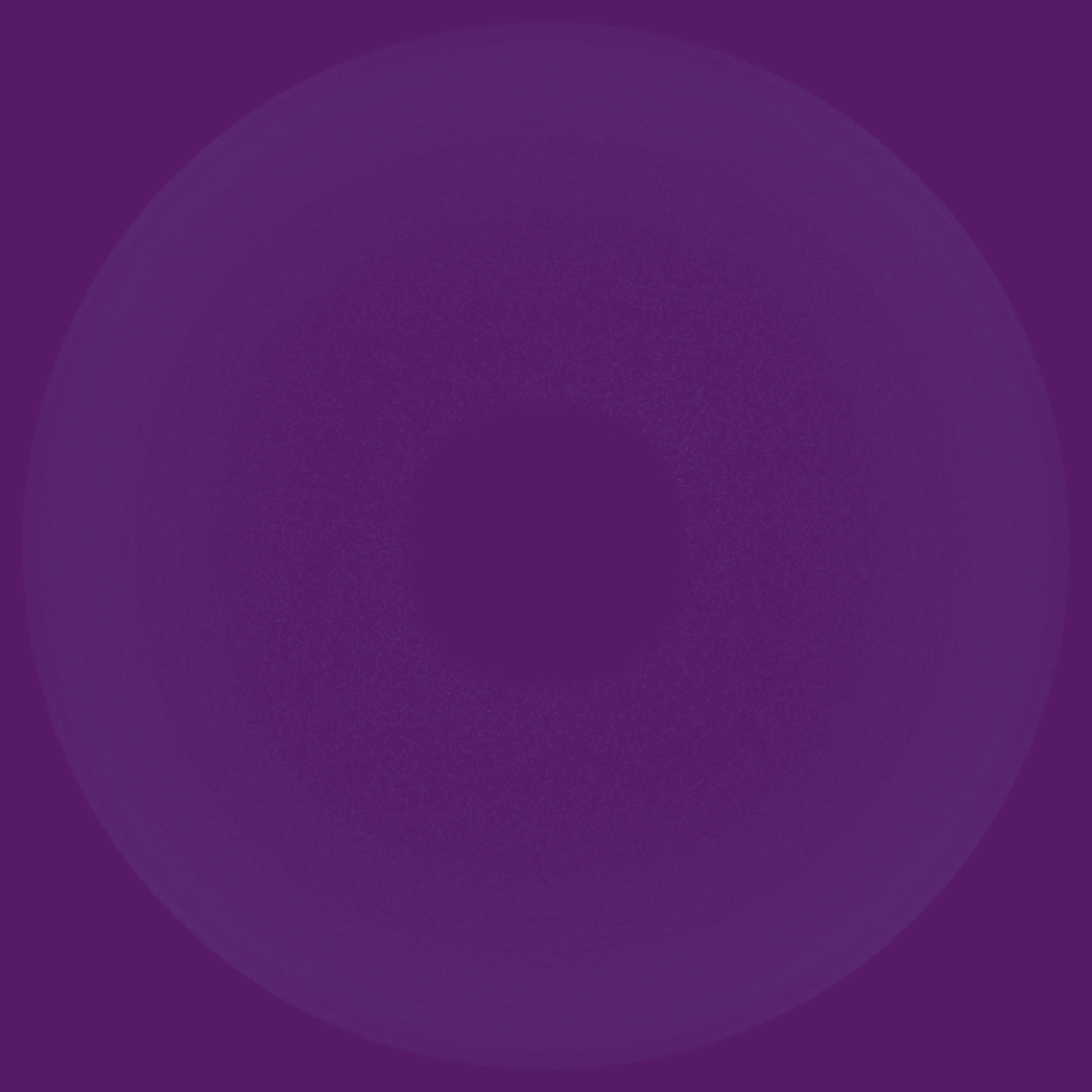}};
        
        \node at (-0.4, -0.2) {\includegraphics[width=0.17\linewidth]{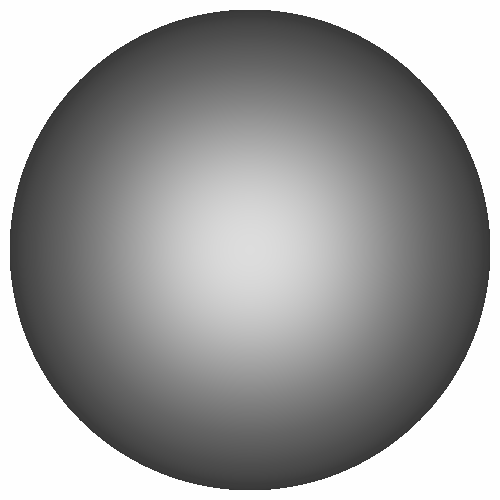}};
        \node at (0.2, -0.7) {\includegraphics[width=0.07\linewidth]{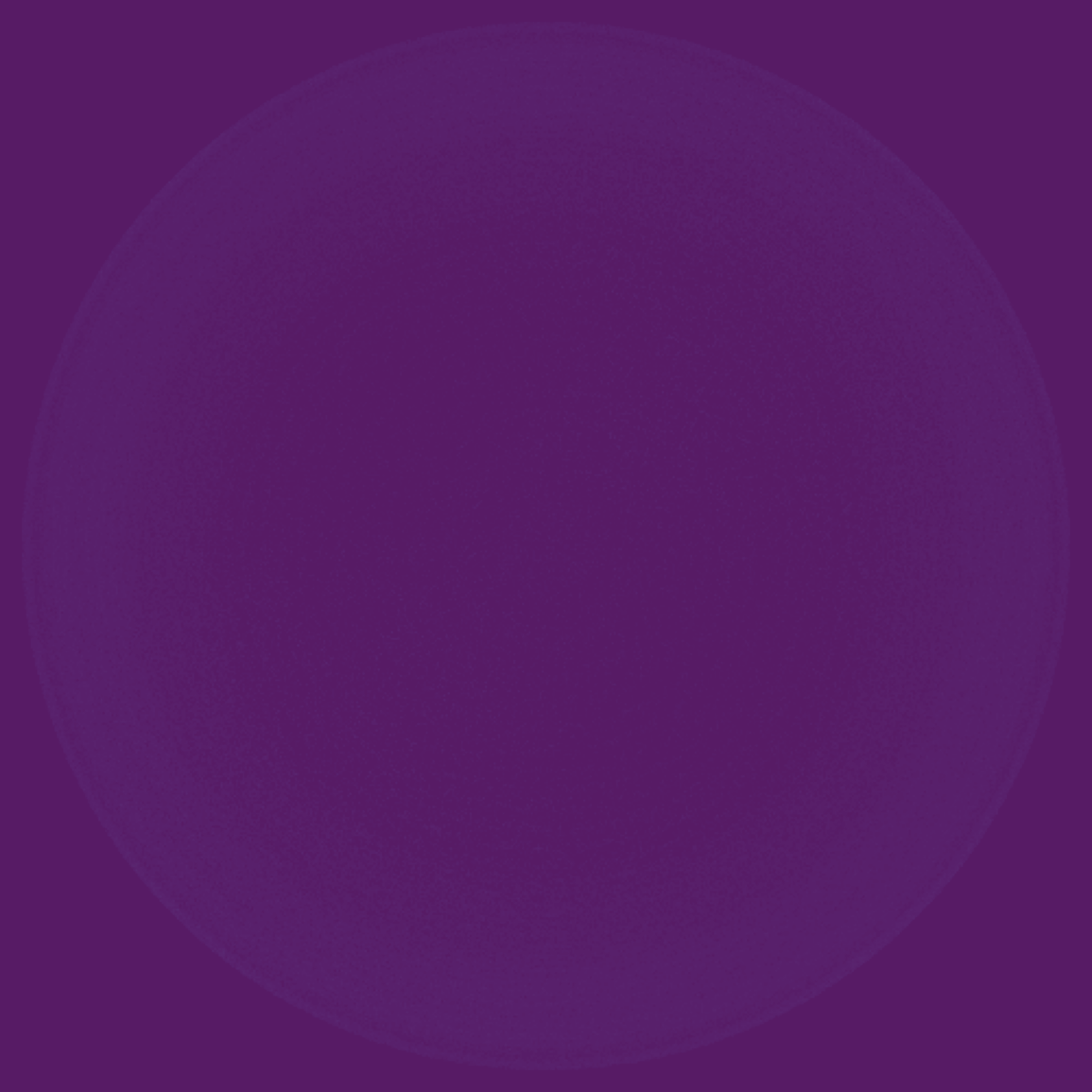}};
        
        \node at (1.4, -0.2) {\includegraphics[width=0.17\linewidth]{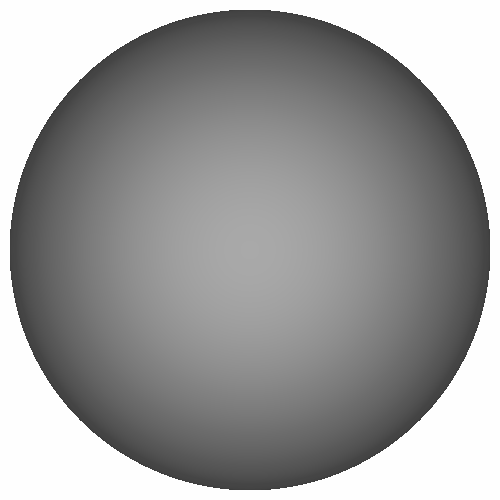}};
        \node at (2.0, -0.7) {\includegraphics[width=0.07\linewidth]{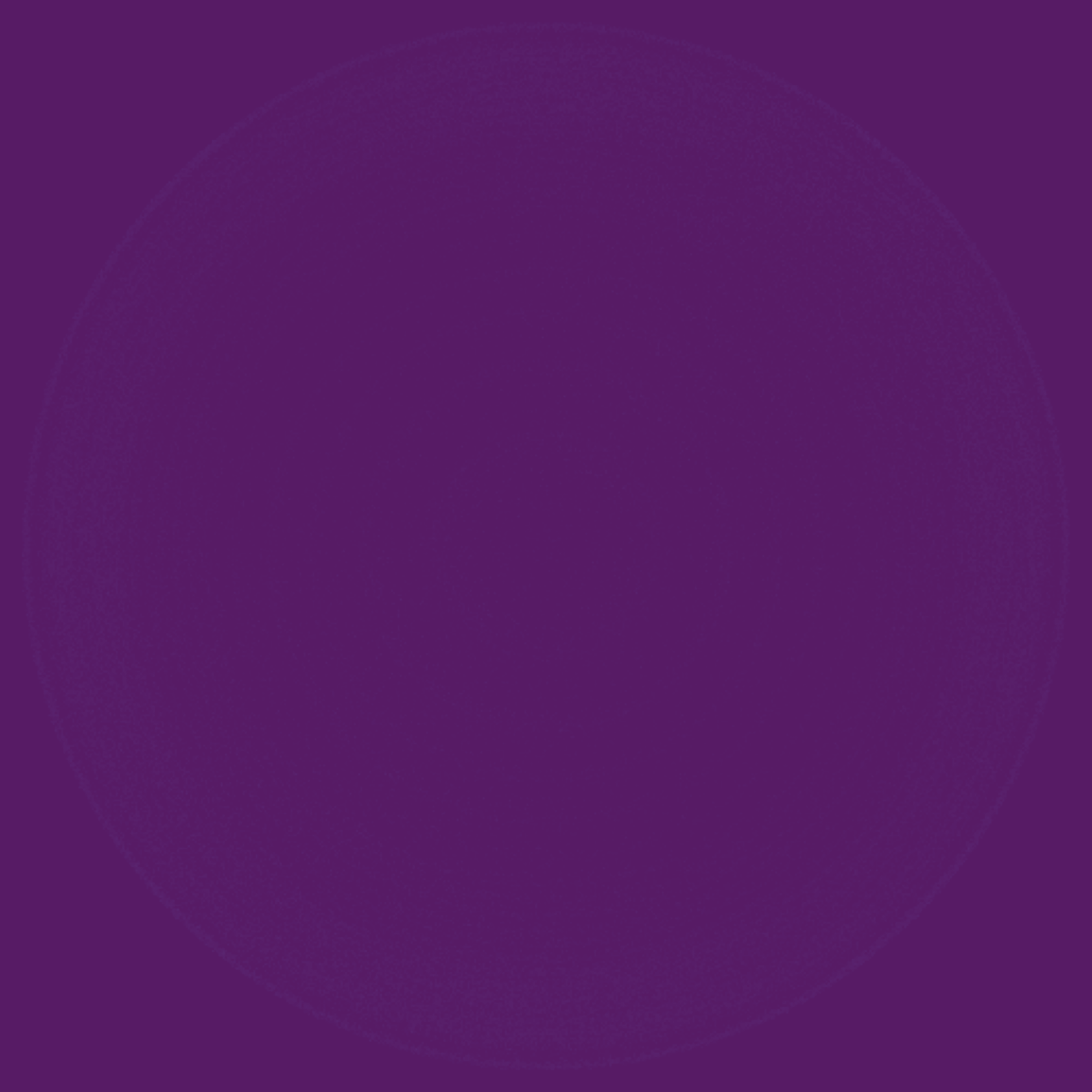}};

        \node at (3.0, -1) {\includegraphics[width=0.04\linewidth]{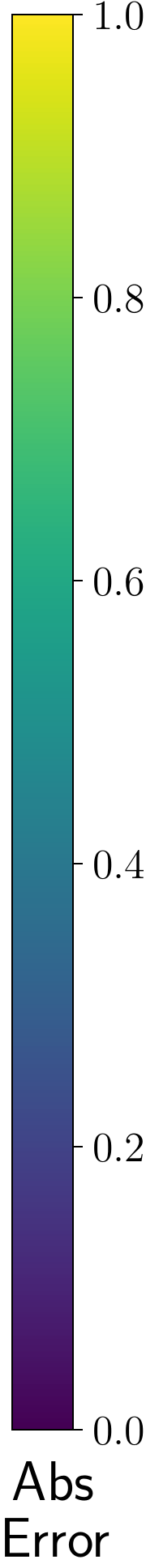}};

        \node at (-5, -2) {\rotatebox{90}{\footnotesize $64^4$ LUT}};
        
        \node at (-4, -2) {\includegraphics[width=0.17\linewidth]{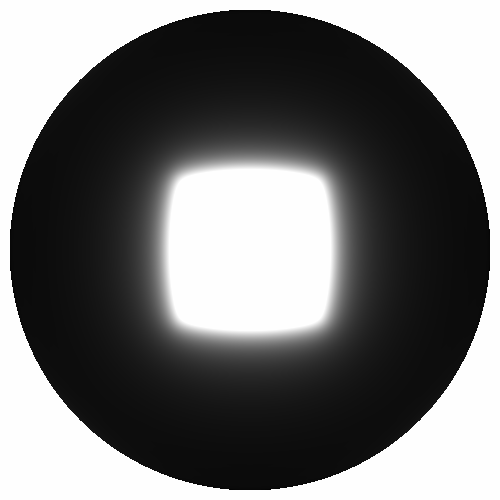}};
        \node at (-3.4, -2.5) {\includegraphics[width=0.07\linewidth]{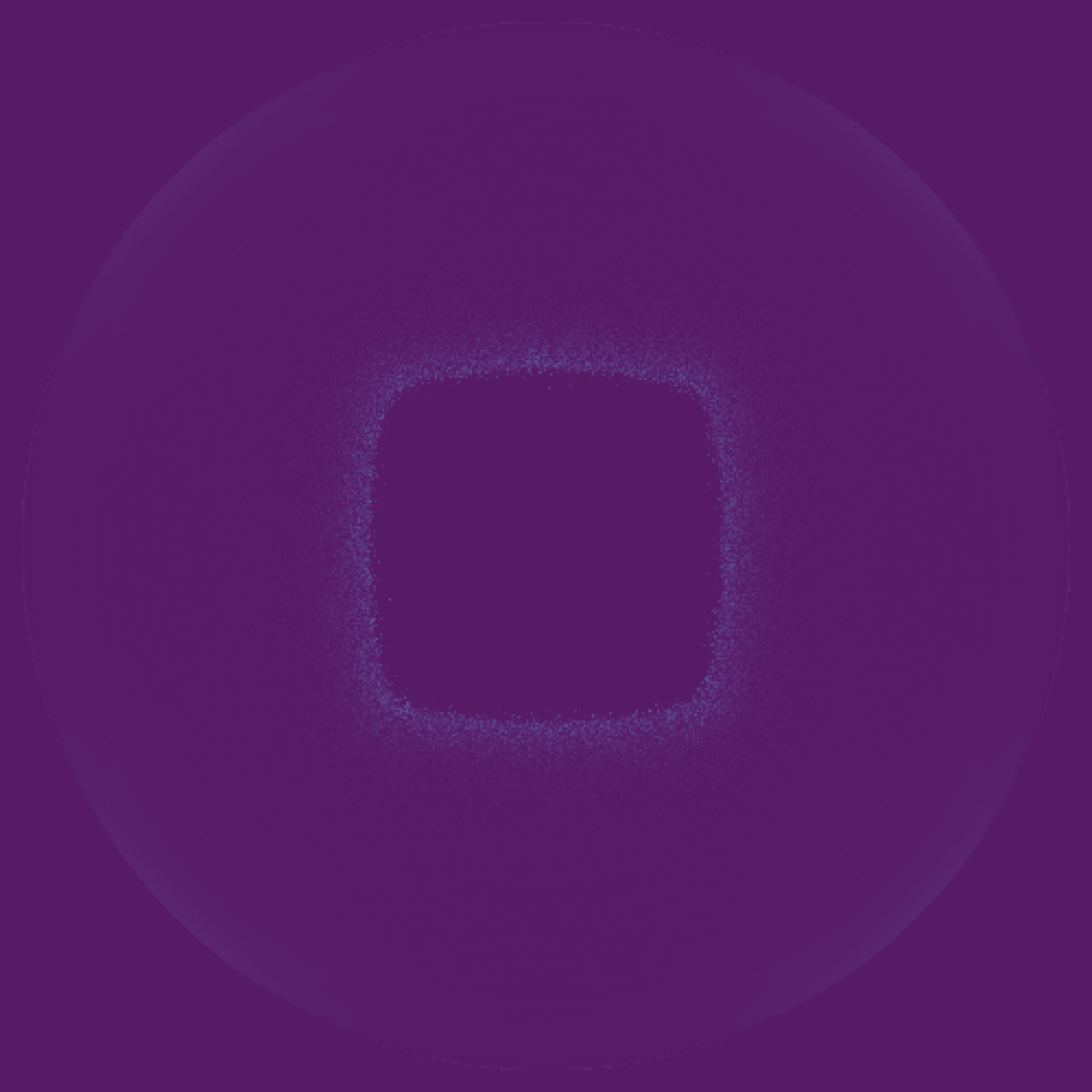}};
        
        \node at (-2.2, -2) {\includegraphics[width=0.17\linewidth]{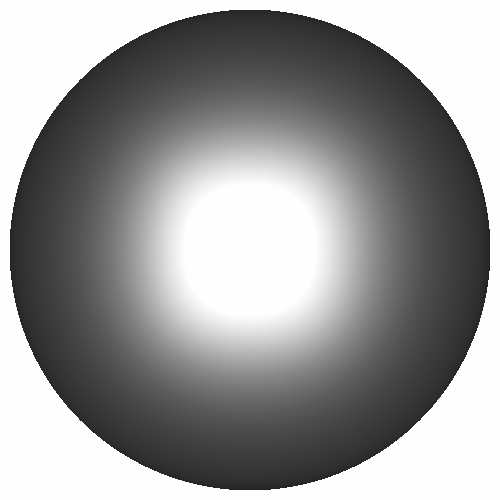}};
        \node at (-1.6, -2.5) {\includegraphics[width=0.07\linewidth]{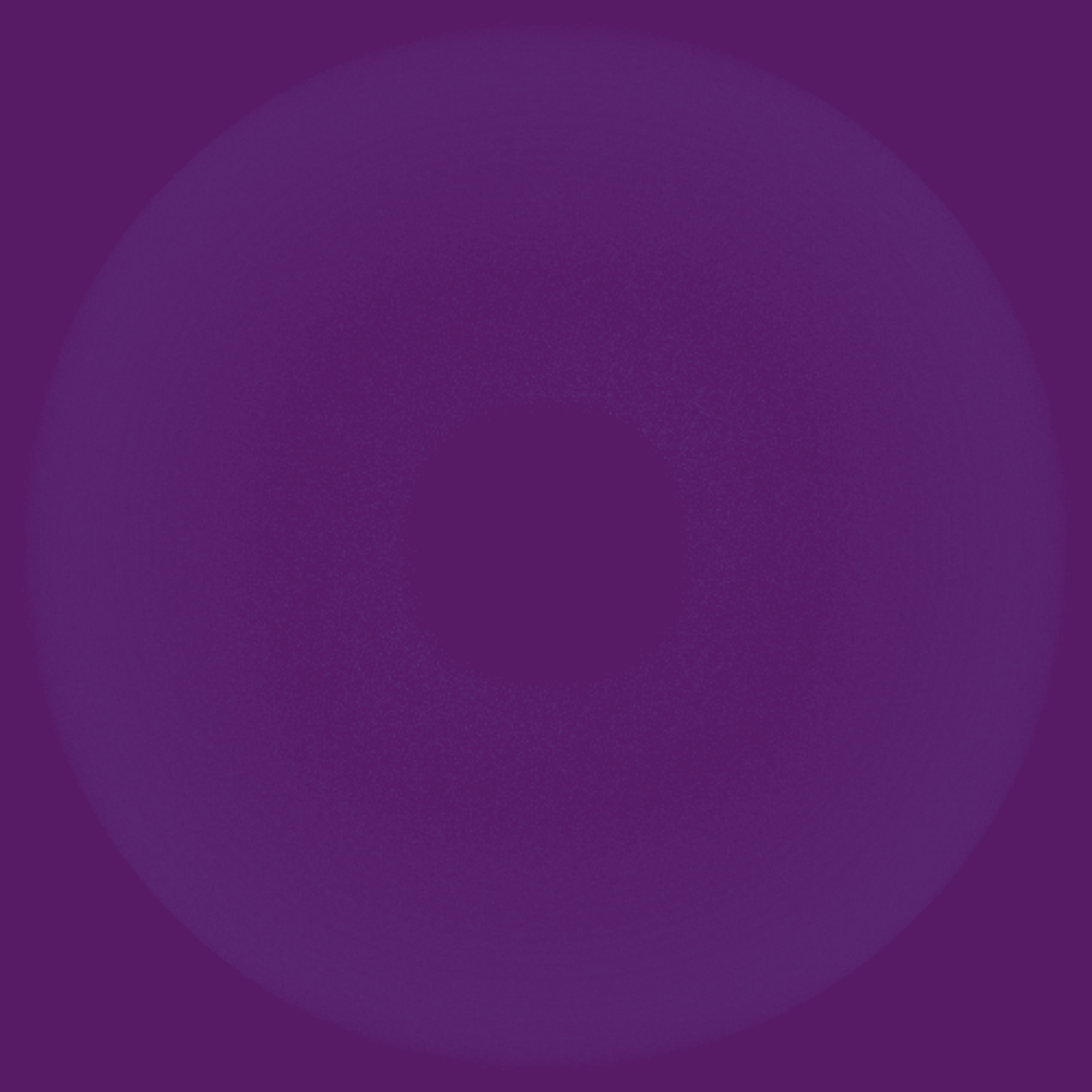}};
        
        \node at (-0.4, -2) {\includegraphics[width=0.17\linewidth]{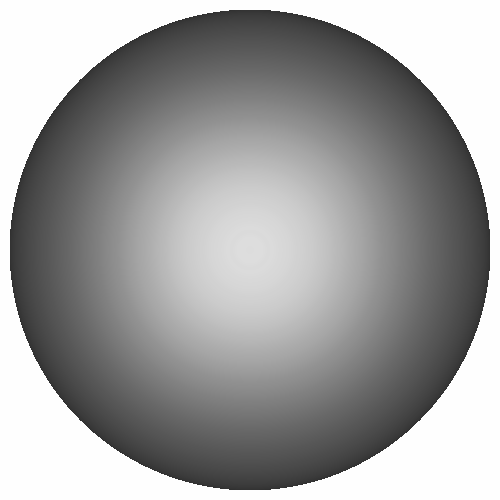}};
        \node at (0.2, -2.5) {\includegraphics[width=0.07\linewidth]{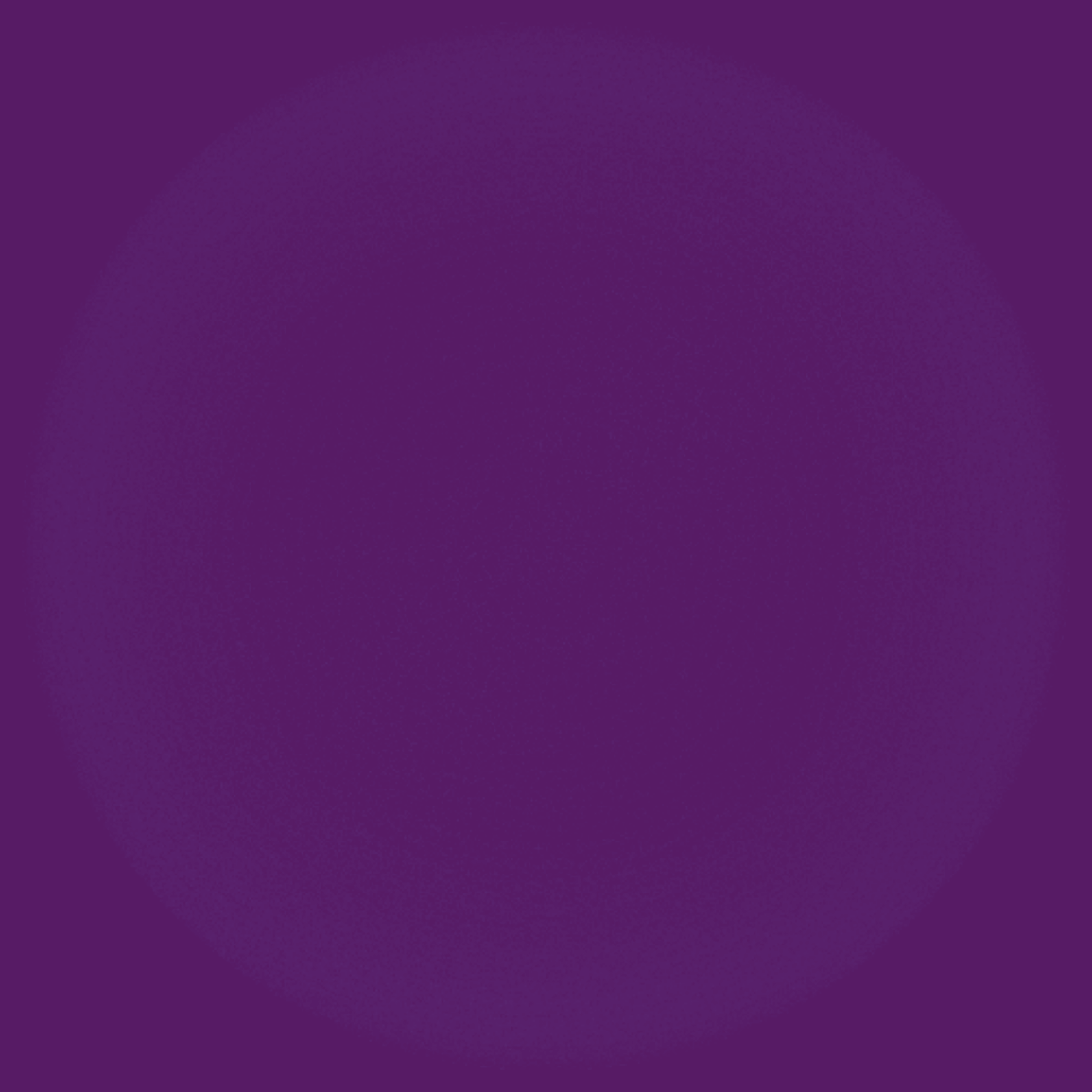}};
        
        \node at (1.4, -2) {\includegraphics[width=0.17\linewidth]{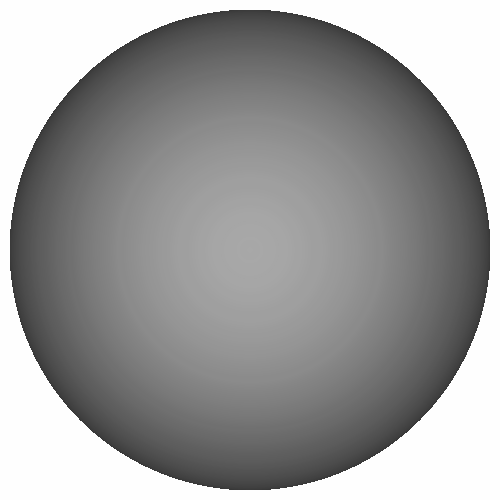}};
        \node at (2.0, -2.5) {\includegraphics[width=0.07\linewidth]{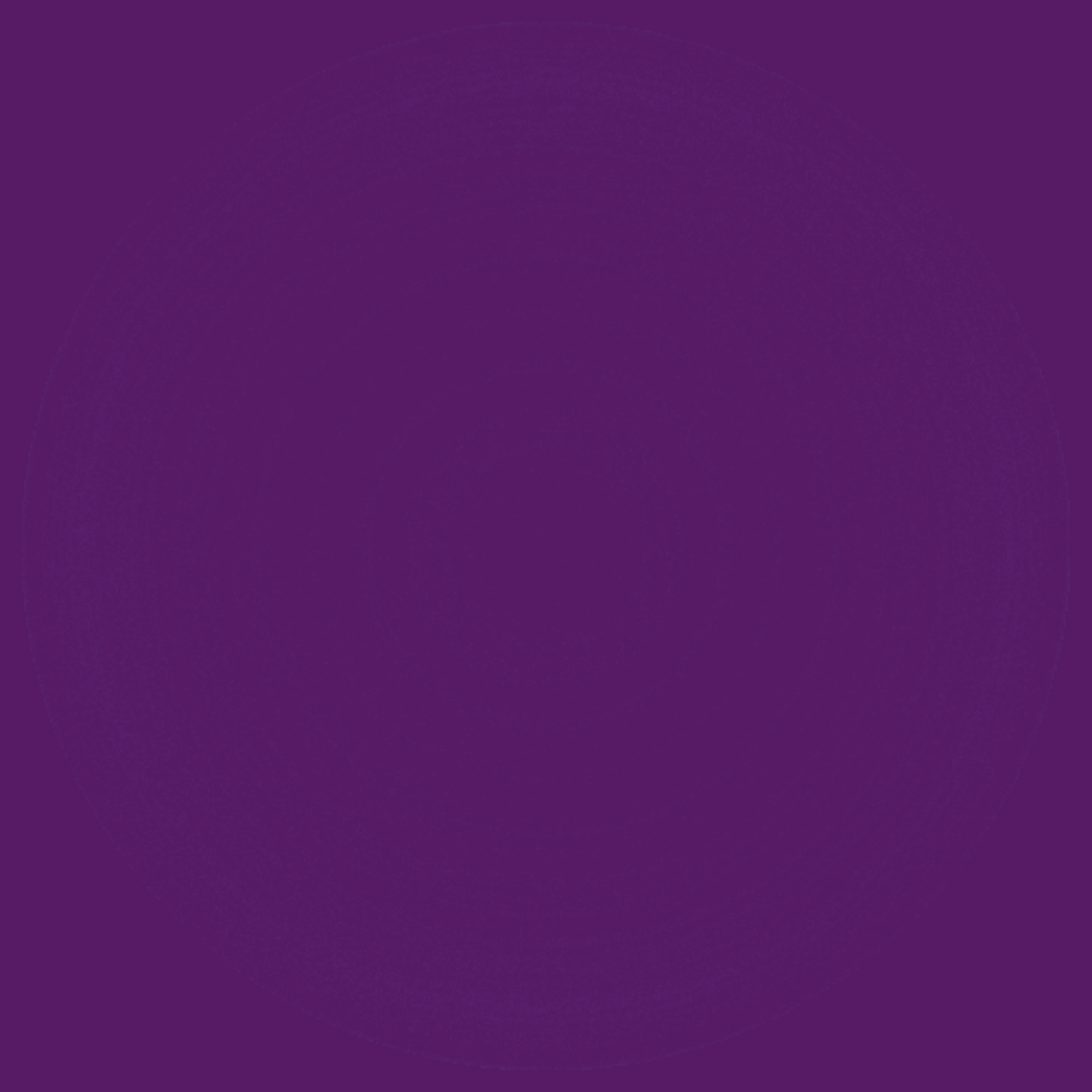}};

    \end{tikzpicture}
    \vspace{-8mm}
    \caption{Renderings using $8^4$ and $64^4$ look-up tables. Inset difference images are with respect to the GGX reference.}
    \label{fig:lut_size_renders}
\end{figure}
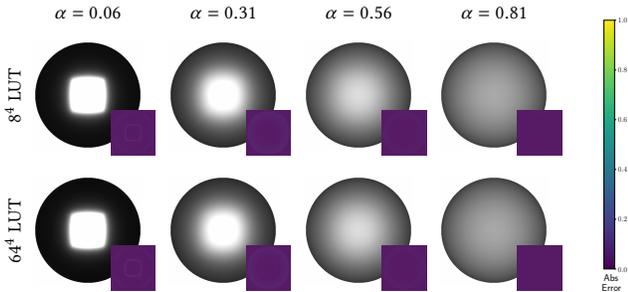

Heitz et al.~\cite{heitz2016} use a $64^2$ resolution for their 2D look-up table $T_{\text{isoGGX}}(\theta,\alpha)$ with five floats per entry (they only use five parameters of the matrix $M$), which represents about 80~KB.
If we were to use the same angular and roughness resolution, i.e., a $64^4$ 4D look-up table with nine floats per entry (we use the full matrix $M$), the memory requirement would be 576~MB.
Thanks to the design choices introduced in the previous sections, we are able to reduce a resolution to $8^4$.
In Figure~\ref{fig:lut_size_renders}, we compare a resolution of $8^4$ and $64^4$ with respect to the GGX reference.
The results show that the additional discretization error at $8^4$ is negligible compared to the LTC approximation.
As such, there is little to be gained by using a larger resolution. 

%% file: section_implementation.tex
\section{Implementation of our Method}
\label{sec:implementation}

In this section, we explain the implementation of our method, and 
summarize its requirements in Table~\ref{table:recap}.

\paragraph{Precomputations.}

We compute an $8^4$ table $T_{\text{anisoGGX}}(\theta, \phi, \alpha, \lambda)$ with a uniform discretization over $\theta \in [0, \frac{\pi}{2}]$, $\phi \in [0, \frac{\pi}{2}]$, $\alpha \in [0, 1]$ and $\lambda \in [0, 1]$.
We compute each entry in the following way:
\begin{itemize}
 \item We fit $M$ to the corresponding anisotropic GGX lobe of roughnesses $\alpha_x=\alpha$ and $\alpha_y=\lambda\,\alpha$ using Algorithm~\ref{alg:fitting} from Section~\ref{sec:fitting}.
 \item We align $M$ by minimizing Equation~(\ref{eq:optim_align}) from Section~\ref{sec:interpolation}, to ensure that interpolation is  well-defined. 
 \item We fix residual errors in $M$ by applying Equation~(\ref{eq:postprocessing}) from Section~\ref{sec:symmetries}.
 \item We make $M$ well-conditioned by dividing it by the length of its last column, as explained in Section~\ref{sec:inversion}.
\end{itemize}
The whole precomputation procedure is implemented in PyTorch and takes around two hours on an NVIDIA GeForce 2080 RTX GPU.

\paragraph{Storage.}

We store the parameters in 3D textures of resolution $64\times8\times8$ to benefit from trilinear hardware interpolation.

\paragraph{Run time.}

In the fragment shader, we proceed as follows:
\begin{itemize}
\item We get $\theta$, $\phi$, $\alpha_x$, and $\alpha_y$.
\item We map $\phi$ to the first quarter $[0, \frac{\pi}{2}]$ using Equation~(\ref{eq:symmetry_azimuthal}) and compute $\alpha$ and $\lambda$ using Equation~(\ref{eq:symmetry_roughness}).
\item  We fetch the 3D textures accordingly, manually interpolate over the last dimension, and apply the flipping and/or rotation matrices following Equations~(\ref{eq:symmetry_azimuthal}) and (\ref{eq:symmetry_roughness}). This provides us with the matrix $M$.
\item We invert $M$ to obtain $M^{-1}$.
\end{itemize}
The cost of these operations is 0.610 ms for a full-screen quad at 1080p resolution with an NVIDIA GeForce RTX 2080 GPU, which is about four times the cost of the isotropic version.

\paragraph{LTC integration.}

Once $M^{-1}$ is obtained, we use the existing LTC integration algorithms as in previous work~\cite{heitz2016,heitz2017a,heitz2017b}.  
Note that the overhead of our method only impacts the obtainment of $M^{-1}$, which can be amortized over the integration of multiple area lights. 

\paragraph{Fresnel.}

We inject the Fresnel term of the GGX BRDF in the same way as Hill et al.~\cite{hill2016}.
We preintegrate the first and second Fresnel terms for each $(\theta,\phi,\alpha,\lambda)$ entry and store them as two additional channels in the look-up table that we interpolate at run time.

\begin{table}[!h]
\centering
\begin{tabular}{@{\hspace{-3mm}}c@{}}
\begin{tabular}{@{} | @{\hspace{1mm}} l @{\hspace{1mm}}|@{\hspace{1mm}} c @{\hspace{1mm}}|@{\hspace{1mm}} c @{\hspace{1mm}} | @{}}
\hline 
& Heitz et al.~\cite{heitz2016,hill2016} & ours \\
\hline 
param. & $M^{-1} = T_{\text{isoGGX}}(\theta,\alpha)$ & $M = T_{\text{anisoGGX}}(\theta,\phi,\alpha,\lambda)$\\
resolution & $64\times64$ & $8\times8\times8\times8$ \\
containers & 2D textures ($64\times64$) & 3D textures ($64\times8\times8$) \\
channels & 5 (for $M$) + 2 (for Fresnel) & 9 (for $M$) + 2 (for Fresnel)  \\
memory & 112 KB & 176 KB \\
interpolation & HW 2D & HW 3D + SW 1D \\
inversion & - & fragment shader \\
total timing & 0.160 ms & 0.610 ms \\
\hline LTC integration & \multicolumn{2}{c|}{0.110 ms/light} \\
\hline 
\end{tabular}
\end{tabular}
\caption{\label{table:recap} Requirements of our method. }
\vspace{-10mm}
\end{table}

%% file: section_results.tex
\section{Results}
\label{sec:results}

In this section, we discuss the results produced by our method. 
Note that our supplemental material covers a dense set of plot and rendering configurations. 

\paragraph{Plots.}

\begin{figure}[!h]
\centering
\begin{tabular}{@{} c @{\hspace{1mm}} c @{}}
\begin{tikzpicture}
\draw (0.0, 0.0) node {\includegraphics[width=0.20\linewidth]{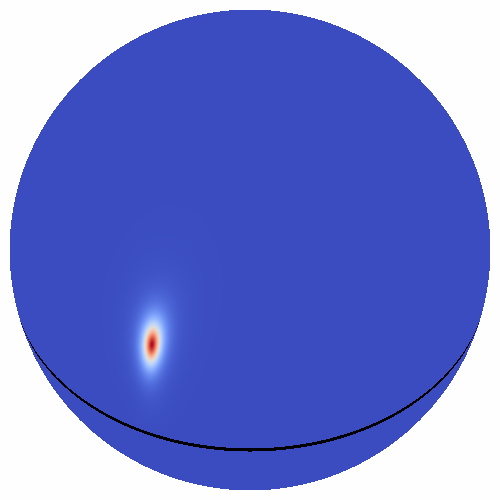}};
\draw (1.8, 0.0) node {\includegraphics[width=0.20\linewidth]{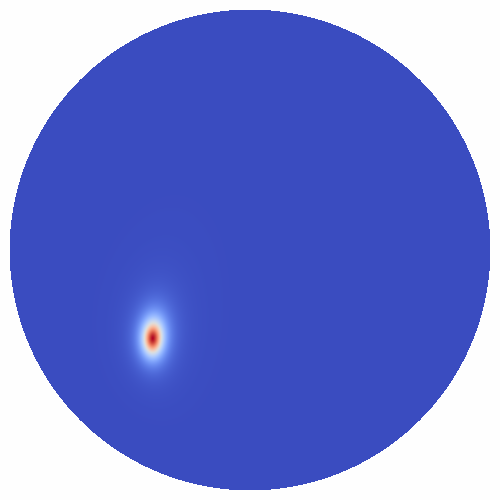}};
\draw (0.9, 1.3) node {$\alpha_x=0.05, \alpha_y=0.05$};
\draw (0.9, 1.0) node {$\theta=66°$, $\phi=294°$};
\end{tikzpicture}
&
\begin{tikzpicture}
\draw (0.0, 0.0) node {\includegraphics[width=0.20\linewidth]{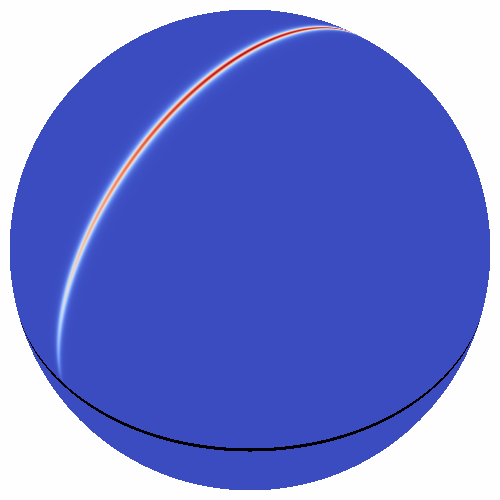}};
\draw (1.8, 0.0) node {\includegraphics[width=0.20\linewidth]{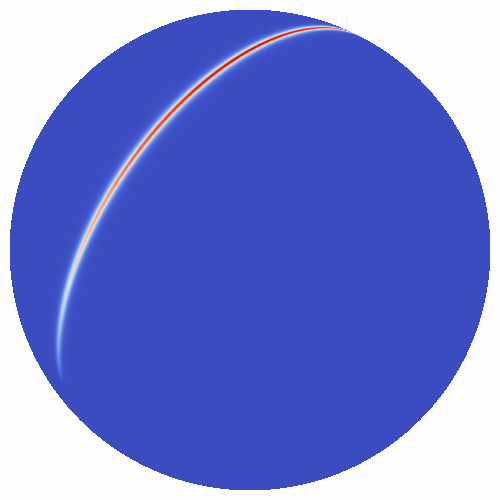}};
\draw (0.9, 1.3) node {$\alpha_x=0.01, \alpha_y=0.80$};
\draw (0.9, 1.0) node {$\theta=0°$, $\phi=272°$};
\end{tikzpicture}
\\
\begin{tikzpicture}
\draw (0.0, 0.0) node {\includegraphics[width=0.20\linewidth]{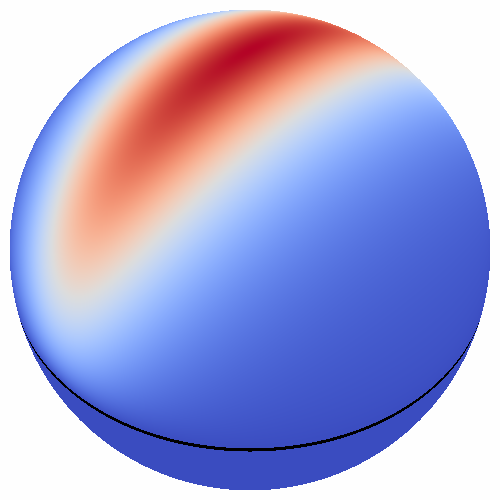}};
\draw (1.8, 0.0) node {\includegraphics[width=0.20\linewidth]{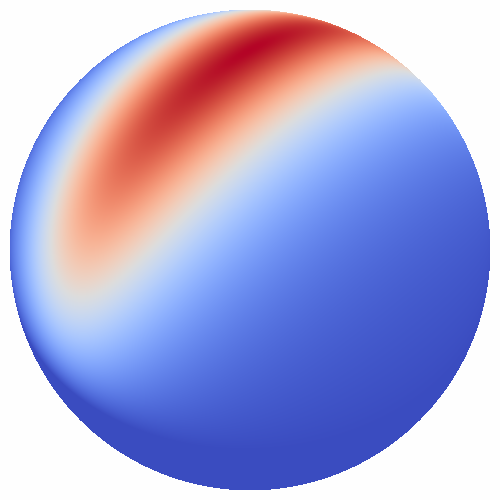}};
\draw (0.9, 1.3) node {$\alpha_x=0.25, \alpha_y=0.80$};
\draw (0.9, 1.0) node {$\theta=0°$, $\phi=316°$};
\end{tikzpicture}
&
\begin{tikzpicture}
\draw (0.0, 0.0) node {\includegraphics[width=0.20\linewidth]{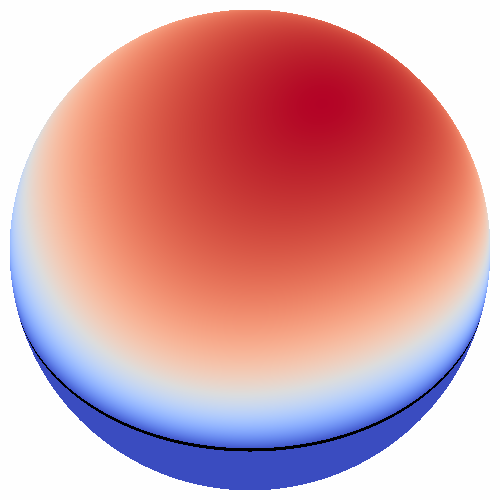}};
\draw (1.8, 0.0) node {\includegraphics[width=0.20\linewidth]{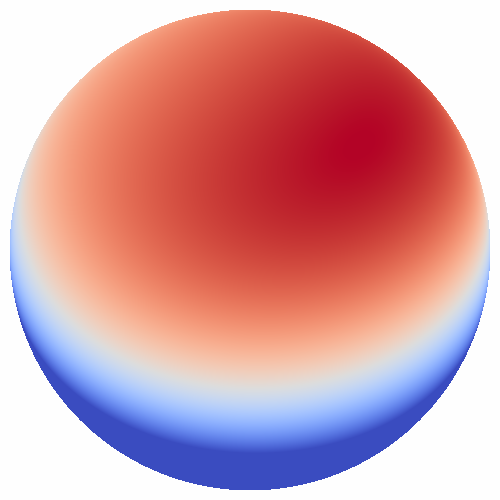}};
\draw (0.9, 1.3) node {$\alpha_x=0.80, \alpha_y=1.00$};
\draw (0.9, 1.0) node {$\theta=44°$, $\phi=316°$};
\end{tikzpicture}
\\
\hline
\\
\begin{tikzpicture}
\draw (0.0, 0.0) node {\includegraphics[width=0.20\linewidth]{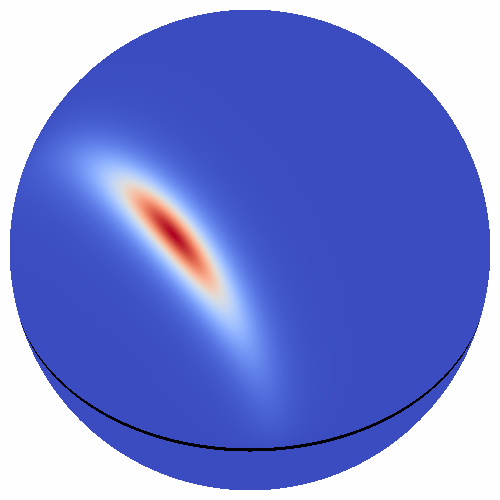}};
\draw (1.8, 0.0) node {\includegraphics[width=0.20\linewidth]{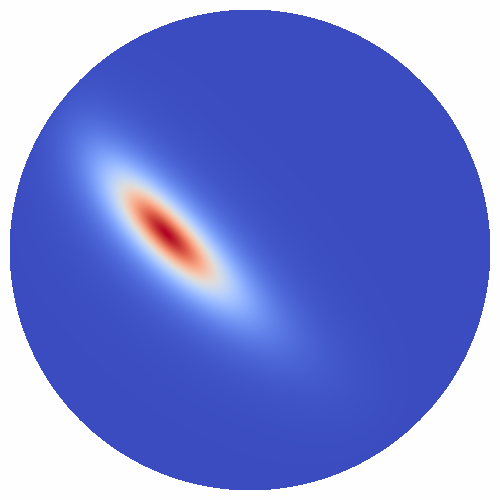}};
\draw (0.9, 1.3) node {$\alpha_x=0.25, \alpha_y=0.05$};
\draw (0.9, 1.0) node {$\theta=44°$, $\phi=294°$};
\end{tikzpicture}
&
\begin{tikzpicture}
\draw (0.0, 0.0) node {\includegraphics[width=0.20\linewidth]{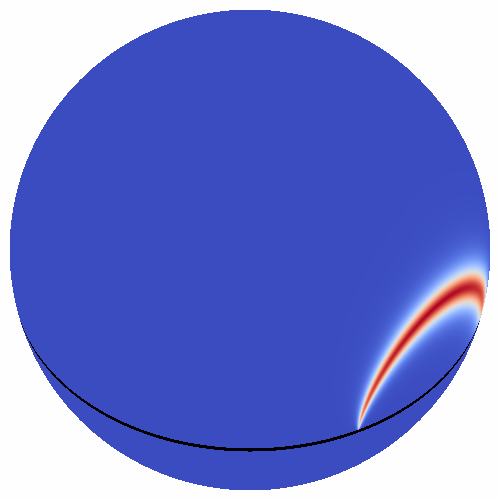}};
\draw (1.8, 0.0) node {\includegraphics[width=0.20\linewidth]{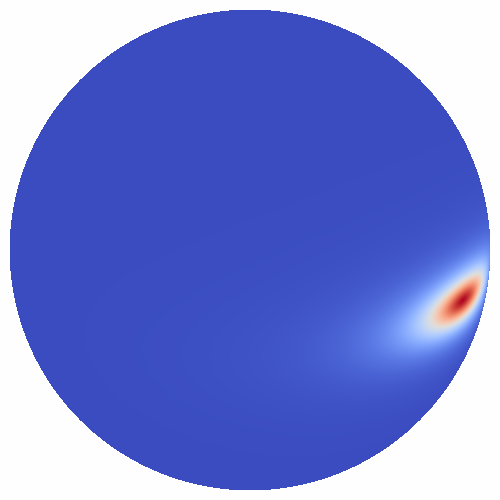}};
\draw (0.9, 1.3) node {$\alpha_x=0.05, \alpha_y=1.00$};
\draw (0.9, 1.0) node {$\theta=88°$, $\phi=338°$};
\end{tikzpicture}
\\
\begin{tikzpicture}
\draw (0.0, 0.0) node {\includegraphics[width=0.20\linewidth]{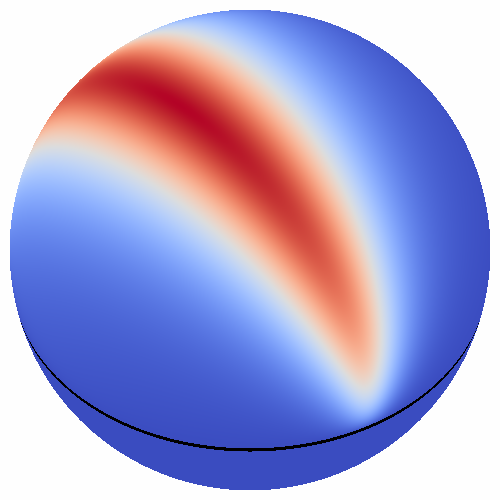}};
\draw (1.8, 0.0) node {\includegraphics[width=0.20\linewidth]{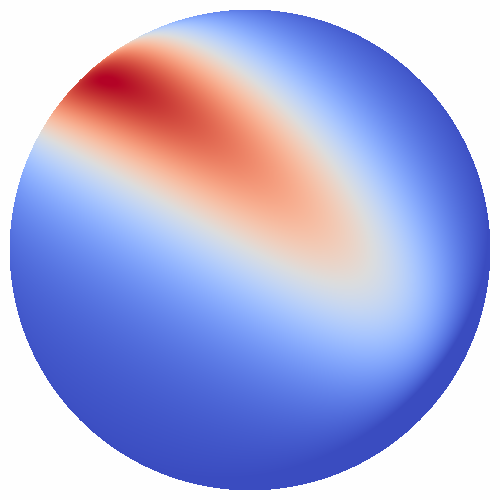}};
\draw (0.9, 1.3) node {$\alpha_x=1.00, \alpha_y=0.25$};
\draw (0.9, 1.0) node {$\theta=66°$, $\phi=338°$};
\end{tikzpicture}
&
\begin{tikzpicture}
\draw (0.0, 0.0) node {\includegraphics[width=0.20\linewidth]{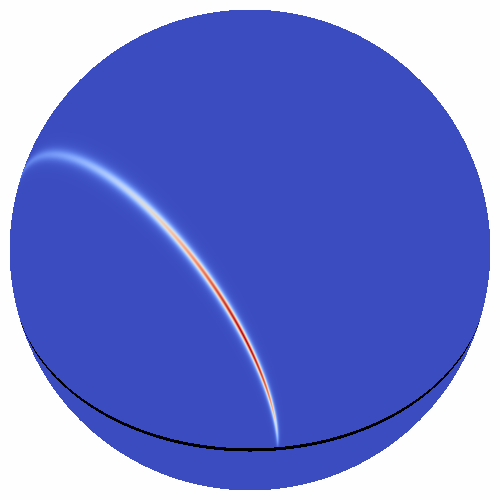}};
\draw (1.8, 0.0) node {\includegraphics[width=0.20\linewidth]{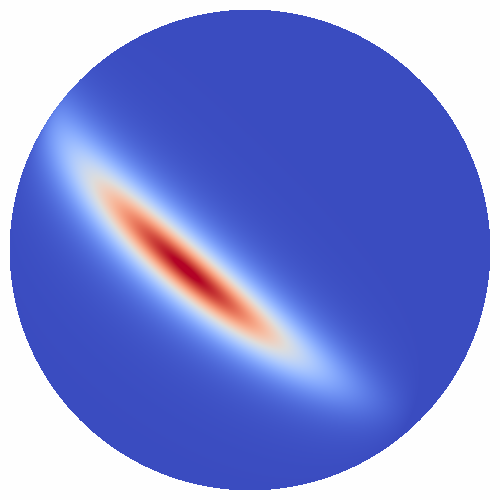}};
\draw (0.9, 1.3) node {$\alpha_x=0.50, \alpha_y=0.01$};
\draw (0.9, 1.0) node {$\theta=66°$, $\phi=316°$};
\end{tikzpicture}
\end{tabular}
\vspace{-5mm}
\caption{\label{fig:results_plots} Plot results. 
\textit{We show the GGX reference (left) and our LTC approximation (right).
More results in our supplemental material.}
}
\end{figure}
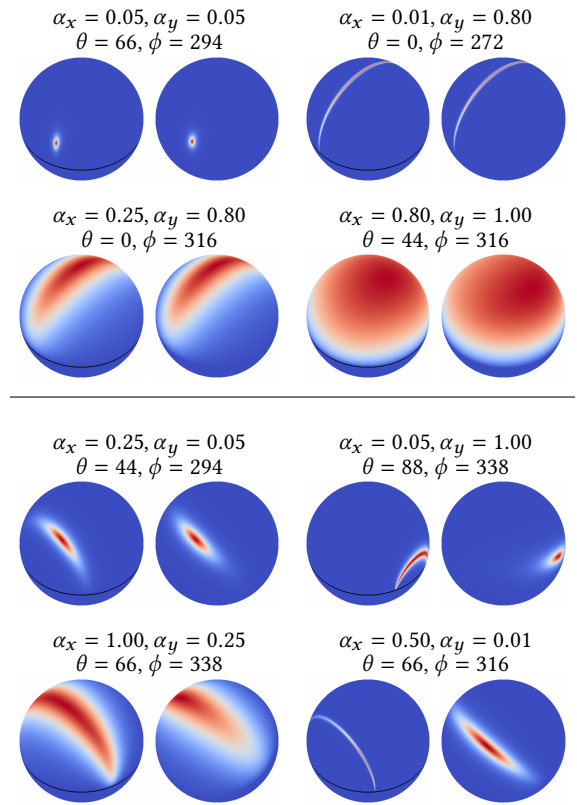

Figure~\ref{fig:results_plots} shows the fitted GGX lobes and our LTC approximation. 
We found out that the main limiting factor of our approximation is not our fitting technique but rather the representation power of LTCs. 
When the shape of the GGX lobe can be closely approximated by an LTC, our fitting technique is always successful (top rows in Fig.~\ref{fig:results_plots}).
However, the GGX lobe can exhibit \textit{lune shapes} in certain configurations (high anisotropy, grazing view angle) and these shapes cannot be represented by LTCs (bottom rows in Fig.~\ref{fig:results_plots}).
Indeed, an LTC is a diffuse distribution transformed by a linear transformation.
This allows for various first-order transformations such as changing the isotropic span of the lobe, elliptic anisotropy or skewness, but excludes lune-shaped lobes. 
In other words, it is not possible to accurately approximate these GGX configurations with LTCs, regardless of the fitting technique.

\paragraph{Renderings.}

\begin{figure}[!h]
\centering
\begin{tabular}{@{} c @{\hspace{3mm}} c @{}}
\begin{tabular}{@{} c @{\hspace{1mm}} c @{}}
\begin{tikzpicture}
\draw (0.0, 0.0) node {\includegraphics[width=0.2\linewidth]{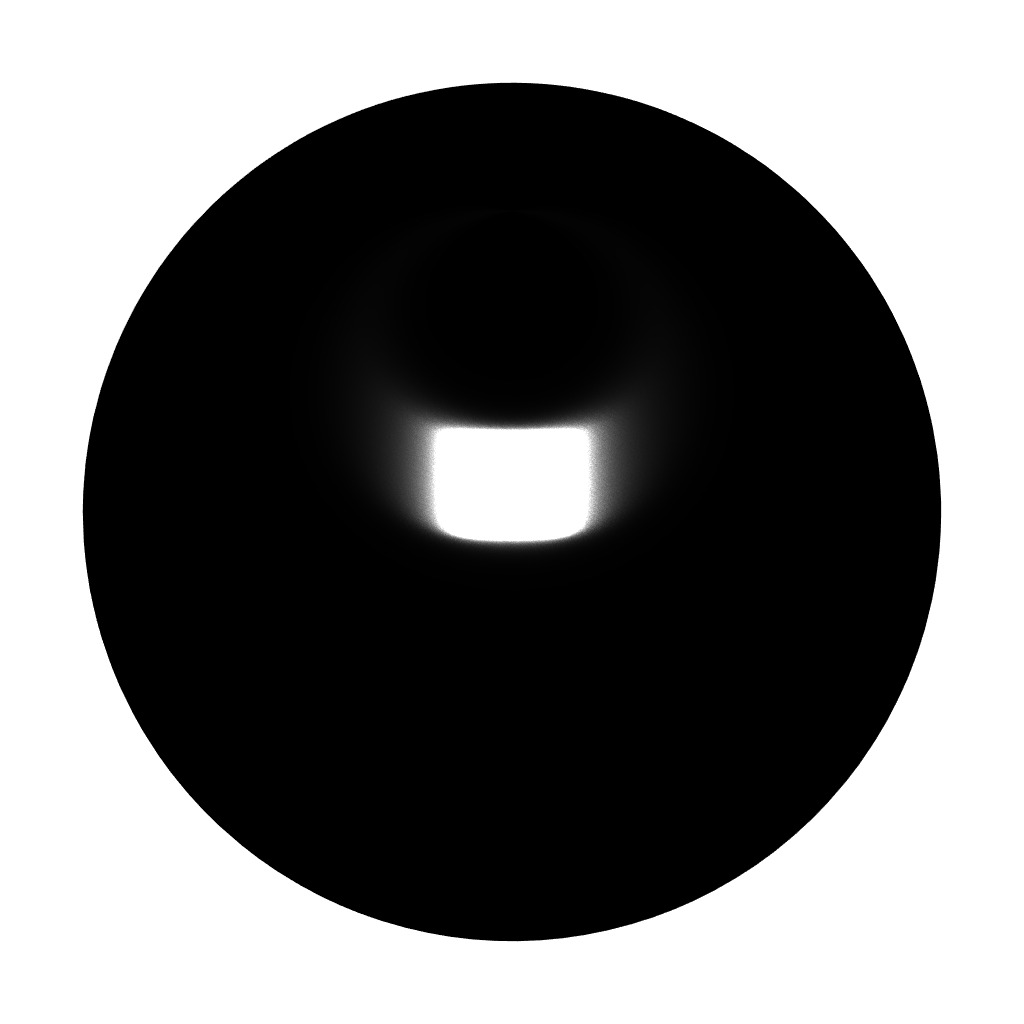}};
\draw (1.6, 0.0) node {\includegraphics[width=0.2\linewidth]{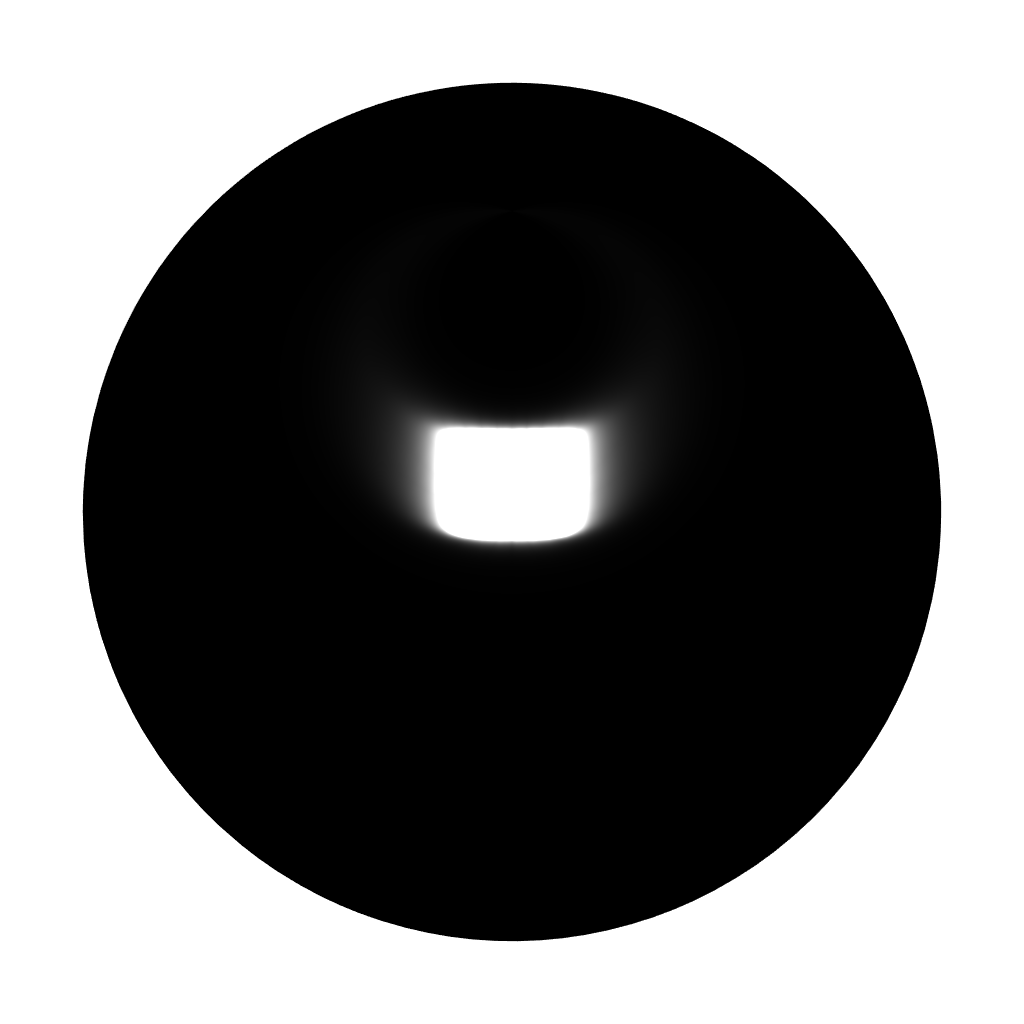}};
\draw (2.3, -0.7) node {\includegraphics[width=0.1\linewidth]{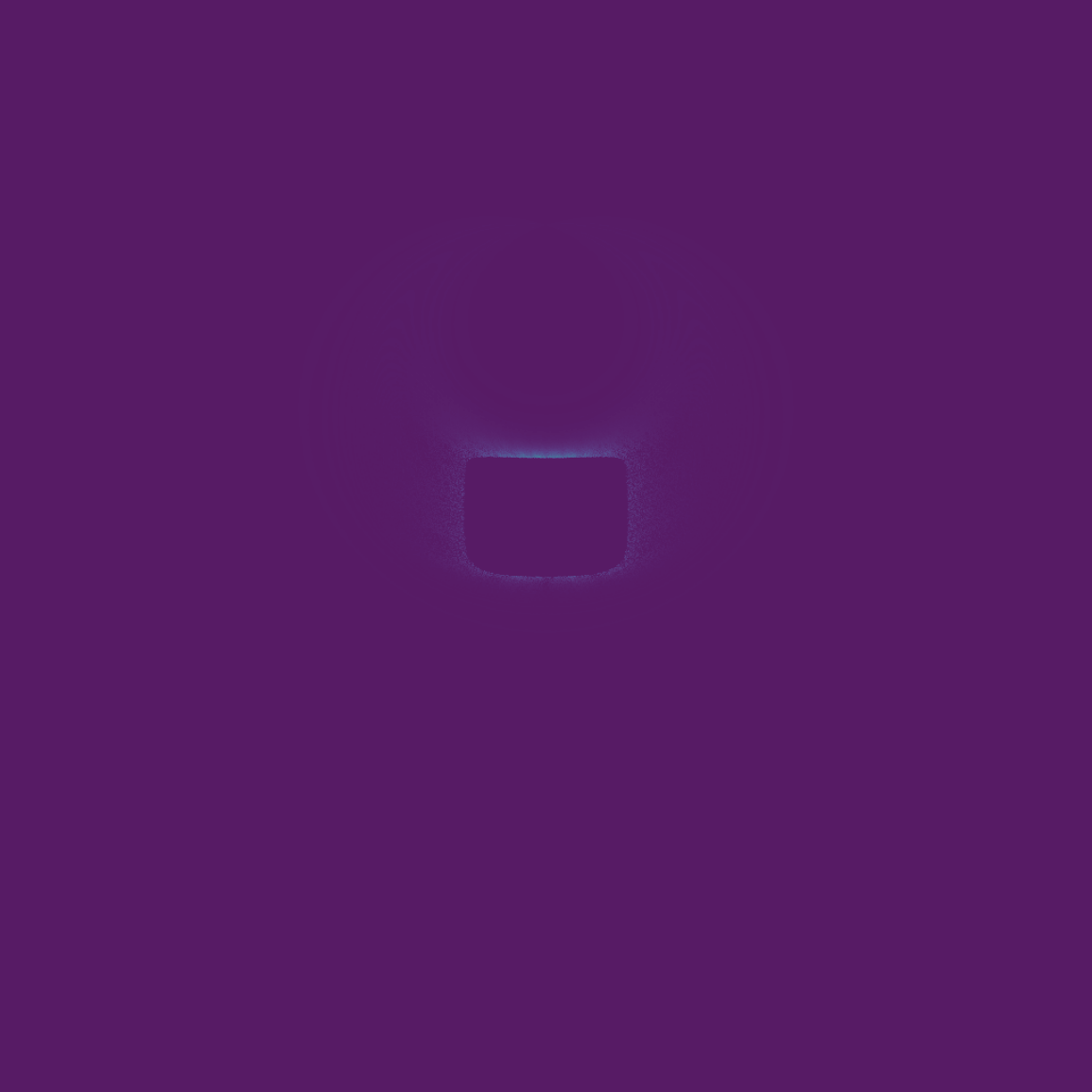}};
\draw (0.8, 1.0) node {$\alpha_x=0.01, \alpha_y=0.05$};
\end{tikzpicture}
&
\begin{tikzpicture}
\draw (0.0, 0.0) node {\includegraphics[width=0.2\linewidth]{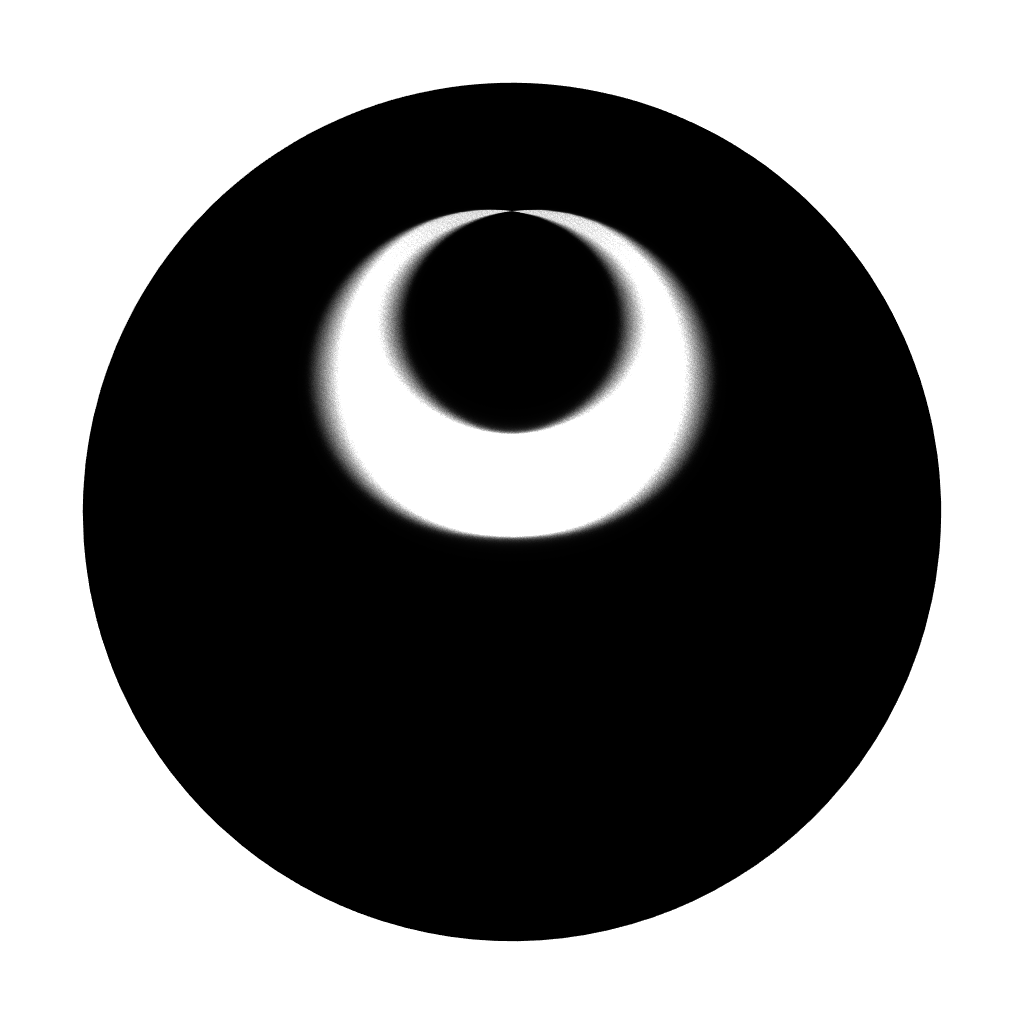}};
\draw (1.6, 0.0) node {\includegraphics[width=0.2\linewidth]{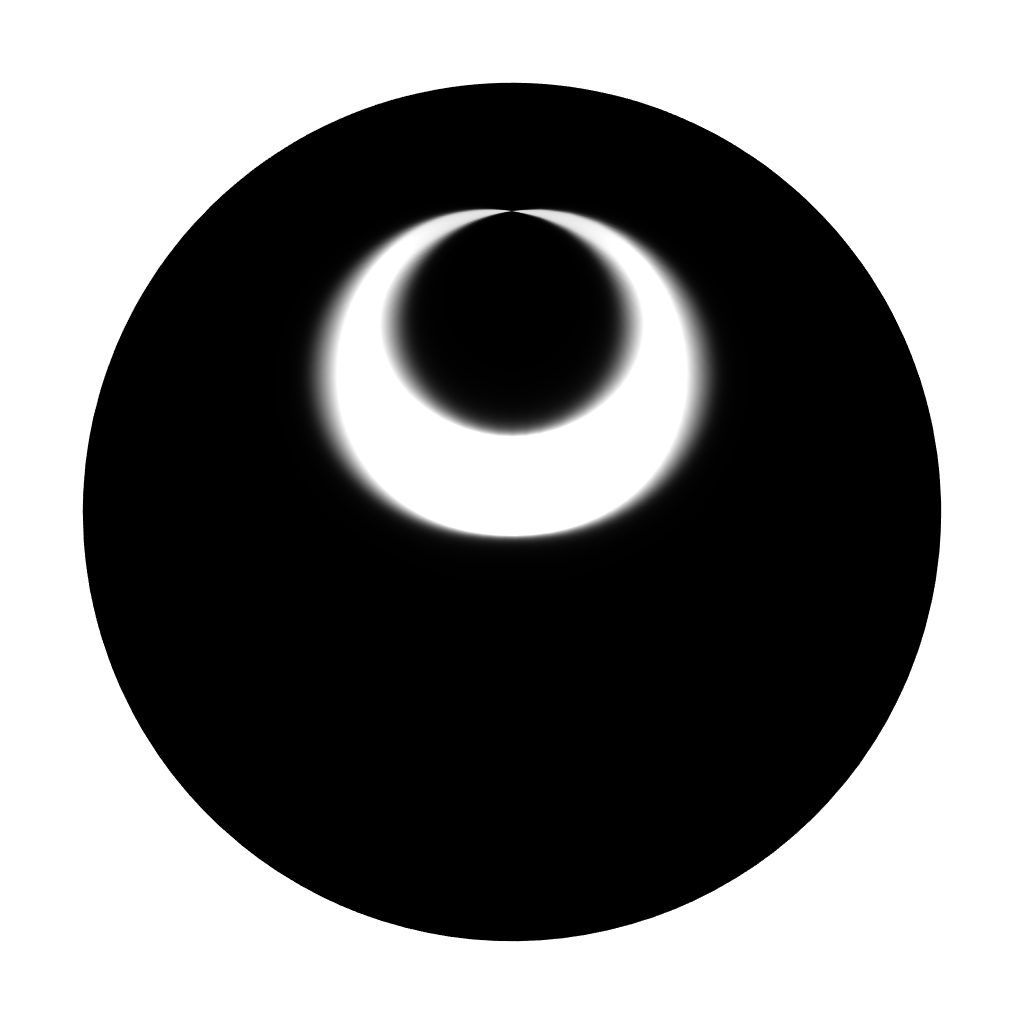}};
\draw (2.3, -0.7) node {\includegraphics[width=0.1\linewidth]{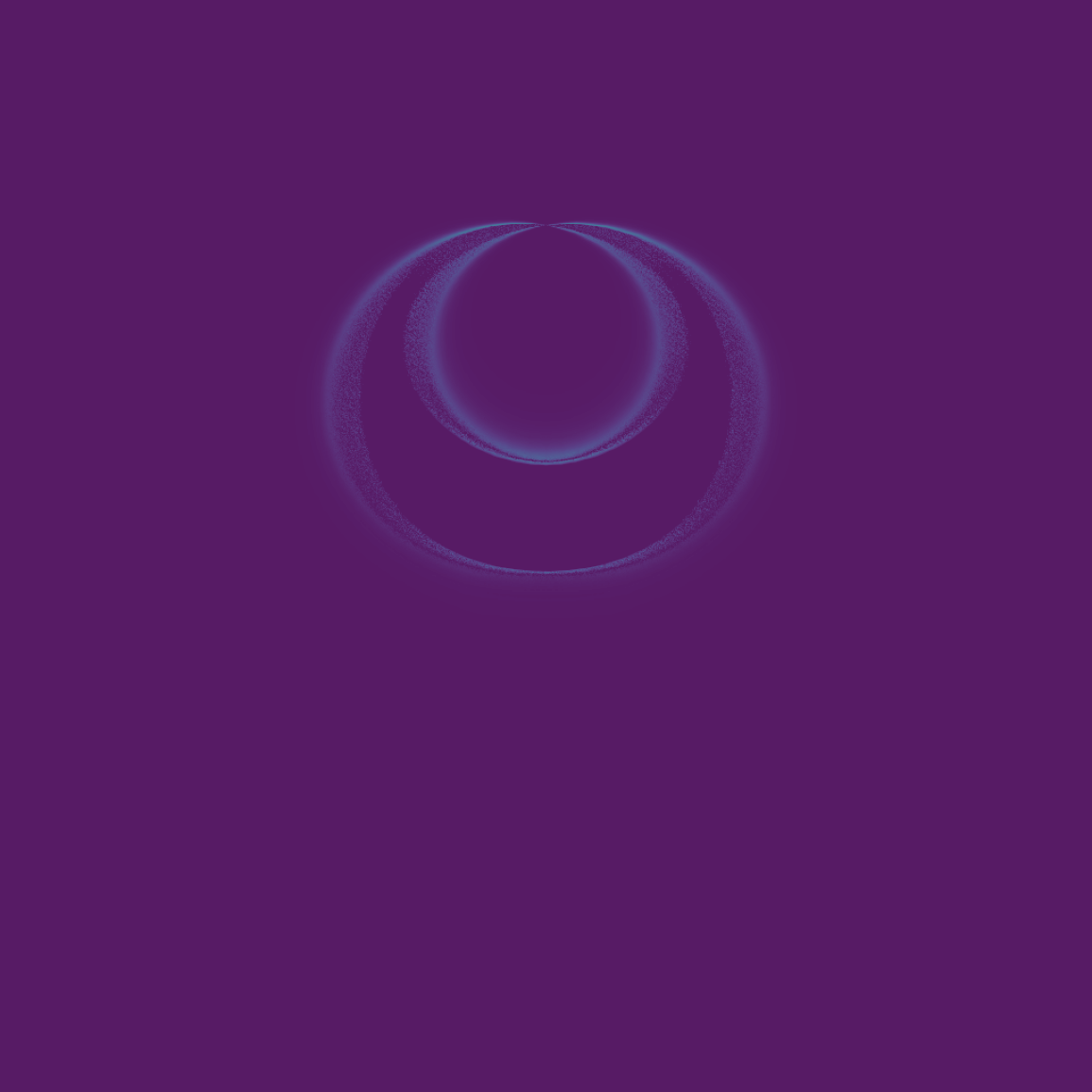}};
\draw (0.8, 1.0) node {$\alpha_x=0.01, \alpha_y=0.50$};
\end{tikzpicture}
\\
\begin{tikzpicture}
\draw (0.0, 0.0) node {\includegraphics[width=0.2\linewidth]{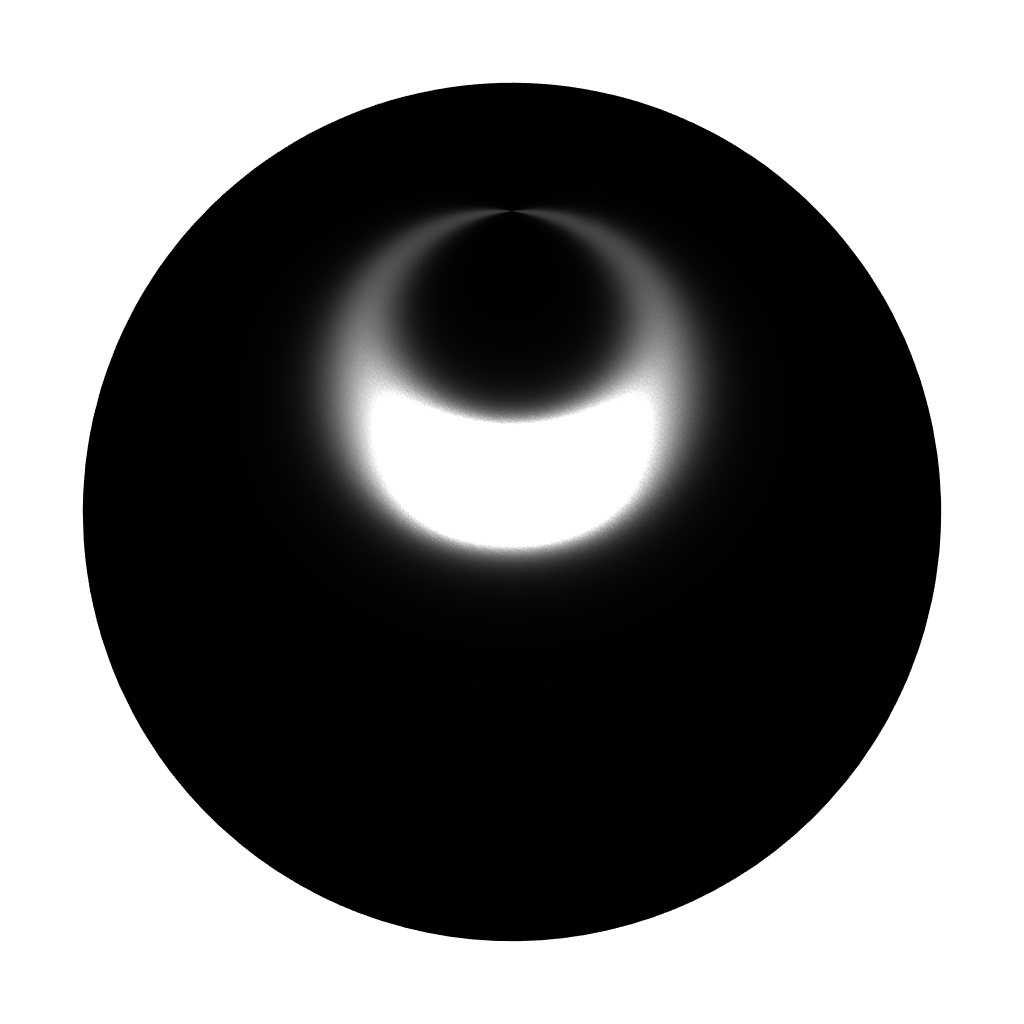}};
\draw (1.6, 0.0) node {\includegraphics[width=0.2\linewidth]{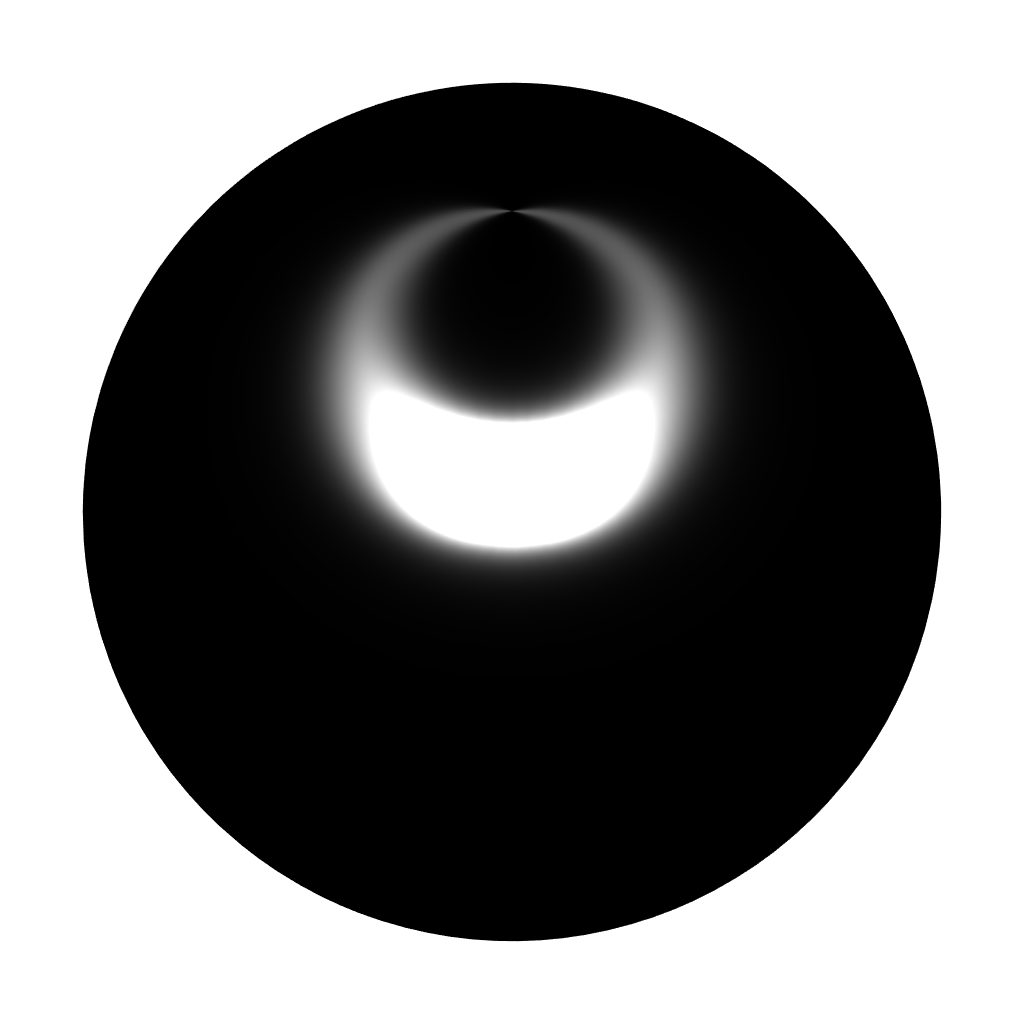}};
\draw (2.3, -0.7) node {\includegraphics[width=0.1\linewidth]{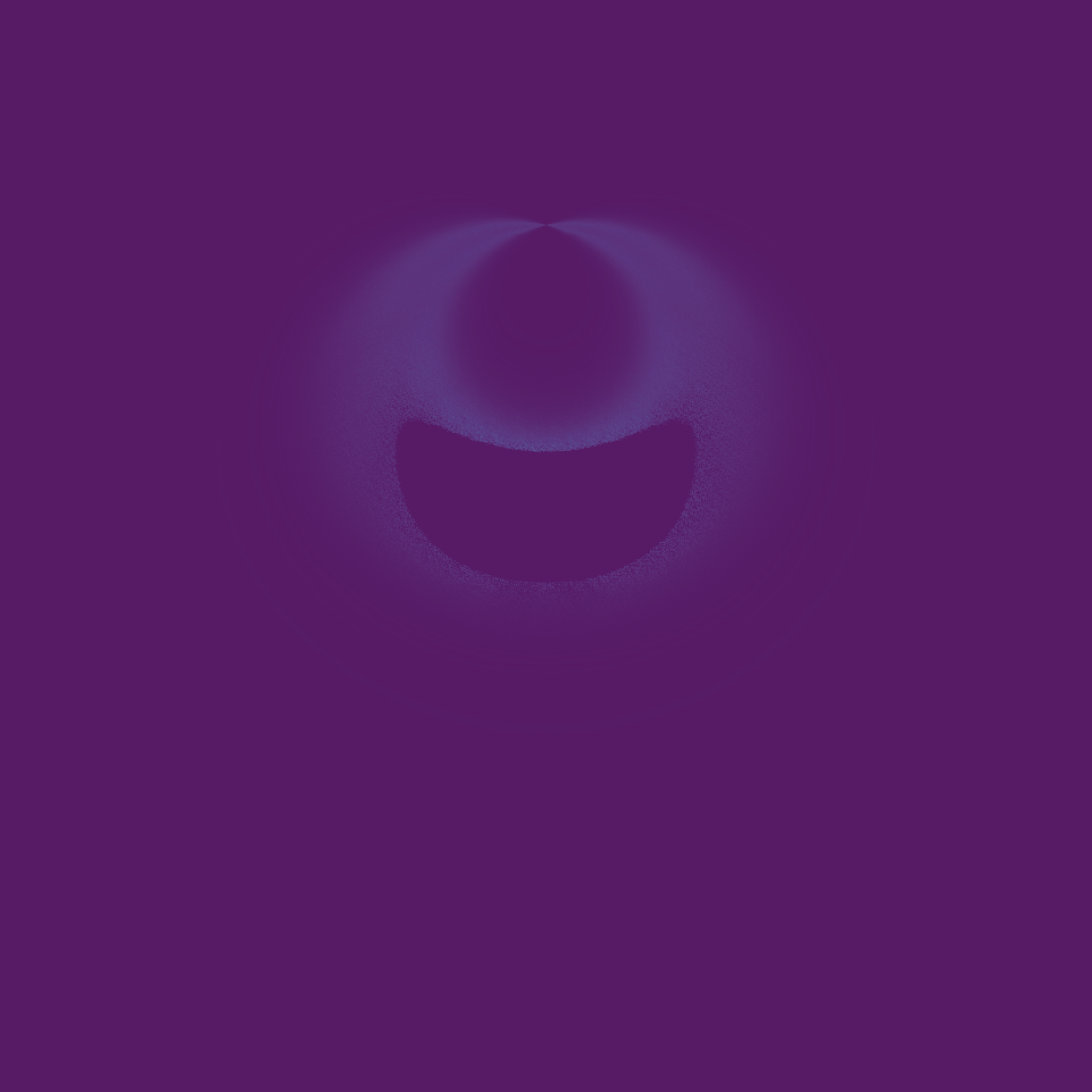}};
\draw (0.8, 1.0) node {$\alpha_x=0.05, \alpha_y=0.25$};
\end{tikzpicture}
&
\begin{tikzpicture}
\draw (0.0, 0.0) node {\includegraphics[width=0.2\linewidth]{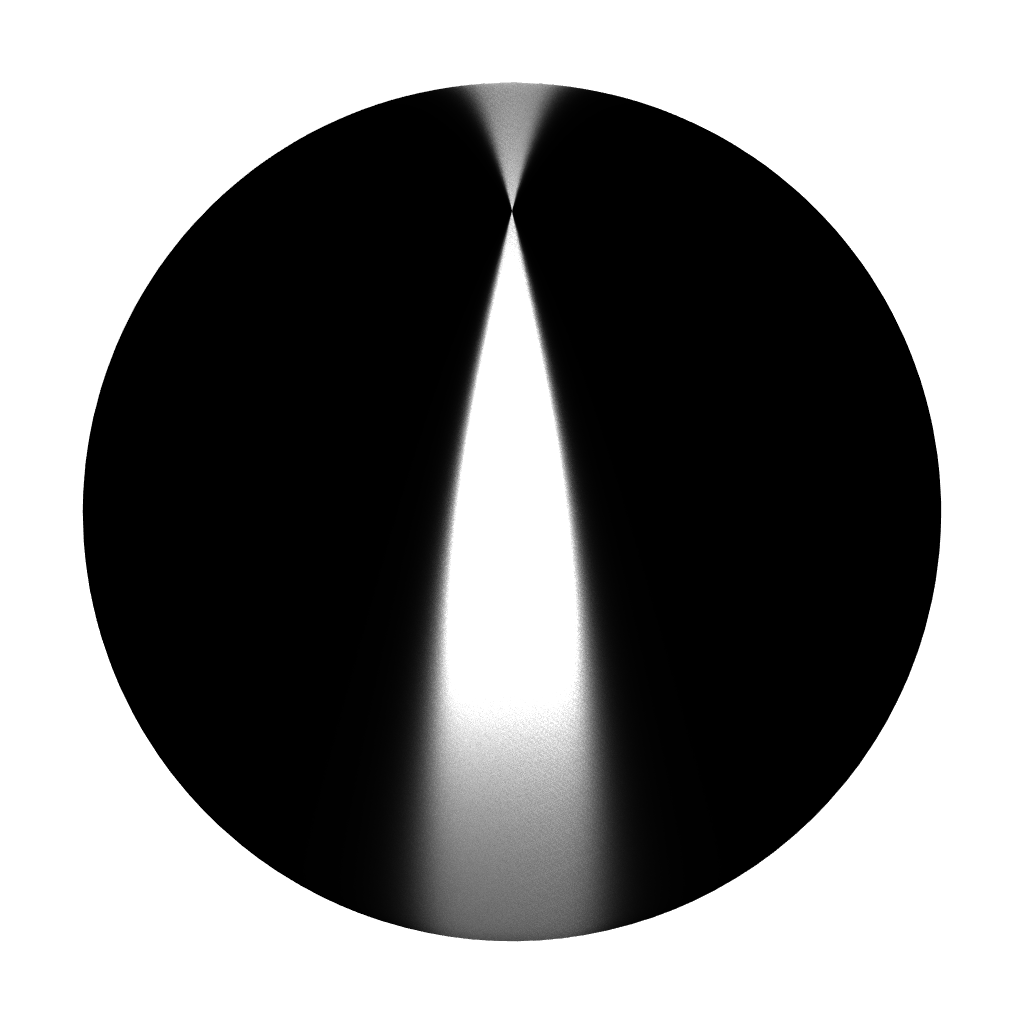}};
\draw (1.6, 0.0) node {\includegraphics[width=0.2\linewidth]{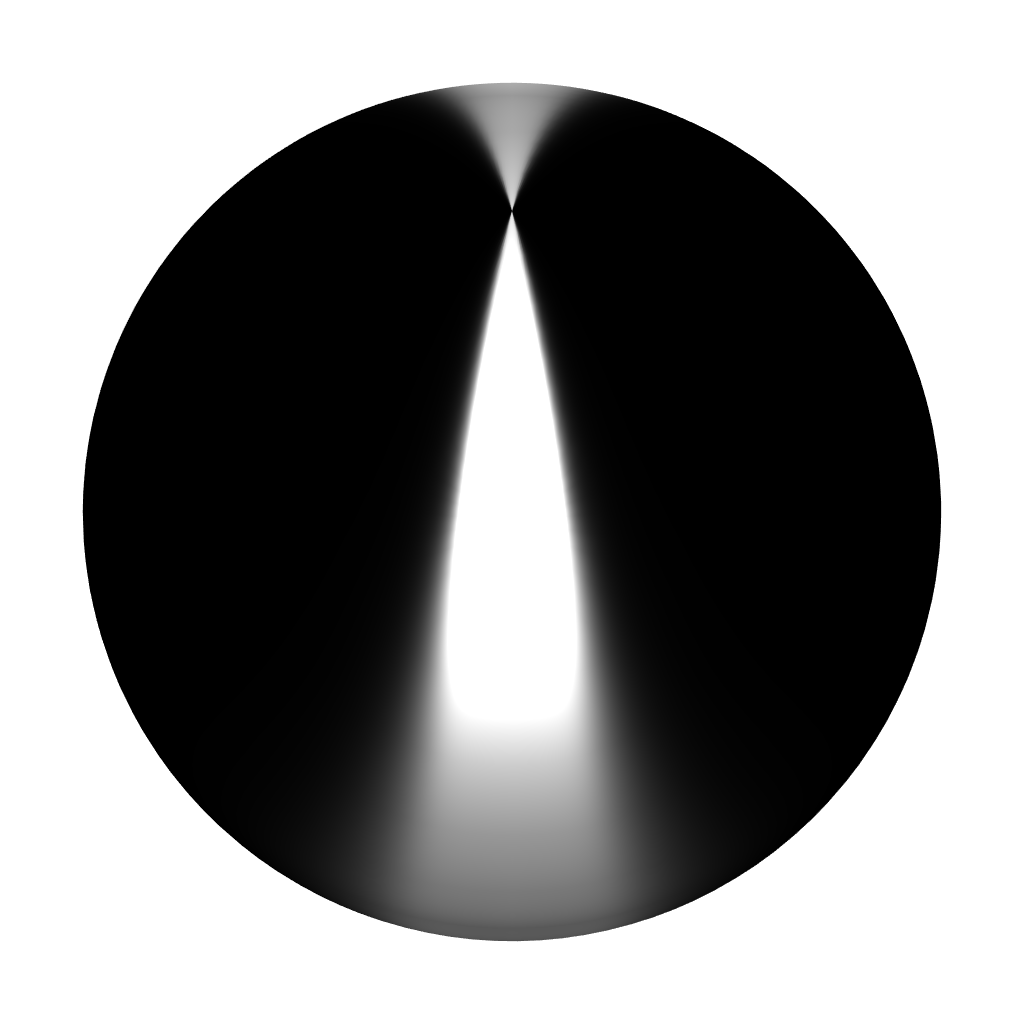}};
\draw (2.3, -0.7) node {\includegraphics[width=0.1\linewidth]{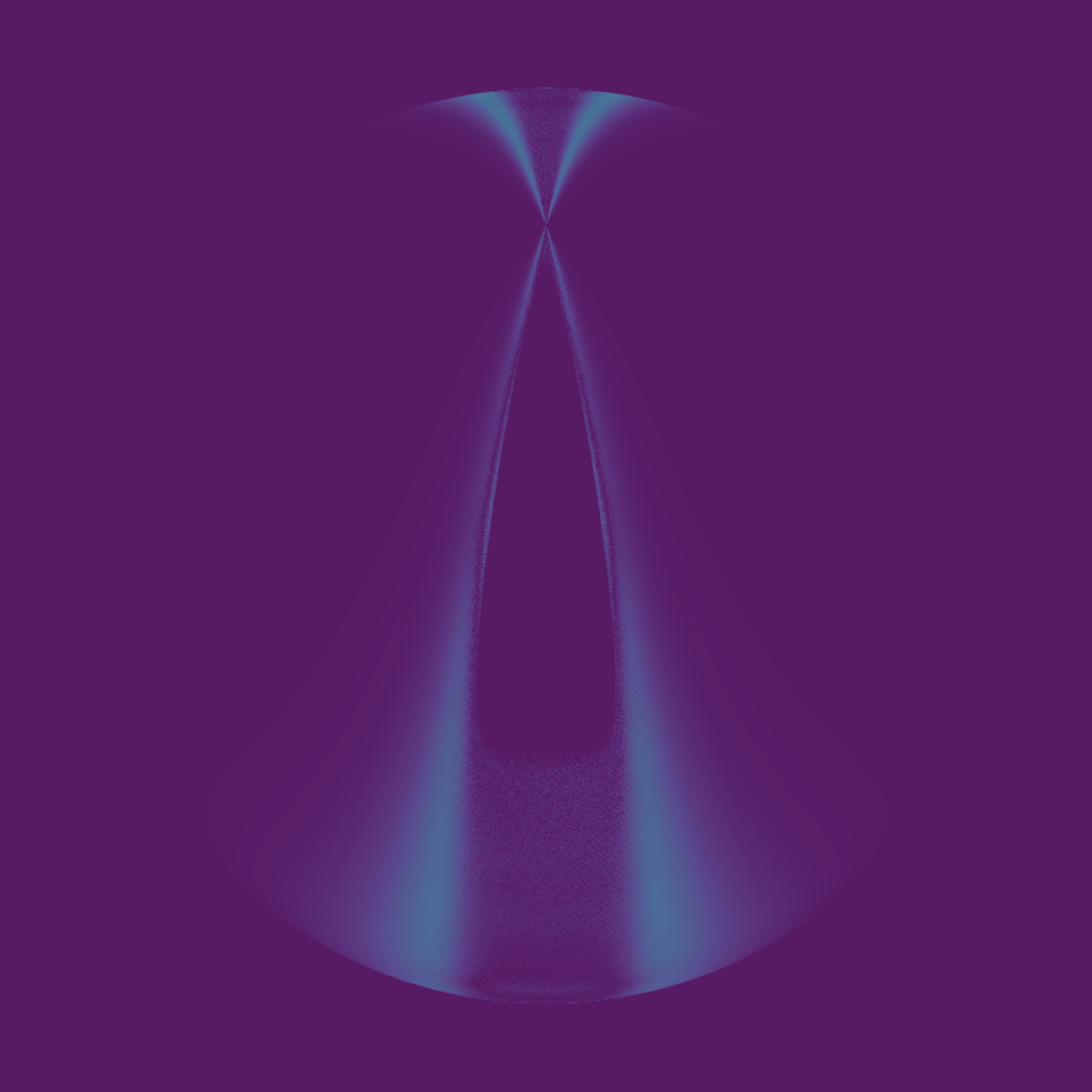}};
\draw (0.8, 1.0) node {$\alpha_x=0.50, \alpha_y=0.02$};
\end{tikzpicture}
\\
\begin{tikzpicture}
\draw (0.0, 0.0) node {\includegraphics[width=0.2\linewidth]{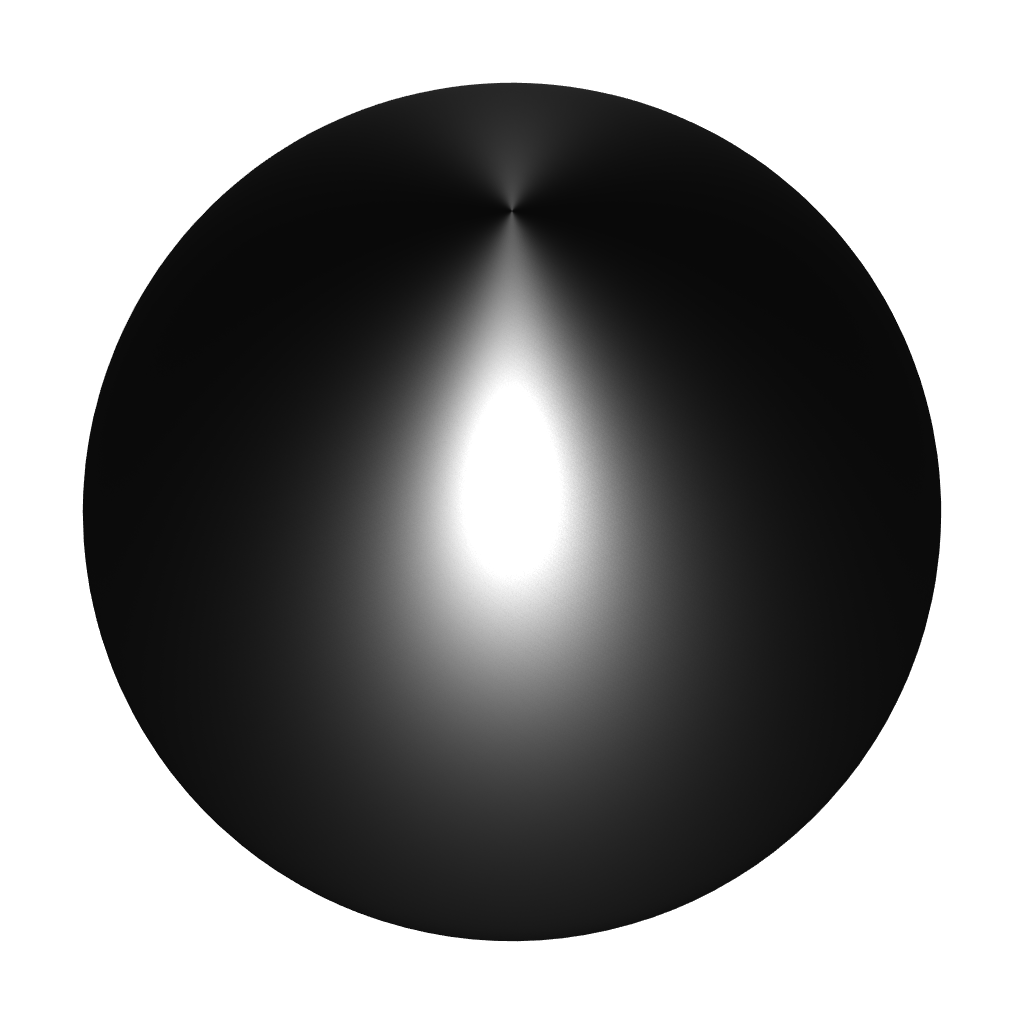}};
\draw (1.6, 0.0) node {\includegraphics[width=0.2\linewidth]{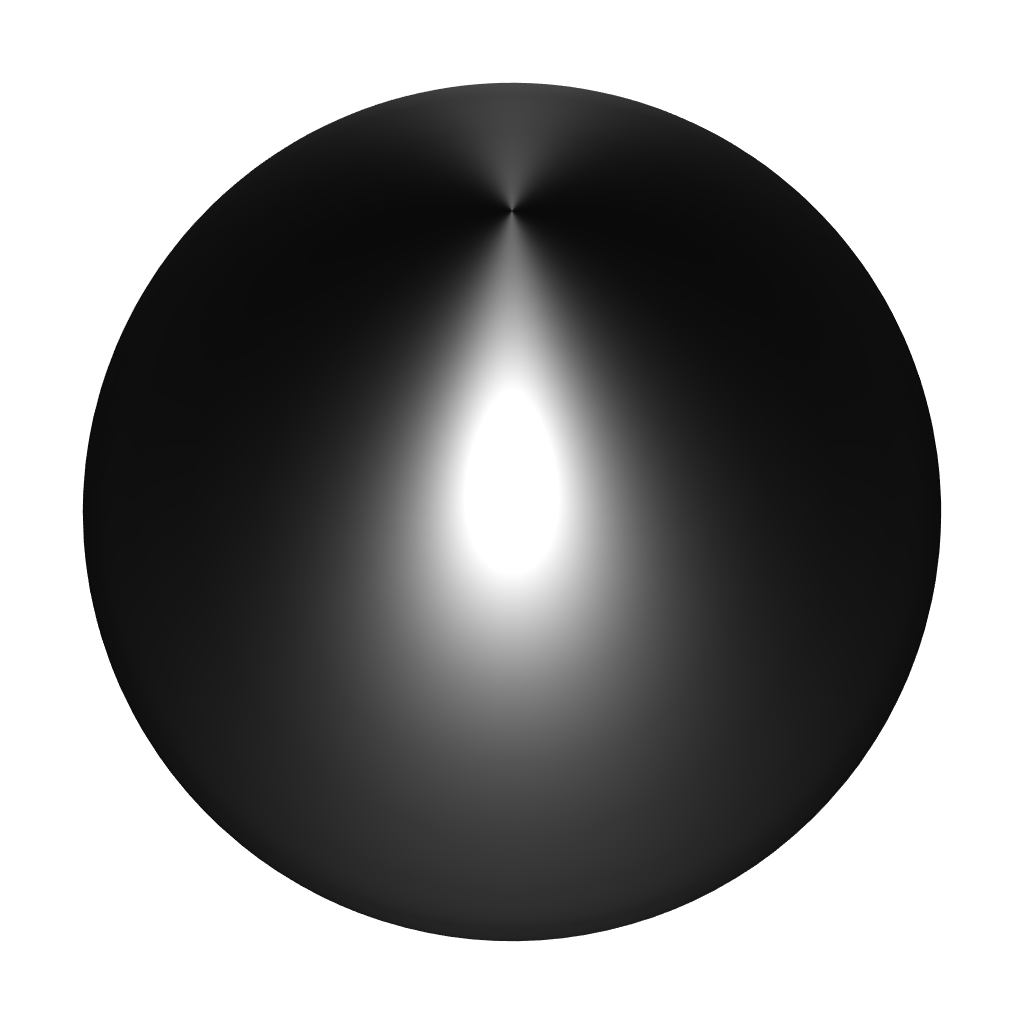}};
\draw (2.3, -0.7) node {\includegraphics[width=0.1\linewidth]{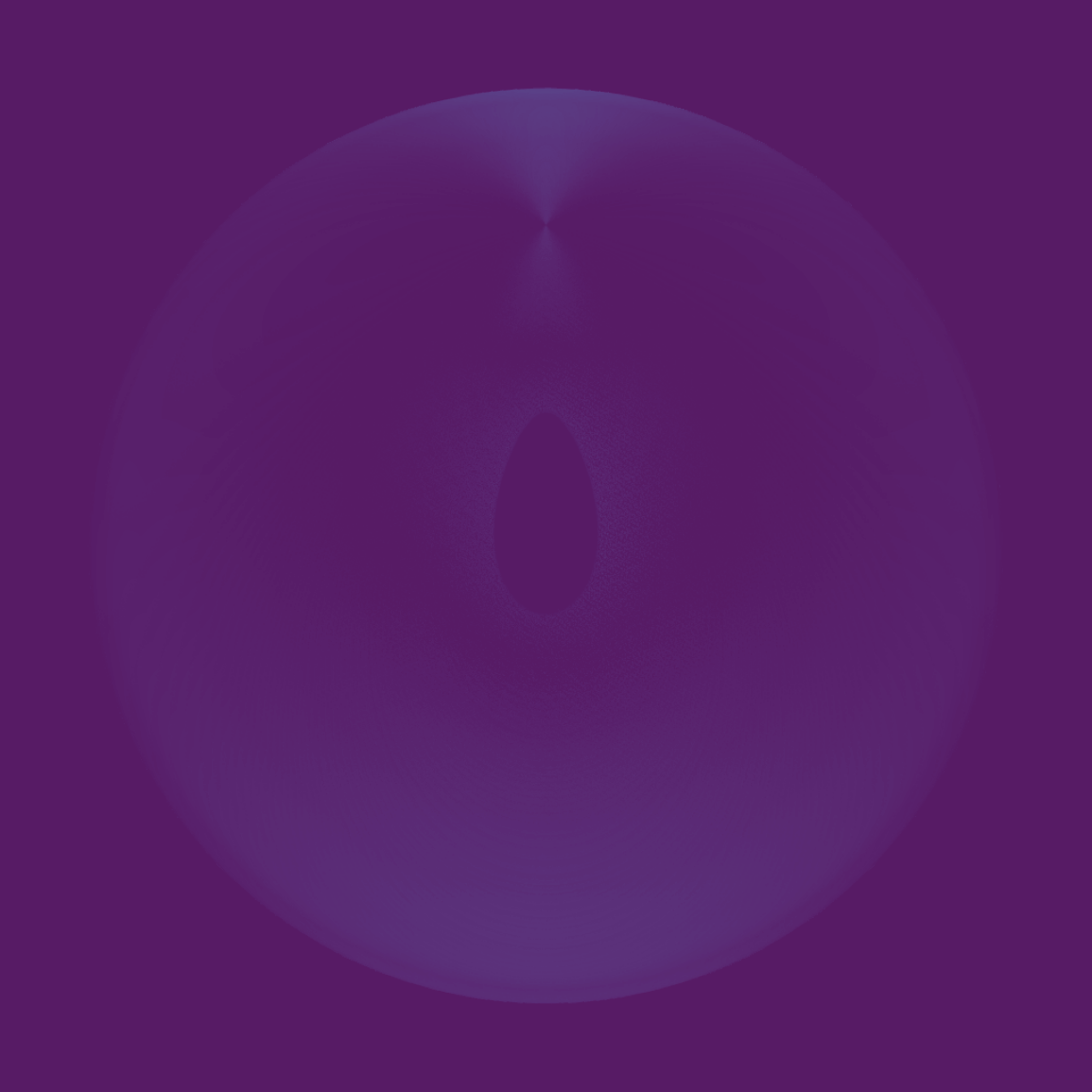}};
\draw (0.8, 1.0) node {$\alpha_x=0.50, \alpha_y=0.25$};
\end{tikzpicture}
&
\begin{tikzpicture}
\draw (0.0, 0.0) node {\includegraphics[width=0.2\linewidth]{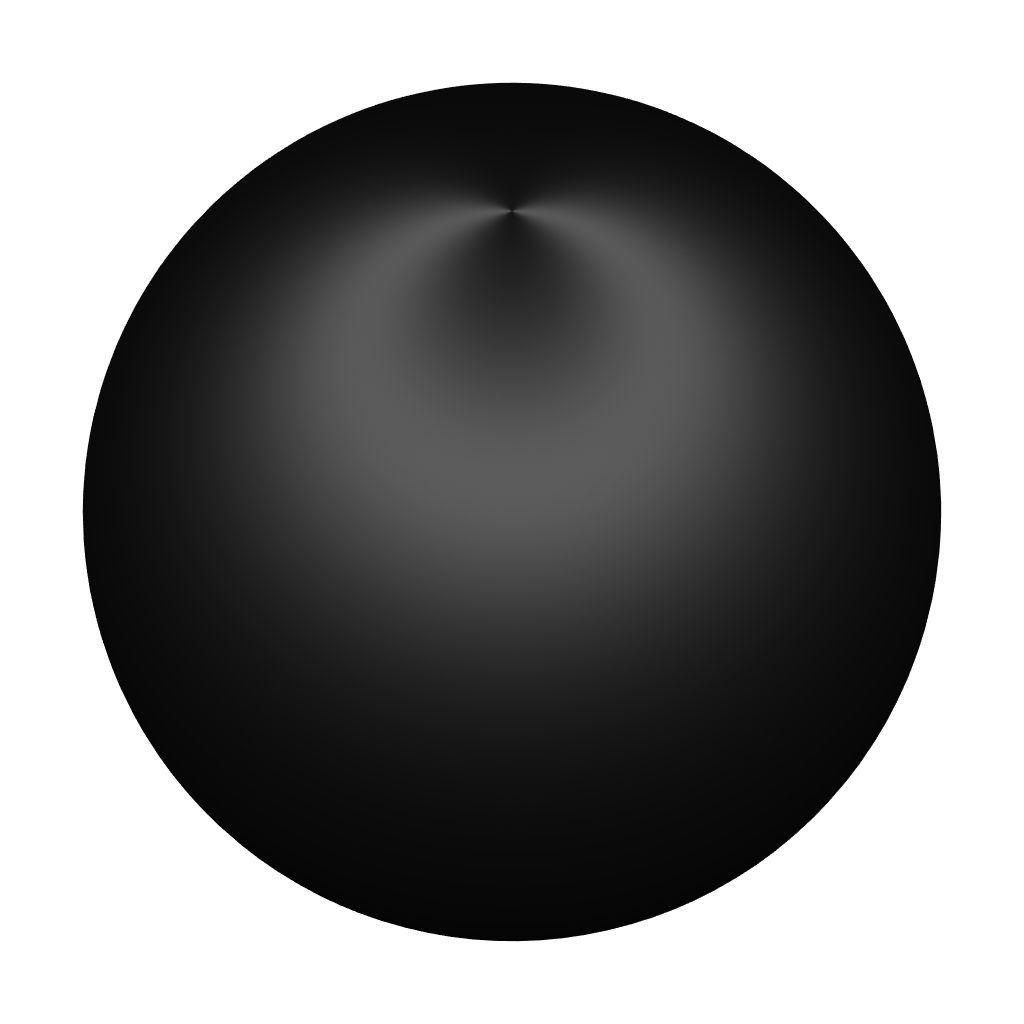}};
\draw (1.6, 0.0) node {\includegraphics[width=0.2\linewidth]{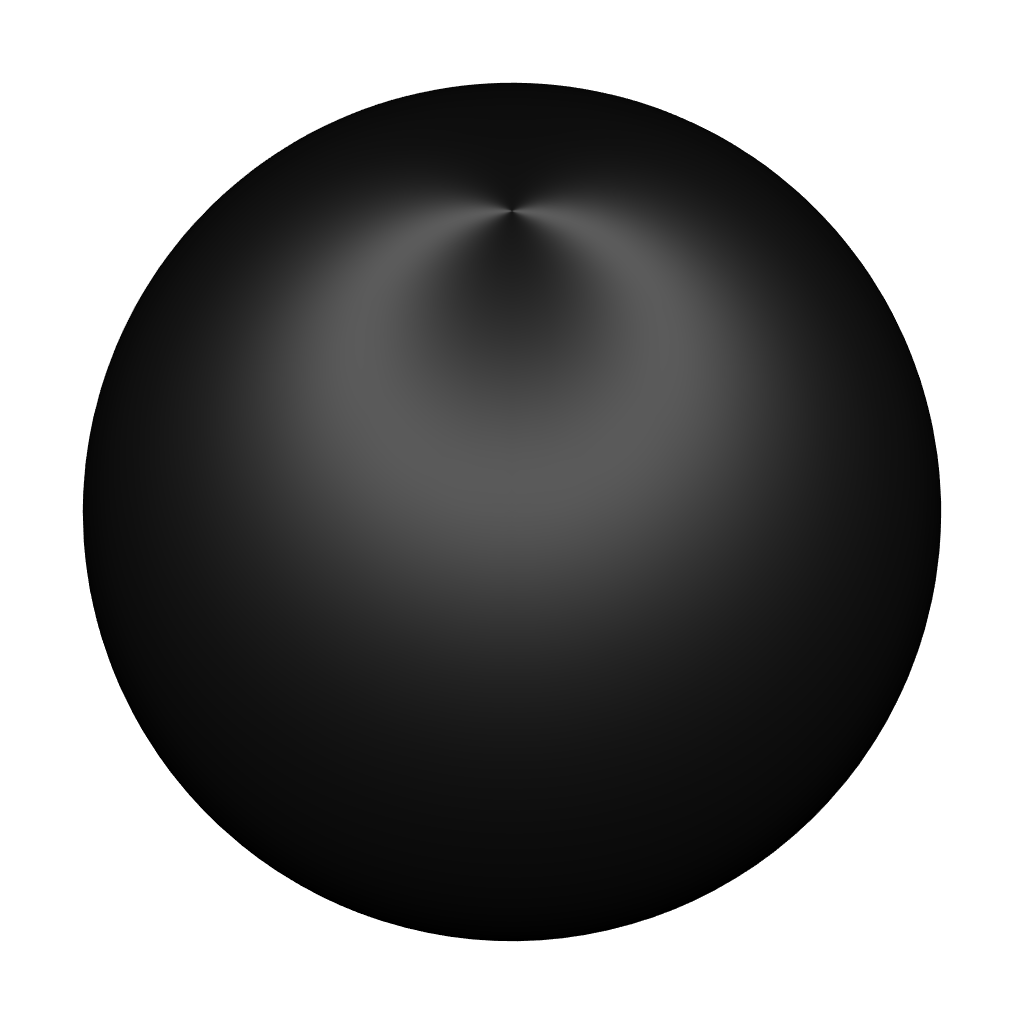}};
\draw (2.3, -0.7) node {\includegraphics[width=0.1\linewidth]{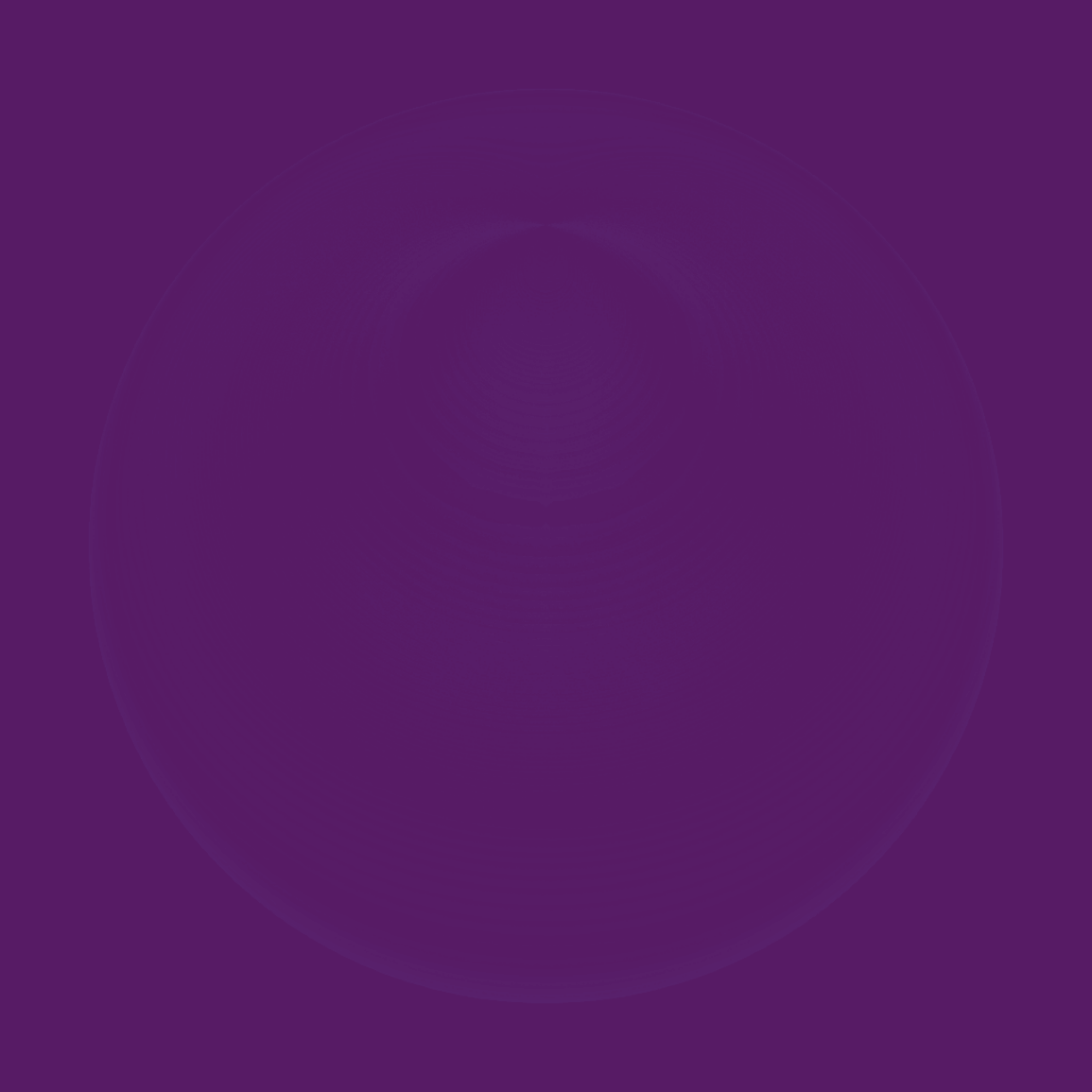}};
\draw (0.8, 1.0) node {$\alpha_x=0.50, \alpha_y=1.00$};
\end{tikzpicture}
\end{tabular}
&
\raisebox{-35mm}{\includegraphics[width=0.04\linewidth]{figures/lut_size_renders/colorbar.pdf}}
\end{tabular}
\vspace{-5mm}
\caption{\label{fig:results_renderings} Rendered spheres with anisotropic materials and rectangular lights. 
\textit{We show the GGX reference (left), our LTC approximation (right), and the difference image. 
More results in our supplemental material.}
}
\end{figure}

Figure~\ref{fig:results_renderings} shows the GGX reference and our LTC approximation. 
As expected from the plots, the approximation might have large errors compared to the reference but it remains plausible, and we did not find configurations where the result is visually unacceptable. Therefore, we believe that our approximation is good enough to be considered for non-predictive real-time rendering, but would discourage its use for more demanding applications.

%% file: section_conclusion.tex
\section{Conclusion}
\label{sec:conclusion}

We have proposed a method to bring Linearly Transformed Cosines to anisotropic GGX. 
It is the product of the experience we gained from many failed attempts.
New insights into the mathematical properties of LTCs, careful design choices, and attention to detail were all crucial elements in ensuring a clean and artifact-free approximation.
We believe that the proposed method provides a plausible approximation that passes the quality bar for game engines.
It has low memory overhead compared to the isotropic version already used by practitioners.
The 4D texture fetch is more expensive than the isotropic version but remains competitive for a real-time technique and can be amortized over multiple area lights. 
Thus, we don't see any barrier to using our anisotropic extension for video games.
Furthermore, the same methodology could be applicable to other anisotropic materials. 
To facilitate reproduction, we plan to release the PyTorch fitting code, the fitted table and a minimalistic OpenGL demo that shows how to use it. 

The main limitation is that our approximation is not accurate enough for all applications.
For instance, we do not advise using it for predictive rendering or as an importance sampling technique for anisotropic GGX.

A notable finding is that the limiting factor of our approximation is the representation power of LTCs, which cannot produce all of the possible GGX shapes, such as lunes.
Therefore, we believe our method reaches the limit of what is possible with LTC approximations of GGX BRDFs. Further improvements should thus be sought with a fundamentally different approach.